\documentclass[11pt,twoside,notitlepage]{article}
\usepackage{one-column}

\newcommand{\kf}[1]{{#1}}

\begin{document}

	%
	%

	\begin{center}
		{\textbf{
			\Large{Transport phenomena in electrolyte solutions:\\
			Non-equilibrium thermodynamics and statistical mechanics}
		}} \\
		\vspace{0.11in}

		{\small
			Kara D. Fong,$^{1,2}$\hyperlink{email1}{$^{\ddag}$} Helen K. Bergstrom,$^{1,2}$\hyperlink{email3}{$^{\dag}$} 
			Bryan D. McCloskey,$^{1,2}$\hyperlink{email4}{$^{*}$}
			and \\
			Kranthi K. Mandadapu$^{1,3}$\hyperlink{email2}{$^{\S}$} \\
		}
		\vspace{0.25in}

		\footnotesize{
			{
				$^1$
				Department of Chemical \& Biomolecular Engineering,
				University of California, Berkeley, CA 94720, USA
				\\
				$^2$
		        Energy Technologies Area, 
		        Lawrence Berkeley National Laboratory, Berkeley, CA 94720, USA
		        \\
				$^3$
				Chemical Sciences Division,
				Lawrence Berkeley National Laboratory, Berkeley, CA 94720, USA
				\\
			}
		}
	\end{center}

	\vspace{13pt}
	%
	%

	\begin{abstract}
		The theory of transport phenomena in multicomponent electrolyte solutions is presented here through the integration of continuum mechanics, electromagnetism, and non-equilibrium thermodynamics. The governing equations of irreversible thermodynamics, including balance laws, Maxwell's equations, internal entropy production, and linear laws relating the thermodynamic forces and fluxes, are derived. Green-Kubo relations for the transport coefficients connecting electrochemical potential gradients and diffusive fluxes are obtained in terms of the flux-flux time correlations. The relationship between the derived transport coefficients and those of the Stefan-Maxwell and infinitely dilute frameworks are presented, and the connection between the transport matrix and experimentally measurable quantities is described. To exemplify application of the derived Green-Kubo relations in molecular simulations, the matrix of transport coefficients for lithium and chloride ions in dimethyl sulfoxide is computed using classical molecular dynamics and compared with experimental measurements.
	\end{abstract}
	\vspace{15pt}


	\noindent\rule{4.6cm}{0.4pt}

	\small
	\noindent{\linkdest{email1}{\hspace{10pt}$^\ddag \,$}karafong\textit{@}berkeley.edu \\
	\linkdest{email3}{\hspace{10pt}$^\dag \,$}helen\_bergstrom\textit{@}berkeley.edu \\
	\linkdest{email4}{\hspace{10pt}$^* \,$}bmcclosk\textit{@}berkeley.edu \\
		\linkdest{email2}{\hspace{10pt}$^\S \,$}kranthi\textit{@}berkeley.edu 
	}

	\vspace{55pt}
	%
	%

	{ \hypersetup{linkcolor=black} \tableofcontents }
	\vspace{20pt}

	%
	%

	\section*{List of Important Symbols}

\begin{longtable}[l]{l l }
$\boldsymbol{b}_i$ & non-electromagnetic body force acting on species $i$\\
$\boldsymbol{B}$ & magnetic field \\
$c_i$ & concentration of species $i$ \\
$c_\mathrm{T}$ & total solution concentration \\
$D_i$ & self-diffusion coefficient of species $i$ \\
$\boldsymbol{D}$ & charge potential \\
$\boldsymbol{D}^{\mathrm{f}}$ & electric displacement\\
$D_{\mathrm{el}}$ & salt diffusion coefficient \\
$D^{ij}$ & Stefan-Maxwell (binary interaction) diffusion coefficients \\
$e$ & energy per unit mass, including kinetic and internal energy\\
$\boldsymbol{E}$ & electric field \\
$\boldsymbol{\mathcal{E}}$ & electromotive intensity \\
$\epsilon$ & dielectric constant \\
$\epsilon_0$ & vacuum permittivity \\
$\varepsilon_{ijk}$ & Levi-Civita symbol \\
$\eta$ & bulk viscosity coefficient \\
$f_{\mathrm{el}}$ & salt activity coefficient \\
$\tilde{f}$ & Helmholtz free energy per unit volume\\
$F$ & Faraday's constant\\
$\boldsymbol{F}$ & Lorentz force\\
$\mathcal{F}$ & Helmholtz free energy \\
$\boldsymbol{g}$ & electromagnetic momentum density \\
$\boldsymbol{H}$ & current potential \\
$\boldsymbol{\mathcal{H}}$ & magnetomotive intensity \\
$\boldsymbol{I}$ & identity tensor\\
$\boldsymbol{j}_i$ & mass flux of species $i$ \\
$\tilde{\boldsymbol{j}}$ & current density \\
$\tilde{\boldsymbol{j}}^{\mathrm{b}}$ & bound current density \\
$\tilde{\boldsymbol{j}}^{\mathrm{f}}$ & free current density \\
$\boldsymbol{J}_i$ & concentration flux of species $i$ \\
$\boldsymbol{J}_i^{\mathrm{s}}$ & concentration flux of species $i$ with respect to solvent velocity \\
$\boldsymbol{J}_{\mathrm{q}}$ & heat flux \\
$\boldsymbol{J}_{\mathrm{s}}$ & entropy flux \\
$\boldsymbol{J}_{\mathrm{el}}$ & salt flux \\
$\boldsymbol{\mathcal{J}}$ & conduction current density \\
$\boldsymbol{\mathcal{J}}^{\mathrm{b}}$ & bound conduction current density \\
$\boldsymbol{\mathcal{J}}^{\mathrm{f}}$ & free conduction current density \\
$\kappa$ & ionic conductivity \\
$k_{\mathrm{B}}$ & Boltzmann constant \\
$K^{ij}$ & Stefan-Maxwell transport coefficients\\
$\boldsymbol{K}_{\mathrm{CC}}$ & covariance matrix for concentration fluctuations\\
$L^{ij}$ & Onsager transport coefficients\\
$L^{{ij}^\mathrm{s}}$ & transport coefficients in solvent reference velocity framework\\
$\lambda$ & shear viscosity coefficient \\
$\boldsymbol{M}$ & magnetization \\
$M_i$ & molecular weight of species $i$  \\ 
$\boldsymbol{\mathcal{M}}$ & Lorentz magnetization \\
$\upmu_0$ & permeability \\ 
$\mu_i$ & chemical potential of species $i$ \\ 
$\overline{\mu}_i$ & electrochemical potential of species $i$  \\ 
$\mu_{\mathrm{el}}$ & salt chemical potential\\ 
$\boldsymbol{n}$ & unit normal vector  \\ 
$N_i$ & number of particles of type $i$ \\ 
$\nu_i$ & stoichiometric coefficient of ion $i$ in a salt\\ 
$\omega_i$ & mass fraction of species $i$ \\
$p$ & pressure \\
$\boldsymbol{P}$ & polarization \\
$\mathcal{P}$ & electrolyte volume under consideration  \\ 
$\partial\mathcal{P}$ & boundary of the electrolyte volume $\mathcal{P}$  \\ 
$\phi$ & electric potential  \\ 
$q$ & total charge density\\
$q^{\mathrm{b}}$ & bound charge density\\
$q^{\mathrm{f}}$ & free charge density\\
$\hat{q}^{\alpha}$ & charge of particle $\alpha$\\
$r$ & heat source or sink per unit mass \\
$\boldsymbol{r}^\alpha$ & position of particle $\alpha$\\
$R$ & ideal gas constant\\
$\rho$ & local mass density\\
$\rho_i$ & local mass density of species $i$\\
$s$ & entropy per unit mass\\
$\tilde{s}$ & entropy per unit volume\\
$\mathcal{S}$ & entropy \\
$\boldsymbol{S}$ & Poynting vector\\
$\sigma_\mathrm{e}$ & external entropy supply per unit mass\\
$\sigma_\mathrm{i}$ & internal entropy production per unit mass\\
$\boldsymbol{t}$ & surface force density\\
$t_i$ & transference number of species $i$ \\ 
$T$ & temperature\\
$\boldsymbol{T}$ & stress tensor from surface forces\\
$\boldsymbol{T}_{\mathrm{M}}$ & Maxwell stress tensor\\
$\bar{\boldsymbol{T}}$ & full stress tensor, including Maxwell stress\\
$u_i$ & electrophoretic mobility of species $i$ \\ 
$\tilde{u}$ & internal energy per unit volume\\
$\tilde{u}_{\mathrm{EM}}$ & electromagnetic energy per volume\\
$\boldsymbol{v}$ & mass-averaged (barycentric) velocity of the electrolyte\\
$\boldsymbol{v}^\alpha$ & velocity of particle $\alpha$\\
$\boldsymbol{v}_i$ & average velocity of all particles of type $i$\\
$V$ & volume\\
$\boldsymbol{x}$ & position within the electrolyte\\
$\boldsymbol{X}_i$ & thermodynamic force\\
$z_i$ & charge valency of species $i$\\
$\dot{(\;\;)}$ & substantial (material) time derivative \\
$\overset{*}{(\;\;)}$ & flux derivative
\end{longtable}

\vspace{25pt}

\section{Introduction}

In this text we present the governing equations for irreversible thermodynamics and transport in multicomponent electrolyte solutions. We use this framework to derive Green-Kubo relations for the transport coefficients in these solutions and contextualize these results relative to experimental measurements and other commonly used transport theories. Finally, we demonstrate application of these equations to compute transport coefficients in a model electrolyte.

Electrolyte solutions play a crucial role in a wide range of systems, with applications ranging from energy technologies such as batteries and fuel cells to biological, geological, and medical systems. Design and optimization of these systems is often contingent on a deep understanding and rigorous formulation of the transport phenomena governing the motion of charged species in solution. Despite nearly a century of progress since the pioneering work of Debye, H\"uckel, and Onsager\cite{debye1923lowering,onsager1926theory,onsager1927theory}, analytical models for predicting electrolyte transport properties remain elusive, particularly at non-dilute concentrations. The complexities induced by long-range electrostatic forces as well as short-range, specific chemical interactions create both conceptual as well as mathematical difficulties in working towards an all-encompassing theory for describing transport phenomena in electrolyte solutions. 

The most ubiquitous framework for understanding transport of ions in concentrated electrolytes is the Stefan-Maxwell equations for multicomponent diffusion\cite{birdtransport,krishna1997maxwell,Newman2004ElectrochemicalSystems}, originally derived from the kinetic theory of gases\cite{hirschfelder1964molecular}. These equations relate the gradient in electrochemical potential $\overline{\mu}_i$ of a species $i$ to the velocities $\boldsymbol{v}_j$ of each of the species $j$ in solution:
\begin{equation}\label{StefanMaxwell}
    c_i\boldsymbol{\nabla}\overline{\mu}_i = \sum_{j\neq i} K^{ij}(\boldsymbol{v}_j - \boldsymbol{v}_i)~,
\end{equation}
where $c_i$ is the concentration of species $i$ and $K^{ij}$ are the Stefan-Maxwell transport coefficients. These equations may be interpreted as a force balance: the thermodynamic force acting on species $i$ (the left side of the equation) is balanced by the frictional forces between species $i$ and each of the other species in solution. It is assumed that this frictional force is proportional to the difference in velocities of the two species. The Stefan-Maxwell transport coefficients may also be expressed in terms of binary interaction diffusion coefficients $D^{ij}$ as 
\begin{equation}
    K^{ij} = \frac{RTc_ic_j}{c_{\mathrm{T}} D^{ij}}~,
\end{equation}
where $R$ is the ideal gas constant, $T$ is temperature, and $c_{\mathrm{T}}$ is the total concentration of the system.\footnote{Throughout this text, the same symbol $c_i$ will be used to denote concentration as both number per volume and mole per volume. It is implied that concentration is in units of mole per volume when appearing with the quantity $RT$, and in units of number per volume when appearing with $k_{\mathrm{B}}T$ where $k_{\mathrm{B}}$ is the Boltzmann constant.}

Alternatively, transport in electrolyte solutions can be analyzed based on the classical theories of thermodynamics of irreversible processes developed by Onsager\cite{onsager_reciprocal_1,onsager_reciprocal_2}, Prigogine\cite{Prigogine1967IntroductionProcesses} and de Groot and Mazur\cite{DeGroot1969Non-EquilibriumThermodynamics}. This framework uses the rate of internal entropy production (dissipation) to relate thermodynamic driving forces and corresponding fluxes with a matrix of transport coefficients:
\begin{equation}
   \boldsymbol{J}_i = \sum_{j} \boldsymbol{L}^{ij} \boldsymbol{X}_j~,
 \end{equation}
 where $\boldsymbol{J}_i$ is the flux of species $i$, $\boldsymbol{X}_j$ is the thermodynamic driving force acting on species $j$, and $\boldsymbol{L}^{ij}$ are the Onsager transport coefficients. The forms of $\boldsymbol{J}_i$ and $\boldsymbol{X}_j$ will be derived herein.

Although both frameworks are consistent with thermodynamics and have been shown to effectively model electrolyte transport, the less common Onsager transport framework possesses several advantages over the Stefan-Maxwell equations. Unlike the Stefan-Maxwell transport coefficients $K^{ij}$ (or $D^{ij}$), the Onsager transport coefficients $L^{ij}$ can be computed directly from molecular simulations using Green-Kubo relations\cite{green1954markoff,kubo1957statistical} (which will be derived herein for multicomponent electrolyte solutions). This allows facile computation of all transport properties even in complex solutions which are challenging to characterize experimentally, such as those with multiple salt species. The Green-Kubo relations also allow direct physical interpretation of $L^{ij}$ as the extent of correlation between the motion of species $i$ and $j$. The Stefan-Maxwell coefficients, $D^{ij}$, however, have a less intuitive meaning and can even diverge to positive or negative infinity under certain conditions\cite{villaluenga2018negative}, making interpretation of transport phenomena challenging. Furthermore, the Onsager transport coefficients can be used directly to solve boundary value problems and obtain concentration profiles in an electrochemical system, while the Stefan-Maxwell transport matrix $\boldsymbol{K}$ must be inverted in order to be used in this manner.

Based on these advantages, we argue that the Onsager transport equations could become a simple and useful framework to study electrolyte transport, as is already the case in other fields such as in the study of membrane permeability\cite{kedem1961physical,kedem1958thermodynamic}. Rigorous formulation of the Onsager transport equations for electrolytes, however, requires integration of the principles of continuum mechanics, electromagnetism, and non-equilibrium thermodynamics. We are unaware of any work which has developed the underlying irreversible thermodynamics in their entirety rather than addressing special limiting cases. The classic texts of de Groot and Mazur\cite{DeGroot1969Non-EquilibriumThermodynamics}; Prigogine\cite{Prigogine1967IntroductionProcesses}; and Hirschfelder, Curtiss, and Bird\cite{hirschfelder1964molecular} each formulate balance laws and the corresponding forces/fluxes in fluid mixtures but do not consider the impact of external electromagnetic fields. \kf{Katchalsky and Curran\cite{katchalsky1965nonequilibrium} and Kjelstrup and Bedeaux\cite{kjelstrup2020non} both present a theory for irreversible processes which accounts for electrostatic work but do not include momentum conservation, yielding an incomplete picture of entropy production in electrolytes. The classic text of Newman and Thomas-Alyea\cite{Newman2004ElectrochemicalSystems} and related works\cite{newman1967transport,bernardi1985general} introduce Stefan-Maxwell and Onsager-like transport equations for electrolytes, albeit without discussion of the momentum, energy, and entropy balances of continuous media upon which these equations are built.} Kovetz\cite{kovetz2000electromagnetic} formulates rigorous balance laws in the presence of electromagnetic fields but does not consider multicomponent systems. None of the aforementioned existing works address problems which require coupling of electromagnetic effects, momentum transport, and multicomponent diffusion.

Herein, we work to develop a more complete theory of electrolyte transport via the following aims:
\begin{enumerate}
\item Integrate the classical frameworks of continuum mechanics and linear irreversible thermodynamics with the theory of electromagnetism to formulate balance laws for electrolyte solutions. Derive the form of the rate of internal entropy production and associated transport laws for electrolyte solutions.
\item Formulate expressions for thermodynamic potentials in systems subject to an electric field, elucidating the role of the electrochemical potential in these thermodynamic relations.
\item Provide rigorous derivation of the Green-Kubo relations for the Onsager transport coefficient matrix.
\item Explicitly relate the Onsager transport coefficients to those of the Stefan-Maxwell and dilute solution equations and to experimentally relevant quantities.
\item Demonstrate the use of the Onsager transport equations and Green-Kubo relations for complete characterization of a simple electrolyte using molecular simulations.
\end{enumerate}

This text is organized as follows. In Sec. \ref{sec:thermo}, we use Maxwell's equations and balance laws of mass, momentum, energy, and entropy to describe the non-equilibrium thermodynamics of electrolyte solutions. In Sec. \ref{sec:linear_irr_thermo}, we simplify these balance laws using linear constitutive relations and introduce the diffusive transport coefficients $L^{ij}$ relating electrochemical potential gradients and diffusive fluxes within the system. In Sec. \ref{sec:statMech}, we build upon the results of Secs. \ref{sec:thermo} and \ref{sec:linear_irr_thermo} to derive the Green-Kubo relations for $L^{ij}$. Next, in Secs. \ref{sec:relating_frameworks} and \ref{infinite_dilution} we relate our derived transport expressions to other commonly used frameworks for analyzing electrolyte transport, including the Stefan-Maxwell equations and the Nernst-Planck equation for infinitely dilute solutions. In Sec. \ref{sec:experiment} we demonstrate the connection between $L^{ij}$ and experimentally-relevant bulk transport properties, namely ionic conductivity, electrophoretic mobility, transference number, and salt diffusion coefficient. Finally, in Sec. \ref{sec:MD} we present results from classical molecular dynamics simulations of a model electrolyte (lithium chloride in dimethyl sulfoxide), in which we use our derived Green-Kubo relations to calculate $L^{ij}$ and show how these transport coefficients can be used to generate a variety of experimental properties of interest for an electrolyte. Additional derivations detailing the effect of electric fields on the formulation and usage of thermodynamic potentials are presented in Appendices \ref{appendix:thermo_potentials} through \ref{appendix:GibbsDuhem}, and methods are presented in Appendices \ref{appendix:Sim_Methods} and \ref{appendix:exp_Methods}.

\section{\label{sec:thermo}Non-equilibrium thermodynamics}

In this section, we derive the governing equations of irreversible thermodynamics in electrolyte solutions. The theories here are built upon on the work of Onsager\cite{onsager_reciprocal_1,onsager_reciprocal_2}, Prigogine\cite{Prigogine1967IntroductionProcesses}, de Groot and Mazur\cite{DeGroot1969Non-EquilibriumThermodynamics}, Katchalsky and Curran\cite{katchalsky1965nonequilibrium}, and Kovetz\cite{kovetz2000electromagnetic}. These classical works are extended to simultaneously describe the phenomena of electromagnetism (Maxwell's equations) and transport of mass, linear and angular momentum, and energy. The resulting balance laws are generally applicable for electrolytes with an arbitrary number of components, without assuming electroneutrality. We then invoke the second law of thermodynamics to analyze entropy production, enabling the formulation of linear laws relating the thermodynamic forces and fluxes within the electrolyte.

\subsection{\label{sec:MassBalance}Mass balance}
The mass balance for electrolyte solutions is identical to that of mixtures of uncharged species. Consider a volume $\mathcal{P}$ with a local density $\rho(\boldsymbol{x},t)$ at any position $\boldsymbol{x}$ and time $t$, defined as the mass per unit volume. Let $\rho_i(\boldsymbol{x},t)$ be the mass of species $i$ per unit volume, such that $\rho\coloneqq \sum_{i} \rho_i$. The linear momentum density of species $i$ is given by $\rho_i \boldsymbol{v}_i$, where $\boldsymbol{v}_i$ is the velocity of species $i$. Let us define the total momentum density at any point in $\mathcal{P}$ as $\rho \boldsymbol{v} \coloneqq \sum_i \rho_i \boldsymbol{v}_i$. The quantity $\boldsymbol{v}=\sum_i \frac{\rho_i}{\rho} \boldsymbol{v}_i$ is the mass-averaged velocity of all species. Utilizing these definitions, the rate of change in total mass of species $i$ in $\mathcal{P}$ is equivalent to the flux of $i$ in and out of the surface area of $\mathcal{P}$, denoted as $\partial\mathcal{P}$. This yields the global form of the balance of mass:
\begin{equation}\label{eq:global_mb}
    \frac{d}{dt}\int_\mathcal{P} \rho_i dv = - \int_{\partial\mathcal{P}} \rho_i (\boldsymbol{v}_i - \boldsymbol{v})\cdot \boldsymbol{n} da~,
\end{equation}
where $\boldsymbol{n}$ is the outward normal of surface $\partial\mathcal{P}$. Based on Eq. \eqref{eq:global_mb}, let the diffusive flux of species $i$ be defined as $\boldsymbol{j}_i\coloneqq\rho_i(\boldsymbol{v}_i-\boldsymbol{v})$, such that
\begin{equation}\label{eq:global_mb_2}
    \frac{d}{dt}\int_\mathcal{P} \rho_i dv = - \int_{\partial\mathcal{P}} \boldsymbol{j}_i\cdot \boldsymbol{n} da~.
\end{equation}
Note that $\sum_i \boldsymbol{j}_i = 0$. Applying the Reynolds transport theorem and the divergence theorem, we obtain
\begin{equation}
    \int_\mathcal{P}(\dot{\rho}_i + \rho_i \boldsymbol{\nabla}\cdot \boldsymbol{v}) dv = -\int_\mathcal{P}\boldsymbol{\nabla}\cdot\boldsymbol{j}_i dv~,
\end{equation}
where the notation $\dot{(\;\;)}$ refers to the substantial or material derivative, $\dot{(\;\;)} = \frac{d}{dt}(\;\;) = \frac{\partial}{\partial t}(\;\;) + \boldsymbol{\nabla}(\;\;)\cdot \boldsymbol{v}$. Further simplification using the localization theorem gives the local form of the species mass balance as 
\begin{equation}\label{species_mass_balance}
\dot{\rho}_i + \rho_i \boldsymbol{\nabla}\cdot\boldsymbol{v} = -\boldsymbol{\nabla}\cdot\boldsymbol{j}_i~.
\end{equation}

Alternatively, Eq. \eqref{species_mass_balance} can be expressed in terms of the concentration of species $i$, $c_i \coloneqq \rho_i/M_i$ where $M_i$ is the molecular weight of species $i$, as
\begin{equation}\label{mass_balance_conc}
    \dot{c}_i + c_i \boldsymbol{\nabla}\cdot\boldsymbol{v} =  -\boldsymbol{\nabla}\cdot(c_i(\boldsymbol{v}_i -  \boldsymbol{v}))= -\boldsymbol{\nabla}\cdot \boldsymbol{J}_i~,
\end{equation}
where $\boldsymbol{J}_i \coloneqq c_i(\boldsymbol{v}_i -  \boldsymbol{v})$. Note that $\boldsymbol{J}_i$ and $\boldsymbol{j}_i$ are simply related by a factor of $M_i$, i.e., $M_i \boldsymbol{J}_i = \boldsymbol{j}_i$.

The species mass balance can be used to obtain the total mass balance of the electrolyte. Summing Eq. \eqref{species_mass_balance} over all species and invoking the relation $\sum_i \boldsymbol{j}_i = 0 $, we obtain
\begin{equation}\label{mass_balance_total}
\dot{\rho} + \rho \boldsymbol{\nabla}\cdot\boldsymbol{v} = 0~.
\end{equation}
For incompressible systems, i.e., constant density, the mass balance leads to
\begin{equation}
\boldsymbol{\nabla}\cdot\boldsymbol{v} = 0~.
\end{equation}

\subsection{Charge balance and Maxwell's equations}

We now review the fundamentals of electromagnetism, generally following the philosophy of Kovetz\cite{kovetz2000electromagnetic}. While most electrolyte applications will involve electroneutral systems, linear dielectrics, and no magnetic effects, in this and the following sections we consider the most general case of both electric and magnetic fields in a non-electroneutral dielectric with arbitrary polarization and magnetization. This general theory enables us to treat more complex electrolyte systems and allows a deeper understanding of the underlying assumptions invoked when we do consider more conventional systems.

Much of electromagnetism is based on the key assumption that electric charge is conserved, i.e., the charge contained in a volume $\mathcal{P}$ changes only via flux of charges through the surface of the volume, $\partial\mathcal{P}$. The charge conservation law can be expresed mathematically as
\begin{equation}
    \int_\mathcal{P} \frac{\partial q}{\partial t}  dv = - \int_\mathcal{\partial P} \tilde{\boldsymbol{j}}\cdot \boldsymbol{n} da~,
\end{equation}
where $q$ is the total charge density and $\tilde{\boldsymbol{j}}\cdot \boldsymbol{n} da$ is the total amount of charge passing through the area element $da$ in the direction of $\boldsymbol{n}$ per unit time. The quantity $\tilde{\boldsymbol{j}}$ is called the current density. Alternatively, this charge balance can be written in terms of the substantial derivative of $q$ as 
\begin{equation}
    \frac{d}{dt}\int_\mathcal{P} q dv = - \int_\mathcal{P} \boldsymbol{\nabla}\cdot (\tilde{\boldsymbol{j}} - q \boldsymbol{v}) dv = - \int_\mathcal{P} \boldsymbol{\nabla}\cdot \boldsymbol{\mathcal{J}} dv
\end{equation}
where $\boldsymbol{\mathcal{J}} \coloneqq \tilde{\boldsymbol{j}} - q \boldsymbol{v}$ is the conduction current density. The corresponding local form of the charge balance law is
\begin{equation}
    \dot{q} + q \boldsymbol{\nabla}\cdot \boldsymbol{v}= - \boldsymbol{\nabla}\cdot \boldsymbol{\mathcal{J}} ~.
\end{equation}

The principle of charge conservation and its invariance to coordinate transformations in four-dimensional space-time motivates the first pair of Maxwell's equations,
\begin{equation}\label{Gauss_law}
    q=\boldsymbol{\nabla}\cdot\boldsymbol{D}~,
\end{equation}
and
\begin{equation}\label{ampere_law}
    \tilde{\boldsymbol{j}} = \boldsymbol{\nabla}\times\boldsymbol{H} - \frac{\partial\boldsymbol{D}}{\partial t}~,
\end{equation}
where $\boldsymbol{D}$ and $\boldsymbol{H}$ are the charge and current potentials, respectively\cite{kovetz2000electromagnetic}.

The second pair of Maxwell's equations are formulated by assuming the existence of two vector fields, the electric field $\boldsymbol{E}$ and magnetic field $\boldsymbol{B}$, which obey the relations
\begin{equation}\label{gauss_magnetism}
    \boldsymbol{\nabla}\cdot\boldsymbol{B} = 0~
\end{equation}
and
\begin{equation}\label{faraday_law}
   \frac{\partial\boldsymbol{B}}{\partial t} = -\boldsymbol{\nabla}\times\boldsymbol{E}~.
\end{equation}
As the electromagnetic field is conservative, $\boldsymbol{E}$ and $\boldsymbol{B}$ can be written in terms of electric and magnetic potentials denoted as $\phi$ and $\boldsymbol{A}$, respectively, as 
\begin{equation}\label{potentials}
\begin{split}
    &\boldsymbol{B} = \boldsymbol{\nabla}\times\boldsymbol{A}~,\\
    &\boldsymbol{E} = - \frac{\partial \boldsymbol{A}}{\partial t} - \boldsymbol{\nabla}\phi~.
\end{split}
\end{equation}
For a system with no magnetic field, as is often the case in physically relevant electrolyte applications, we may simply write $\boldsymbol{E} = - \boldsymbol{\nabla}\phi$.

The two pairs of Maxwell's equations are related by the aether constitutive relations\cite{kovetz2000electromagnetic}, 
\begin{equation}\label{aether_1}
    \boldsymbol{D} =\epsilon_0 \boldsymbol{E}~,
\end{equation}
where $\epsilon_0$ is the vacuum permittivity, and
\begin{equation}\label{aether_2}
    \boldsymbol{B} =\upmu_0 \boldsymbol{H}~,
\end{equation}
where $\upmu_0$ is the permeability.

The quantities $\tilde{\boldsymbol{j}}$, $\boldsymbol{E}$, and $\boldsymbol{H}$ depend on the choice of reference frame. In merging the theory of electromagnetism with continuum mechanics, it is convenient to re-cast Maxwell's equations in terms of quantities that are invariant under Galilean transformations (note that the charge, charge potential, magnetic field, and conduction current density are Galilean invariants). For a material with velocity $\boldsymbol{v}$, we can define the Galilean invariants $\boldsymbol{\mathcal{E}}$, called the electromotive intensity, and $\boldsymbol{\mathcal{H}}$, the magnetomotive intensity, as
\begin{equation}\label{e_field_invariant}
    \boldsymbol{\mathcal{E}} = \boldsymbol{E} + \boldsymbol{v} \times \boldsymbol{B}
\end{equation}
and
\begin{equation}\label{m_field_invariant}
    \boldsymbol{\mathcal{H}} = \boldsymbol{H} - \boldsymbol{v} \times \boldsymbol{D}~.
\end{equation}
In terms of these Galilean invariants, Eq. \eqref{ampere_law} and Eq. \eqref{faraday_law} become
\begin{equation}
    \boldsymbol{\mathcal{J}} = \boldsymbol{\nabla}\times\boldsymbol{\mathcal{H}}- \overset{*}{\boldsymbol{D}}
\end{equation}
 and
\begin{equation}\label{faraday_invariant}
    \overset{*}{\boldsymbol{B}} = -\boldsymbol{\nabla}\times\boldsymbol{\mathcal{E}}
\end{equation}
respectively, where we use the notation $\overset{*}{\boldsymbol{A}}$ to denote the flux derivative\cite{kovetz2000electromagnetic}, i.e. $\overset{*}{\boldsymbol{A}} = \frac{\partial\boldsymbol{A}}{\partial t} + (\boldsymbol{\nabla}\cdot \boldsymbol{A})\boldsymbol{v}-\boldsymbol{\nabla}\times(\boldsymbol{v}\times\boldsymbol{A})$. 

All of the electromagnetism equations introduced thus far provide a microscopic picture of the system by accounting for the charge of each individual particle comprising the body. However, when considering charges in a dielectric medium rather than in vacuum, it is typically more convenient to decompose the total charge density of the system $q$ into the free charge density ($q^{\mathrm{f}}$) and bound charge density ($q^{\mathrm{b}}$). This decomposition yields
\begin{equation}\label{total_free_bound_charge}
    q = q^{\mathrm{f}} + q^{\mathrm{b}}
\end{equation}
and correspondingly
\begin{equation}\label{total_current}
    \tilde{\boldsymbol{j}} = \tilde{\boldsymbol{j}}^{\mathrm{f}} + \tilde{\boldsymbol{j}}^{\mathrm{b}}~.
\end{equation}
Equation~\eqref{total_current} leads naturally to the quantities $\boldsymbol{\mathcal{J}}^{\mathrm{b}} = \tilde{\boldsymbol{j}}^{\mathrm{b}} - q^{\mathrm{b}} \boldsymbol{v}$ and $\boldsymbol{\mathcal{J}}^{\mathrm{f}} = \tilde{\boldsymbol{j}}^{\mathrm{f}} - q^{\mathrm{f}} \boldsymbol{v}$. In an electrolyte, free charges correspond to mobile ions in solution, while bound charges are those of the solvent molecules comprising the dielectric medium, which result in polarization $\boldsymbol{P}$ and magnetization $\boldsymbol{M}$. The quantities $\boldsymbol{P}$ and $\boldsymbol{M}$ are related to the bound charge and current density by
\begin{equation}
    q^{\mathrm{b}} = -\boldsymbol{\nabla}\cdot\boldsymbol{P}
\end{equation}
and
\begin{equation}
    \tilde{\boldsymbol{j}}^{\mathrm{b}}=\boldsymbol{\nabla}\times\boldsymbol{M}+\frac{\partial\boldsymbol{P}}{\partial t}~.
\end{equation}
Defining the Lorentz magnetization (a Galilean invariant) as $\boldsymbol{\mathcal{M}} = \boldsymbol{M} + \boldsymbol{v}\times\boldsymbol{P}$, we can also write
\begin{equation}\label{J_bound}
    \boldsymbol{\mathcal{J}}^{\mathrm{b}}=\boldsymbol{\nabla}\times\boldsymbol{\mathcal{M}}+\overset{*}{\boldsymbol{P}}~.
\end{equation}

From the distinction between free and bound charge we can write the first pair of Maxwell's equations (Eq. \eqref{Gauss_law} and \eqref{ampere_law}) in matter as
\begin{equation}\label{gauss_medium}
    q^{\mathrm{f}} = \boldsymbol{\nabla}\cdot\boldsymbol{D}^{\mathrm{f}}~,
\end{equation}
and
\begin{equation}\label{ampere_medium}
    \tilde{\boldsymbol{j}}^{\mathrm{f}} = \boldsymbol{\nabla}\times\boldsymbol{H}^{\mathrm{f}}-\frac{\partial\boldsymbol{D}^{\mathrm{f}}}{\partial t}~,
\end{equation}
where
\begin{equation}\label{displacement_polarization}
    \boldsymbol{D}^{\mathrm{f}} = \boldsymbol{D} + \boldsymbol{P}
\end{equation}
and
\begin{equation}
    \boldsymbol{H}^{\mathrm{f}} = \boldsymbol{H}  - \boldsymbol{M}~.
\end{equation}
In terms of Galilean invariants, Eq. \eqref{ampere_medium} is
\begin{equation}
    \boldsymbol{\mathcal{J}}^{\mathrm{f}} = \boldsymbol{\nabla}\times\boldsymbol{\mathcal{H}}^{\mathrm{f}} - \overset{*}{\boldsymbol{D}^{\mathrm{f}}}~,
\end{equation}
where
\begin{equation}\label{h_free_charge}
    \boldsymbol{\mathcal{H}}^{\mathrm{f}} = \boldsymbol{\mathcal{H}}  - \boldsymbol{\mathcal{M}}~.
\end{equation}

In the following sections, we derive the balances of linear momentum, angular momentum and energy of a body in the presence of an electromagnetic field. 
The most rigorous approach for doing so is based on knowing the conserved quantities of the electromagentic field. To this end, it is known that solutions to Maxwell's equations
Eqs.~\eqref{Gauss_law}-\eqref{faraday_law}
with the aether constitutive relations Eqs.~\eqref{aether_1} and \eqref{aether_2} in vacuum can also be expressed as stationary points of an action functional in space-time corresponding to a Maxwell Lagrangian $\mathcal{L}$ \cite{jackson2007classical, cohen1997photons, bjorken1964relativistic}. 
The existence of such a Lagrangian and its invariance under translations in space-time and Lorentz transformations allows us to apply Noether's theorem to identify the conserved quantities of the electromagnetic field, namely the linear momentum  $\epsilon_0 \boldsymbol{E}\times \boldsymbol{B}$, angular momentum $\epsilon_0 \boldsymbol{x} \times (\boldsymbol{E}\times \boldsymbol{B})$, and energy $\frac{1}{2}\Big(\epsilon_0 E^2 + \frac{1}{\upmu_0} B^2\Big)$ \cite{bjorken1964relativistic}. 
Given these expressions, one may express the total linear and angular momentum, and energy per unit volume of a body in the presence of an electromagnetic field to be $\rho \boldsymbol{g} = \rho \boldsymbol{v} + \epsilon_0\boldsymbol{E}\times \boldsymbol{B}$, $\boldsymbol{x} \times \rho \boldsymbol{g}$, and $\rho \bar{e} = \rho e + \frac{1}{2}\Big(\epsilon_0 E^2 + \frac{1}{\upmu_0} B^2\Big)$, respectively, where $e$ is the energy per unit mass of the body (including kinetic and interatomic potential energies) but without the energy of the electromagnetic field. Ideally, one would formulate the balance laws based on the time changes of these compound quantities, which is the approach followed by Kovetz \cite{kovetz2000electromagnetic}. However, in light of familiarity of the principles of momentum and energy transport in chemical engineering and continuum mechanics, in what follows, we proceed to derive the local forms of balance laws starting from a physically intuitive perspective by capturing the effects of the electromagentic field through the Lorentz force, and then end with the forms of momentum and energy balances in terms of the compound fields.

\subsection{Linear momentum balance}
In this section, we derive equations for the balance of linear momentum in an electrolyte. As mentioned before, we begin with a physically intuitive derivation in which the influence of the electromagnetic field is captured through the Lorentz force. This is the form conventionally presented in electromagnetism texts\cite{schwinger2019classical,griffiths1999introduction} and provides a valid description of momentum transport within a body. We will argue, however, that this approach is less convenient when describing the boundary conditions of a system and may lead to incorrect interpretations of surface forces at a material boundary. We will end with an alternate form of the momentum balance which is more generally applicable. The formulation of these two forms of the linear momentum balance as well as the corresponding forms of the angular momentum balance largely follows the approach of Steigmann\cite{steigmann2009formulation}, who has reinterpreted Kovetz's work from a continuum mechanics perspective.

The global balance of momentum says that changes in total momentum in a body $\mathcal{P}$ must be balanced by the sum of all forces acting on the body: 
\begin{equation}\label{momentum_1}
    \frac{d}{dt}\int_\mathcal{P} \rho \boldsymbol{v} dv = \int_\mathcal{P}  \sum_i\rho_i \boldsymbol{b}_i dv + \int_{\partial\mathcal{P}} \boldsymbol{t}da + \boldsymbol{F}~,
\end{equation}
where $\boldsymbol{b}_i$ denotes non-electromagnetic body forces (such as gravity) acting on species $i$ and $\boldsymbol{t}$ is a surface force density. By Cauchy's lemma and tetrahedron argument, $\boldsymbol{t}$ may be rewritten in terms of the stress tensor $\boldsymbol{T}$ as $\boldsymbol{t} =\boldsymbol{T}\boldsymbol{n}$\cite{kovetz2000electromagnetic,steigmann2009formulation}, where, recall, $\boldsymbol{n}$ is the outward normal vector. The quantity $\boldsymbol{F}$ is the Lorentz force exerted on the body from the electromagnetic field:
\begin{equation}\label{Lorentz_Force}
    \boldsymbol{F} = \int_\mathcal{P}(q\boldsymbol{E} + \tilde{\boldsymbol{j}}\times \boldsymbol{B})dv~.
\end{equation}
Recall that in writing the Lorentz force in terms of the total charge and current density, we are capturing effects of the electromagnetic field on both the free ions in solution as well as the solvent medium.

The Lorentz force can be rewritten in terms of Galilean invariants, $\boldsymbol{\mathcal{E}}$ and $\boldsymbol{\mathcal{J}}$, using Eq. \eqref{e_field_invariant} and $\boldsymbol{\mathcal{J}} = \tilde{\boldsymbol{j}} - q \boldsymbol{v}$:
\begin{equation}
    \boldsymbol{F} =\int_\mathcal{P}(q\boldsymbol{E} + \tilde{\boldsymbol{j}}\times \boldsymbol{B})dv= \int_\mathcal{P}(q\boldsymbol{\mathcal{E}} + \boldsymbol{\mathcal{J}}\times \boldsymbol{B})dv~.
\end{equation}
The local form of Eq. \eqref{momentum_1} is then given by
\begin{equation}\label{local_momentum_1}
    \rho\dot{\boldsymbol{v}} = \boldsymbol{\nabla}\cdot\boldsymbol{T} + \sum_i\rho_i \boldsymbol{b}_i+q\boldsymbol{\mathcal{E}} + \boldsymbol{\mathcal{J}}\times \boldsymbol{B}~.
 \end{equation}
 
In a dielectric medium, Eq. \eqref{local_momentum_1} is more useful if written in terms of free (rather than total) charges. After some manipulation using Eqs. \eqref{total_free_bound_charge} through \eqref{h_free_charge}, it can be shown that Eq. \eqref{local_momentum_1} can alternatively be expressed as
\begin{equation}\label{local_momentum_free_charge}
\begin{split}
    \rho\dot{\boldsymbol{v}} = \boldsymbol{\nabla}\cdot[\boldsymbol{T} -& \boldsymbol{\mathcal{E}}\otimes\boldsymbol{P} - (\boldsymbol{\mathcal{M}}\cdot\boldsymbol{B})\boldsymbol{I} + \boldsymbol{\mathcal{M}}\otimes\boldsymbol{B}] + \sum_i\rho_i \boldsymbol{b}_i\\&+q^{\mathrm{f}}\boldsymbol{\mathcal{E}} + \boldsymbol{\mathcal{J}}^{\mathrm{f}}\times \boldsymbol{B}+\boldsymbol{P}\cdot\boldsymbol{\nabla}\boldsymbol{\mathcal{E}} + \boldsymbol{\mathcal{M}}\cdot\boldsymbol{\nabla}\boldsymbol{B} + \boldsymbol{\mathcal{M}}\times(\boldsymbol{\nabla}\times\boldsymbol{B})+\overset{*}{\boldsymbol{P}}\times\boldsymbol{B}~,
\end{split}
 \end{equation}
where $\boldsymbol{I}$ is the identity tensor. In most physically-relevant scenarios, $q^{\mathrm{f}}$ and $\tilde{\boldsymbol{j}}^{\mathrm{f}}$ are typically zero due to the condition of electroneutrality, a consequence of the substantial energy requirements for separating charges by a macroscopic distance. As can be seen from Eq. \eqref{local_momentum_free_charge}, however, even under this condition the electric field still alters the momentum of the system via the polarization and magnetization of the dielectric medium. In some situations, electroneutrality may be violated, namely within the electric double layers at charged interfaces. The violation of electroneutrality may also be important for nanoconfined systems where the size of the double layer is comparable to the length scale of the fluid region\cite{luo2015electroneutrality, levy2019breakdown}. In what follows, we aim to maintain generality and carry out the majority of derivations without assuming electroneutrality whenever possible. 
 
It is important to note that the the Lorentz force $\boldsymbol{F}$ need not vanish at a material boundary. Thus, the surface force $\boldsymbol{t}$ in Eq. \eqref{momentum_1} and corresponding stress tensor $\boldsymbol{T}$ do not describe the overall traction on the surface of a body. In order to quantify the overall surface forces and formulate boundary conditions, it is necessary to rewrite the Lorentz force in terms of the divergence of some quantity $\boldsymbol{T}_\mathrm{M}$ (called the Maxwell stress tensor) representing the surface stress induced by the electromagnetic field. In the remainder of this section, we use Maxwell's equations to derive the form of $\boldsymbol{T}_\mathrm{M}$ and rewrite the linear momentum balance in a form more amenable to boundary condition analysis. Let us begin by revisiting the quantity $q\boldsymbol{E} + \tilde{\boldsymbol{j}}\times \boldsymbol{B}$. We can rewrite this quantity using Eqs. \eqref{Gauss_law}, \eqref{ampere_law}, \eqref{aether_1}, and \eqref{aether_2} as
 \begin{equation}\label{Lorentz_1}
     q\boldsymbol{E} + \tilde{\boldsymbol{j}}\times \boldsymbol{B} = \epsilon_0(\boldsymbol{\nabla}\cdot\boldsymbol{E})\boldsymbol{E} + \frac{1}{\upmu_0}(\boldsymbol{\nabla}\times\boldsymbol{B})\times\boldsymbol{B}-\epsilon_0\frac{\partial\boldsymbol{E}}{\partial t}\times\boldsymbol{B}~.
 \end{equation}
The last term on the right side of Eq. \eqref{Lorentz_1} can be rewritten as
\begin{equation}\label{poynting}
\begin{split}
    \epsilon_0\frac{\partial\boldsymbol{E}}{\partial t}\times\boldsymbol{B}& = \epsilon_0\frac{\partial}{\partial t}(\boldsymbol{E}\times\boldsymbol{B}) - \epsilon_0\boldsymbol{E}\times \frac{\partial\boldsymbol{B}}{\partial t}\\&=\epsilon_0\frac{\partial}{\partial t}(\boldsymbol{E}\times\boldsymbol{B}) + \epsilon_0\boldsymbol{E}\times(\boldsymbol{\nabla}\times\boldsymbol{E})~.
\end{split}
\end{equation}
In the last equality we have used Eq. \eqref{faraday_law}. Substituting Eq. \eqref{poynting} into Eq. \eqref{Lorentz_1} yields
\begin{equation}\label{lorentz_5}
\begin{split}
    q\boldsymbol{E} + \tilde{\boldsymbol{j}}\times \boldsymbol{B} =\epsilon_0\bigg[(\boldsymbol{\nabla}&\cdot\boldsymbol{E})\boldsymbol{E} -\boldsymbol{E}\times(\boldsymbol{\nabla}\times\boldsymbol{E})\bigg]\\&+\frac{1}{\upmu_0}\bigg[(\boldsymbol{\nabla}\cdot\boldsymbol{B})\boldsymbol{B}-\boldsymbol{B}\times(\boldsymbol{\nabla}\times\boldsymbol{B})\bigg]-\epsilon_0\frac{\partial}{\partial t}(\boldsymbol{E}\times\boldsymbol{B})~.
\end{split}
\end{equation}
Note that the term $(\boldsymbol{\nabla}\cdot\boldsymbol{B})\boldsymbol{B}$ is equal to zero by Eq. \eqref{gauss_magnetism} and is only added such that the electric and magnetic field terms appear symmetrically in the equation. We can further simplify Eq. \eqref{lorentz_5} by using the vector identity $\boldsymbol{A}\times(\boldsymbol{\nabla}\times\boldsymbol{A}) = \frac{1}{2}\boldsymbol{\nabla}A^2 - \boldsymbol{A}\cdot \boldsymbol{\nabla}\boldsymbol{A}$, which yields
\begin{equation}\label{lorentz_2}
\begin{split}
    q\boldsymbol{E} + \tilde{\boldsymbol{j}}\times \boldsymbol{B} =\epsilon_0\bigg[(\boldsymbol{\nabla}&\cdot\boldsymbol{E})\boldsymbol{E} +\boldsymbol{E}\cdot(\boldsymbol{\nabla}\boldsymbol{E})\bigg]+\frac{1}{\upmu_0}\bigg[(\boldsymbol{\nabla}\cdot\boldsymbol{B})\boldsymbol{B}+\boldsymbol{B}\cdot(\boldsymbol{\nabla}\boldsymbol{B})\bigg]\\&-\frac{1}{2}\boldsymbol{\nabla}\bigg[\epsilon_0 E^2 + \frac{1}{\upmu_0} B^2 \bigg]-\epsilon_0\frac{\partial}{\partial t}(\boldsymbol{E}\times\boldsymbol{B})~.
\end{split}
\end{equation}
We may now define the Maxwell stress tensor $\boldsymbol{T}_\mathrm{M}$ as
\begin{equation}\label{maxwell_stress}
    \boldsymbol{T}_\mathrm{M} =  \bigg[\epsilon_0\boldsymbol{E}\otimes\boldsymbol{E} + \frac{1}{\upmu_0}\boldsymbol{B}\otimes\boldsymbol{B}\bigg]-\frac{1}{2}\bigg[\epsilon_0 E^2 + \frac{1}{\upmu_0} B^2\bigg]\boldsymbol{I}
\end{equation}
and express Eq. \eqref{lorentz_2} as
\begin{equation}\label{lorentz_22}
    \frac{\partial}{\partial t}(\epsilon_0\boldsymbol{E}\times\boldsymbol{B}) -\boldsymbol{\nabla}\cdot \boldsymbol{T}_\mathrm{M} + q\boldsymbol{E} + \tilde{\boldsymbol{j}}\times \boldsymbol{B} = 0~.
\end{equation}
Eq. \eqref{lorentz_22} is the local statement of conservation of momentum for the electromagnetic field itself, and the quantity $\epsilon_0\boldsymbol{E}\times\boldsymbol{B}$ is the momentum density of an electromagnetic field\cite{schwinger2019classical}. The overall Lorentz force thus becomes
\begin{equation}\label{lorentz_3}
    \boldsymbol{F} = \int_\mathcal{P}(\boldsymbol{\nabla}\cdot \boldsymbol{T}_\mathrm{M}) dv - \int_\mathcal{P}\frac{\partial}{\partial t}(\epsilon_0\boldsymbol{E}\times\boldsymbol{B})dv~.
\end{equation}
For the case of time-independent fields, Eq. \eqref{lorentz_3} shows how the Lorentz force may be equivalently interpreted in terms of surface forces or traction. Using the definition of the substantial derivative and the Reynolds transport theorem, Eq. \eqref{lorentz_3} can be rewritten as
\begin{equation}
    \boldsymbol{F} =\int_\mathcal{P}(\boldsymbol{\nabla}\cdot \boldsymbol{T}_\mathrm{M}) dv-\frac{d}{dt} \int_\mathcal{P}\big(\epsilon_0\boldsymbol{E}\times\boldsymbol{B}\big)dv +\int_\mathcal{P}\boldsymbol{\nabla}\cdot\big(\epsilon_0(\boldsymbol{E}\times\boldsymbol{B})\otimes\boldsymbol{v}\big)dv~.
\end{equation}
Defining a new quantity, 
\begin{equation}
    \boldsymbol{\hat{T}} = \boldsymbol{T}_\mathrm{M} +\epsilon_0(\boldsymbol{E}\times\boldsymbol{B})\otimes\boldsymbol{v}~,
\end{equation}
the Lorentz Force can be rewritten as
\begin{equation}\label{lorentz_4}
    \boldsymbol{F} =\int_\mathcal{P}(\boldsymbol{\nabla}\cdot \boldsymbol{\hat{T}}) dv-\frac{d}{dt} \int_\mathcal{P}\big(\epsilon_0\boldsymbol{E}\times\boldsymbol{B}\big)dv~.
\end{equation}
In Eq. \eqref{lorentz_4} we have decomposed the Lorentz force into two contributions. The first is from surface stresses induced by the electromagnetic field, and the second represents the electromagnetic contributions to the total momentum. Given Eq. \eqref{lorentz_4}, we can now reformulate the global form of the momentum balance in Eq. \eqref{momentum_1} as
\begin{equation}\label{momentum_2}
    \frac{d}{dt}\int_\mathcal{P} \rho (\boldsymbol{v}+\epsilon_0 \boldsymbol{E}\times\boldsymbol{B}/\rho) dv = \int_\mathcal{P}  \sum_i\rho_i \boldsymbol{b}_i dv + \int_{\mathcal{P}} \boldsymbol{\nabla}\cdot(\boldsymbol{T}+\boldsymbol{\hat{T}})dv~.
\end{equation}
Finally, defining the composite stress tensor $\boldsymbol{\bar{T}} = \boldsymbol{T}+\boldsymbol{\hat{T}}$ and a modified momentum density $\boldsymbol{g} = \boldsymbol{v} + \epsilon_0\boldsymbol{E}\times\boldsymbol{B}/\rho$, Eq. \eqref{momentum_2} reduces to
\begin{equation}
    \frac{d}{dt}\int_\mathcal{P} \rho \boldsymbol{g} dv =\int_\mathcal{P}  \sum_i\rho_i \boldsymbol{b}_i dv + \int_{\mathcal{P}} (\boldsymbol{\nabla}\cdot\boldsymbol{\bar{T}})dv,
\end{equation}
with the local form of the balance
\begin{equation}\label{local_momentum_2}
    \rho\dot{\boldsymbol{g}} = \boldsymbol{\nabla}\cdot\boldsymbol{\bar{T}} + \sum_i\rho_i \boldsymbol{b}_i~.
 \end{equation}
The quantity $\boldsymbol{g}$, as mentioned before, is a more general representation of the momentum per unit mass in a body subject to electromagnetic fields, as it captures the momentum of the electrolyte body as well as that of the electromagnetic field itself \cite{kovetz2000electromagnetic}. Thus, we can interpret our original momentum balance (Eq. \eqref{local_momentum_1}) as accounting for changes in momentum of only the body $\mathcal{P}$ (considering the electromagnetic field only as an external force), while in Eq. \eqref{local_momentum_2} we account for momentum changes in both the body $\mathcal{P}$ and the electromagnetic field together. Either form is valid within the bulk of the body, but only Eq. \eqref{local_momentum_2} provides a transparent description of behavior at a material boundary, where the overall surface force per unit area or traction is given by $\bar{\boldsymbol{t}} = \boldsymbol{\bar{T}}\boldsymbol{n}$ (and not $\boldsymbol{T}\boldsymbol{n}$, as may be incorrectly concluded from Eq. \eqref{local_momentum_1}).

The forms of the linear momentum balances derived in this section are generally applicable to any body subject to an electromagnetic field. In Sec. \ref{sec:linear_irr_thermo}, we will assume linear constitutive relations and derive the forms of the momentum balance that are applicable to most liquid electrolyte solutions.

\subsection{\label{sec:angular_momentum}Angular momentum balance}
We now present the balance of angular momentum in two forms. The first is based on the linear momentum balance of the form Eq. \eqref{local_momentum_1}, which considers the momentum of the body to be $\rho \boldsymbol{v}$. The second treats the momentum of both the body and the electromagnetic field, captured in the quantity $\boldsymbol{g}$, as in Eq. \eqref{local_momentum_2}. We will analyze the implications that these angular momentum balances have on the symmetry of the stress tensor in each of these forms. 

Based on Eq. \eqref{local_momentum_1}, the angular momentum balance can be written as
\begin{equation}\label{angular_momentum_1}
    \frac{d}{dt}\int_{\mathcal{P}} (\boldsymbol{x}\times\rho\boldsymbol{v}) dv = \int_{\partial \mathcal{P}} (\boldsymbol{x}\times \boldsymbol{T}\boldsymbol{n}) da + \int_{\mathcal{P}}  \boldsymbol{x}\times \bigg(\sum_i\rho_i \boldsymbol{b}_i+q\boldsymbol{E} + \tilde{\boldsymbol{j}}\times \boldsymbol{B} \bigg) dv~.
\end{equation}
Incorporating the Reynolds transport theorem and overall mass balance (Eq. \eqref{mass_balance_total}) to simplify the left side of Eq. \eqref{angular_momentum_1} gives
\begin{equation}
    \int_{\mathcal{P}} \rho \boldsymbol{x}\times\dot{\boldsymbol{v}} dv= \int_{\partial \mathcal{P}} (\boldsymbol{x}\times \boldsymbol{T}\boldsymbol{n}) da + \int_{\mathcal{P}}  \boldsymbol{x}\times \bigg(\sum_i\rho_i \boldsymbol{b}_i+q\boldsymbol{E} + \tilde{\boldsymbol{j}}\times \boldsymbol{B}\bigg) dv~.
\end{equation}
Comparing with Eq. \eqref{local_momentum_1}, we may eliminate the second term on the right side and write
\begin{equation}
    \int_{\mathcal{P}} \boldsymbol{x}\times(\boldsymbol{\nabla}\cdot \boldsymbol{T}) dv = \int_{\partial \mathcal{P}} (\boldsymbol{x}\times \boldsymbol{T}\boldsymbol{n}) da~.
\end{equation}
Rearranging and applying the localization theorem yields
\begin{equation}\label{t_symmetry_indicial}
    \varepsilon_{ijk} T_{kj} = 0~,
\end{equation}
where $\varepsilon_{ijk}$ is the Levi-Civita symbol. This leads to the familiar result that the stress tensor is symmetric,
\begin{equation}\label{t_symmetry}
    \boldsymbol{T} = \boldsymbol{T}^T ~.
\end{equation}

Now let us write the overall angular momentum balance based on Eq. \eqref{local_momentum_2}:
\begin{equation}
    \frac{d}{dt}\int_{\mathcal{P}} (\boldsymbol{x}\times\rho\boldsymbol{g}) dv = \int_{\partial \mathcal{P}} (\boldsymbol{x}\times \bar{\boldsymbol{T}}\boldsymbol{n}) da + \int_{\mathcal{P}}  \boldsymbol{x}\times \bigg(\sum_i\rho_i \boldsymbol{b}_i \bigg) dv~.
\end{equation}
Analogous simplifications allow us to conclude that
\begin{equation}
    \int_{\mathcal{P}}(\boldsymbol{v}\times\rho\boldsymbol{g} + \boldsymbol{x} \times (\boldsymbol{\nabla}\cdot \bar{\boldsymbol{T}}))dv = \int_{\partial \mathcal{P}} (\boldsymbol{x}\times \bar{\boldsymbol{T}}\boldsymbol{n}) da~,
\end{equation}
or
\begin{equation}\label{t_bar_symmetry}
    \varepsilon_{ijk}(\bar{T}_{kj} + \rho v_k g_j) = 0~.
\end{equation}
Thus, the overall stress tensor $\bar{\boldsymbol{T}}$ which is relevant in the boundary conditions of the body is only symmetric in the case where $v_k g_j = v_j g_k$. This will in general only be true if their is no electric or magnetic field. However, we can show that the result in Eq.~\eqref{t_bar_symmetry} is equivalent to Eq. \eqref{t_symmetry} by incorporating the anti-symmetric portions of $\bar{\boldsymbol{T}}$ and $\rho\boldsymbol{v}\otimes\boldsymbol{g}$ into Eq. \eqref{t_bar_symmetry}:
\begin{equation}
     \varepsilon_{ijk}(\bar{T}_{kj} + \rho v_k g_j) = \varepsilon_{ijk}(T_{kj}+ \epsilon_0\varepsilon_{klm}E_lB_mv_j+ \epsilon_0 \varepsilon_{jlm}E_l B_m v_k) = 0~.
\end{equation}
The last two terms on the right side cancel, leading once again to Eqs. \eqref{t_symmetry_indicial} and \eqref{t_symmetry}.
 
\subsection{Energy balance}
We now develop expressions for conservation of energy in electrolyte systems. As with the momentum balances, we develop two forms of this balance law: one which considers the energy of only the electrolyte body and one which includes the energy of both the body and the electromagnetic field.

The first of these forms of the global energy balance can be formulated by balancing the total change in energy with all of the sources of heat and work on the system. This yields
\begin{equation}\label{energy_1}
    \frac{d}{dt}\int_\mathcal{P} \rho e dv = \int_\mathcal{P} \rho r dv - \int_{\partial\mathcal{P}} \boldsymbol{J}_\mathrm{q}\cdot \boldsymbol{n} da + \int_{\partial\mathcal{P}} \boldsymbol{t}\cdot \boldsymbol{v} da + \int_\mathcal{P} \sum_i\rho_i \boldsymbol{b}_i\cdot\boldsymbol{v}_idv + \int_\mathcal{P} \mathbb{P}^{\mathrm{EM}}dv~,
\end{equation}
where $e$ is total energy per unit mass, $r$ is energy per mass produced through body heating, and $\boldsymbol{J}_\mathrm{q}$ is the heat flux vector. Again we emphasize that $e$ is the energy of the body $\mathcal{P}$, which is affected by the electromagnetic field, but it does not include the energy of the electromagnetic field itself. The quantities $\boldsymbol{t}\cdot \boldsymbol{v}$ and $\sum_i\rho_i \boldsymbol{b}_i\cdot\boldsymbol{v}_i$ give the rate of work done by surface and body forces, respectively. In the last term of Eq. \eqref{energy_1}, $\mathbb{P}^{\mathrm{EM}}$ represents the rate of work done on the body by the electromagnetic field via the Lorentz force. Evaluating this power requires introducing a microscopic picture of charge transport in terms of the positions and velocities of individual particles, $\boldsymbol{r}^\alpha$ and $\boldsymbol{v}^\alpha$, respectively.\footnote{Throughout the text, superscript Greek indices ($\alpha$, $\beta$) denote individual particles, while subscript Latin indices ($i$, $j$) denote species of a given type.} The current density $\boldsymbol{\tilde{j}}$, for example, may be written on a microscopic level as $\boldsymbol{\tilde{j}} = \sum_\alpha \hat{q}^{\alpha} \boldsymbol{v}^\alpha \Delta(\boldsymbol{x} - \boldsymbol{r}^\alpha)$, where $\hat{q}^\alpha$ is the charge (not the charge density) of particle $\alpha$ and $\Delta(\boldsymbol{x} - \boldsymbol{r}^\alpha)$ is a coarse-graining function connecting the microscopic particle picture to the continuum level. The Lorentz force acting on an individual particle $\boldsymbol{F}^{\alpha}$ is
\begin{equation}\label{Lorentz_Force_per_speces}
    \boldsymbol{F}^{\alpha} = \hat{q}^{\alpha}\boldsymbol{E} + \hat{q}^{\alpha} \boldsymbol{v}^{\alpha}\times \boldsymbol{B}~.
\end{equation}
The power corresponding to this Lorentz force is
\begin{equation}
    \mathbb{P}^{\mathrm{EM}} = \sum_{\alpha}(\boldsymbol{F}^{\alpha}\cdot\boldsymbol{v}^{\alpha})\Delta(\boldsymbol{x} - \boldsymbol{r}^\alpha) = \bigg[\sum_{\alpha} \hat{q}^{\alpha}\boldsymbol{v}^{\alpha} \cdot \boldsymbol{E}+ \sum_{\alpha}  \hat{q}^{\alpha} (\boldsymbol{v}^{\alpha}\times \boldsymbol{B})\cdot\boldsymbol{v}^{\alpha}\bigg]\Delta(\boldsymbol{x} - \boldsymbol{r}^\alpha) = \boldsymbol{\tilde{j}}\cdot\boldsymbol{E}~.
\end{equation}

The global energy balance can thus be expressed as 
\begin{equation}\label{global_energy_1}
    \frac{d}{dt}\int_\mathcal{P} \rho e dv = \int_\mathcal{P} \rho r dv - \int_{\mathcal{P}} \boldsymbol{\nabla}\cdot\boldsymbol{J}_\mathrm{q}dv + \int_{\mathcal{P}} \boldsymbol{\nabla}\cdot(\boldsymbol{T}^T\boldsymbol{v}) dv + \int_\mathcal{P} \sum_i\rho_i \boldsymbol{b}_i\cdot\boldsymbol{v}_idv + \int_{\mathcal{P}}\boldsymbol{\tilde{j}}\cdot\boldsymbol{E}dv~,
\end{equation}
with the corresponding local form
\begin{equation}
    \rho \dot{e} = \sum_i\rho_i \boldsymbol{b}_i\cdot\boldsymbol{v}_i + \boldsymbol{\nabla}\cdot(\boldsymbol{T}^T\boldsymbol{v})+\rho r - \boldsymbol{\nabla}\cdot\boldsymbol{J}_\mathrm{q} + \boldsymbol{\tilde{j}}\cdot\boldsymbol{E}~.
\end{equation}
We can simplify this energy balance by incorporating the momentum balance as written in Eq. \eqref{local_momentum_1}, yielding
\begin{equation}\label{local_energy_1}
    \rho \dot{e} = \rho \dot{\boldsymbol{v}}\cdot\boldsymbol{v} + \boldsymbol{T}:\boldsymbol{\nabla}\boldsymbol{v}+\rho r - \boldsymbol{\nabla}\cdot\boldsymbol{J}_\mathrm{q} + \sum_i \boldsymbol{j}_i \cdot \boldsymbol{b}_i + \boldsymbol{\mathcal{J}}\cdot\boldsymbol{\mathcal{E}}~.
\end{equation}
Note that for body forces such as gravity which act uniformly on all species, $\sum_i \boldsymbol{j}_i \cdot \boldsymbol{b}_i = 0$.

Let us rewrite the local energy balance Eq.~\eqref{local_energy_1} in terms of free and bound charges, rather than the total charge. Substituting Eq. \eqref{J_bound}, the quantity $\boldsymbol{\mathcal{J}}\cdot\boldsymbol{\mathcal{E}}$ in Eq. \eqref{local_energy_1} can be rewritten as
\begin{equation}
    \boldsymbol{\mathcal{J}}\cdot\boldsymbol{\mathcal{E}} = \boldsymbol{\mathcal{J}}^{\mathrm{f}}\cdot\boldsymbol{\mathcal{E}}+\boldsymbol{\mathcal{J}}^{\mathrm{b}}\cdot\boldsymbol{\mathcal{E}} = \boldsymbol{\mathcal{J}}^{\mathrm{f}}\cdot\boldsymbol{\mathcal{E}} + \boldsymbol{\mathcal{E}}\cdot\overset{*}{\boldsymbol{P}} + \boldsymbol{\mathcal{E}}\cdot\boldsymbol{\nabla}\times\boldsymbol{\mathcal{M}}~.
\end{equation}
Using Eq. \eqref{faraday_invariant}, the last term in this equation can be rewritten as $\boldsymbol{\mathcal{E}}\cdot\boldsymbol{\nabla}\times\boldsymbol{\mathcal{M}} = \boldsymbol{\nabla}\cdot(\boldsymbol{\mathcal{M}}\times\boldsymbol{\mathcal{E}})-\boldsymbol{\mathcal{M}}\cdot\overset{*}{\boldsymbol{B}}$. Thus, we have
\begin{equation}
    \boldsymbol{\mathcal{J}}\cdot\boldsymbol{\mathcal{E}} = \boldsymbol{\mathcal{J}}^{\mathrm{f}}\cdot\boldsymbol{\mathcal{E}} + \boldsymbol{\mathcal{E}}\cdot\overset{*}{\boldsymbol{P}} + \boldsymbol{\nabla}\cdot(\boldsymbol{\mathcal{M}}\times\boldsymbol{\mathcal{E}})-\boldsymbol{\mathcal{M}}\cdot\overset{*}{\boldsymbol{B}}~.
\end{equation}
We can further simplify by noting that for two vectors $\boldsymbol{A}$ and $\boldsymbol{B}$, $\boldsymbol{A}\cdot\overset{*}{\boldsymbol{B}} = \boldsymbol{A}\cdot\dot{\boldsymbol{B}} + \big[(\boldsymbol{A}\cdot\boldsymbol{B})\boldsymbol{I}-\boldsymbol{A}\otimes\boldsymbol{B}]:\boldsymbol{\nabla}\boldsymbol{v}$. Equation \eqref{local_energy_1} is thus
\begin{equation}\label{energy_free_charge}
\begin{split}
        \rho \dot{e} = \rho \dot{\boldsymbol{v}}\cdot\boldsymbol{v} + [\boldsymbol{T}+&(\boldsymbol{\mathcal{E}}\cdot\boldsymbol{P})\boldsymbol{I} - \boldsymbol{\mathcal{E}}\otimes\boldsymbol{P} - (\boldsymbol{\mathcal{M}}\cdot\boldsymbol{B})\boldsymbol{I} + \boldsymbol{\mathcal{M}}\otimes\boldsymbol{B}]:\boldsymbol{\nabla}\boldsymbol{v}\\&+\rho r - \boldsymbol{\nabla}\cdot\bar{\boldsymbol{J}}_\mathrm{q} + \sum_i \boldsymbol{j}_i \cdot \boldsymbol{b}_i+ \boldsymbol{\mathcal{J}}^{\mathrm{f}}\cdot\boldsymbol{\mathcal{E}} + \boldsymbol{\mathcal{E}}\cdot\dot{\boldsymbol{P}} -\boldsymbol{\mathcal{M}}\cdot\dot{\boldsymbol{B}}~,
\end{split}
\end{equation}
where we have defined the quantity $\bar{\boldsymbol{J}}_\mathrm{q} = \boldsymbol{J}_\mathrm{q} + \boldsymbol{\mathcal{E}}\times\boldsymbol{\mathcal{M}}$ as a modified heat flux vector.

Equation~\eqref{energy_free_charge} is the form of the energy balance which will be useful in deriving internal entropy production. However, 
we can also express the energy balance in terms of $\boldsymbol{g}$, the momentum of the body including the electromagnetic field, and $\boldsymbol{\bar{T}}$, the composite stress tensor used in the momentum balance as written in Eq. \eqref{local_momentum_2}.
To do so, let us rewrite the quantity $(\boldsymbol{\tilde{j}}\cdot\boldsymbol{E})$ in terms of the applied electric and magnetic fields. Using Eqs. \eqref{ampere_law}, \eqref{aether_1}, and \eqref{aether_2}, we can write
\begin{equation}
\begin{split}
   \boldsymbol{\tilde{j}}\cdot\boldsymbol{E} &= \frac{1}{\upmu_0}\boldsymbol{E}\cdot(\boldsymbol{\nabla}\times\boldsymbol{B})-\boldsymbol{E}\cdot \epsilon_0 \frac{\partial\boldsymbol{E}}{\partial t} \\&= \frac{1}{\upmu_0}\big(-\boldsymbol{\nabla}\cdot(\boldsymbol{E}\times\boldsymbol{B}) + \boldsymbol{B}\cdot(\boldsymbol{\nabla}\times\boldsymbol{E})\big)-\boldsymbol{E}\cdot \epsilon_0 \frac{\partial\boldsymbol{E}}{\partial t}~.
\end{split}
\end{equation}
Applying Eq. \eqref{faraday_law}, we obtain
\begin{equation}\label{pi_em_plus_heating}
   \boldsymbol{\tilde{j}}\cdot\boldsymbol{E} = -\boldsymbol{\nabla}\cdot\bigg(\frac{1}{\upmu_0}\boldsymbol{E}\times\boldsymbol{B}\bigg) -\frac{1}{\upmu_0} \boldsymbol{B}\cdot\frac{\partial\boldsymbol{B}}{\partial t}-\epsilon_0\boldsymbol{E}\cdot  \frac{\partial\boldsymbol{E}}{\partial t}~.
\end{equation}
Thus, 
\begin{equation}\label{j_dot_e_0}
    \int_\mathcal{P} \boldsymbol{\tilde{j}}\cdot\boldsymbol{E} dv = -\int_\mathcal{P}\bigg[\boldsymbol{\nabla}\cdot\bigg(\frac{1}{\upmu_0}\boldsymbol{E}\times\boldsymbol{B}\bigg) +\frac{1}{\upmu_0} \boldsymbol{B}\cdot\frac{\partial\boldsymbol{B}}{\partial t}+\epsilon_0\boldsymbol{E}\cdot  \frac{\partial\boldsymbol{E}}{\partial t}\bigg]dv~.
\end{equation}
Recall that the electromagnetic energy per volume $\tilde{u}_{\mathrm{EM}}$ can be expressed as $\tilde{u}_{\mathrm{EM}} = \frac{1}{2}(\epsilon_0 E^2+\frac{1}{\upmu_0}B^2)$, and consequently $\frac{\partial}{\partial t}(\tilde{u}_{\mathrm{EM}}) = \epsilon_0 \boldsymbol{E}\cdot\frac{\partial\boldsymbol{E}}{\partial t}+ \frac{1}{\upmu_0}\boldsymbol{B}\cdot\frac{\partial\boldsymbol{B}}{\partial t}$. Thus, we can rewrite Eq. \eqref{j_dot_e_0} as
\begin{equation}\label{j_dot_E_1}
    \int_\mathcal{P} \boldsymbol{\tilde{j}}\cdot\boldsymbol{E} dv =-\int_\mathcal{P}\bigg[\boldsymbol{\nabla}\cdot\bigg(\frac{1}{\upmu_0}\boldsymbol{E}\times\boldsymbol{B}\bigg) - \boldsymbol{\nabla}\cdot(\tilde{u}_{\mathrm{EM}}\boldsymbol{v})\bigg]dv-\frac{d}{dt}\int_\mathcal{P}\tilde{u}_{\mathrm{EM}}dv~.
\end{equation}
When written in terms of partial (as opposed to substantial) derivatives, the local form of Eq. \eqref{j_dot_E_1} is
\begin{equation}\label{em_energy_balance}
    \frac{\partial \tilde{u}_{\mathrm{EM}}}{\partial t} = -\boldsymbol{\tilde{j}}\cdot\boldsymbol{E} - \boldsymbol{\nabla}\cdot\bigg(\frac{1}{\upmu_0}\boldsymbol{E}\times\boldsymbol{B}\bigg)~.
\end{equation}
Equation \eqref{em_energy_balance} is the energy balance of the electromagnetic field alone\cite{feynman1964feynman}, where the change in energy of the electromagnetic field is balanced by the work done by the Lorentz force (the first term on the right side) and the energy flux of the field (the second term). The latter quantity, $\frac{1}{\upmu_0}\boldsymbol{E}\times\boldsymbol{B} \eqqcolon \boldsymbol{S}$, is referred to as the Poynting vector.

We may now proceed by integrating the electromagnetic energy balance in Eq.~\eqref{j_dot_E_1} with the energy balance for the system as a whole. To this end, incorporating the definition of $\tilde{u}_{\mathrm{EM}}$ and applying the divergence theorem to the first term on the right side of Equation \eqref{j_dot_E_1}, we obtain
\begin{equation}
\begin{split}
    \int_\mathcal{P} \boldsymbol{\tilde{j}}\cdot\boldsymbol{E} dv =\int_{\
    \partial\mathcal{P}}\bigg[\frac{1}{2}\bigg(\epsilon_0 E^2+\frac{1}{\upmu_0}B^2\bigg)\boldsymbol{v}-&\bigg(\frac{1}{\upmu_0}\boldsymbol{E}\times\boldsymbol{B}\bigg) \bigg]\cdot \boldsymbol{n} da\\&-\frac{d}{dt}\int_\mathcal{P}\bigg[\frac{1}{2}\bigg(\epsilon_0 E^2+\frac{1}{\upmu_0}B^2\bigg)\bigg]dv~.
\end{split}
\end{equation}
After some manipulation using Eqs. \eqref{aether_2}-\eqref{m_field_invariant} and \eqref{maxwell_stress}, it can be shown that the integrand of the first term on the right-hand side can be rewritten as 
\begin{equation}
\begin{split}
    \frac{1}{2}\bigg(\epsilon_0 E^2+\frac{1}{\upmu_0}B^2\bigg)\boldsymbol{v}\cdot\boldsymbol{n}-\bigg(\frac{1}{\upmu_0}\boldsymbol{E}\times\boldsymbol{B}\bigg)\cdot\boldsymbol{n} &= -\boldsymbol{\mathcal{E}}\times\boldsymbol{\mathcal{H}}\cdot\boldsymbol{n} + \boldsymbol{T}_{\mathrm{M}}\boldsymbol{v}\cdot\boldsymbol{n} + \epsilon_0[(\boldsymbol{v}\otimes\boldsymbol{E}\times\boldsymbol{B})\boldsymbol{v}]\cdot\boldsymbol{n}\\&=-\boldsymbol{\mathcal{E}}\times\boldsymbol{\mathcal{H}}\cdot\boldsymbol{n} +\hat{\boldsymbol{T}}\boldsymbol{n}\cdot\boldsymbol{v}~.
\end{split}
\end{equation}
In the second equality we have made use of the symmetry of $\boldsymbol{T}_{\mathrm{M}}$. Therefore,
\begin{equation}
    \int_\mathcal{P} \boldsymbol{\tilde{j}}\cdot\boldsymbol{E} dv =\int_\mathcal{P}\boldsymbol{\nabla}\cdot(\boldsymbol{\hat{T}}^T\boldsymbol{v}- \boldsymbol{\mathcal{E}}\times\boldsymbol{\mathcal{H}})dv-\frac{d}{dt}\int_\mathcal{P}\bigg[\frac{1}{2}\bigg(\epsilon_0 E^2+\frac{1}{\upmu_0}B^2\bigg)\bigg]dv~.
\end{equation}
Substituting this expression into the global energy balance (Eq. \eqref{global_energy_1}) yields
\begin{equation}\label{global_energy_2}
    \frac{d}{dt}\int_\mathcal{P} \rho \bar{e} dv = \int_\mathcal{P} \rho r dv - \int_{\mathcal{P}} \boldsymbol{\nabla}\cdot(\boldsymbol{J}_\mathrm{q}+ \boldsymbol{\mathcal{E}}\times\boldsymbol{\mathcal{H}})dv + \int_{\mathcal{P}} \boldsymbol{\nabla}\cdot(\boldsymbol{\bar{T}}^T\boldsymbol{v}) dv + \int_\mathcal{P} \sum_i\rho_i \boldsymbol{b}_i\cdot\boldsymbol{v}_i dv ~.
\end{equation}
Here we have defined the quantity $\bar{e} = e + \frac{1}{2}(\epsilon_0 E^2+\frac{1}{\upmu_0}B^2)/\rho$, which is the energy per unit mass of the system including the vacuum energy of the electromagnetic field. The quantity $\boldsymbol{\mathcal{E}}\times\boldsymbol{\mathcal{H}}$ may be interpreted as an additional flux of energy from the electromagnetic field; this term is the Galilean invariant analogue of the Poynting vector introduced in Eq. \eqref{em_energy_balance}. The corresponding local form of Eq. \eqref{global_energy_2} is
\begin{equation}\label{local_energy_3}
    \rho \dot{\bar{e}} =  \boldsymbol{\nabla}\cdot(\boldsymbol{\bar{T}}^T\boldsymbol{v})+\rho r +\sum_i\rho_i \boldsymbol{b}_i\cdot\boldsymbol{v}_i - \boldsymbol{\nabla}\cdot(\boldsymbol{J}_\mathrm{q}+\boldsymbol{\mathcal{E}}\times\boldsymbol{\mathcal{H}})~.
\end{equation}
We can now incorporate the momentum balance. Taking the dot product of $\boldsymbol{v}$ with both sides of Eq. \eqref{local_momentum_2} and subtracting the resulting equation from Eq. \eqref{local_energy_3}, we obtain
\begin{equation}\label{local_energy_2}
    \rho \dot{\bar{e}} =  \rho \dot{\boldsymbol{g}}\cdot\boldsymbol{v} + \boldsymbol{\bar{T}}:\boldsymbol{\nabla}\boldsymbol{v}+\rho r  +\sum_i\boldsymbol{j}_i\cdot\boldsymbol{b}_i - \boldsymbol{\nabla}\cdot(\boldsymbol{J}_\mathrm{q}+\boldsymbol{\mathcal{E}}\times\boldsymbol{\mathcal{H}})~.
\end{equation}

We may analogously rewrite the alternate form of the local energy balance, Eq. \eqref{local_energy_2}, in terms of free charges. Using Eq. \eqref{h_free_charge}, the quantity $\boldsymbol{\mathcal{E}}\times\boldsymbol{\mathcal{H}}$ becomes $\boldsymbol{\mathcal{E}}\times\boldsymbol{\mathcal{H}}^{\mathrm{f}} + \boldsymbol{\mathcal{E}}\times\boldsymbol{\mathcal{M}}$. As in Eq. \eqref{energy_free_charge}, we may utilize the definition of the modified heat flux vector $\bar{\boldsymbol{J}}_\mathrm{q} = \boldsymbol{J}_\mathrm{q} + \boldsymbol{\mathcal{E}}\times\boldsymbol{\mathcal{M}}$ to write 
\begin{equation}
    \rho \dot{\bar{e}} = \rho \dot{\boldsymbol{g}}\cdot\boldsymbol{v} + \boldsymbol{\bar{T}}:\boldsymbol{\nabla}\boldsymbol{v}+\rho r +\sum_i\boldsymbol{j}_i\cdot\boldsymbol{b}_i - \boldsymbol{\nabla}\cdot(\bar{\boldsymbol{J}}_\mathrm{q}+\boldsymbol{\mathcal{E}}\times\boldsymbol{\mathcal{H}}^{\mathrm{f}})~.
\end{equation}

\subsection{Entropy balance}\label{sec:entropy_bal}
In this section, we will introduce the entropy balance and the second law of thermodynamics for electrolyte solutions. In doing so we provide a rigorous derivation for the rate of internal entropy production for multicomponent systems in the presence of electromagnetic fields. We will ultimately simplify this result specifically for an electrolyte with no applied magnetic field. This section is an extension of the work of de Groot and Mazur\cite{DeGroot1969Non-EquilibriumThermodynamics} to charged systems in the presence of electromagnetic fields. 

To begin, we postulate that the total change in entropy in the system can be written as
\begin{equation}\label{entropy_balance_form}
    \rho \dot{s} = -\boldsymbol{\nabla}\cdot\boldsymbol{J}_{\mathrm{s}} + \rho \sigma_\mathrm{e} + \rho \sigma_\mathrm{i}~,
\end{equation}
where $s$ is the entropy per unit mass, $\boldsymbol{J}_{\mathrm{s}}$ is entropy flux, $\sigma_\mathrm{e}$ is entropy production from body forces, and $\sigma_\mathrm{i}$ is internal entropy production ($\sigma_\mathrm{i} \geq 0$ by the second law of thermodynamics). Following Sahu et al.\cite{sahu2017irreversible} and Mandadapu\cite{mandadapu2011homogeneous}, the components of this entropy balance can be obtained by working with the Helmholtz free energy. The Helmholtz free energy per volume, $\tilde{f}$, can be written as\cite{kovetz2000electromagnetic,landau2013electrodynamics}
\begin{equation}\label{f_tilde_1}
     \tilde{f} = \rho e - \frac{1}{2}\rho\boldsymbol{v}\cdot \boldsymbol{v}-\rho Ts - \boldsymbol{\mathcal{E}}\cdot\boldsymbol{P}~.
\end{equation}
Taking the substantial derivative of both sides and incorporating the mass balance (Eq. \eqref{mass_balance_total}), Eq. \eqref{f_tilde_1} becomes
\begin{equation}\label{entropy_balance_0}
    \rho \dot{s} = \frac{1}{T}\bigg[\rho\dot{e}-\rho \dot{\boldsymbol{v}}\cdot\boldsymbol{v}-\dot{\tilde{f}} - \tilde{f}(\boldsymbol{I}:\boldsymbol{\nabla}\boldsymbol{v})  - \rho s \dot{T}  - \boldsymbol{\mathcal{E}}\cdot\dot{\boldsymbol{P}} - \dot{\boldsymbol{\mathcal{E}}}\cdot\boldsymbol{P} - (\boldsymbol{\mathcal{E}}\cdot\boldsymbol{P})(\boldsymbol{I}:\boldsymbol{\nabla}\boldsymbol{v})\bigg]~.
\end{equation}
Incorporating the energy balance in Eq. \eqref{energy_free_charge} allows us to rewrite Eq. \eqref{entropy_balance_0} as
\begin{equation}\label{entropy_balance_1}
\begin{split}
    \rho \dot{s} = \frac{1}{T}\bigg[-\dot{\tilde{f}} - \tilde{f}(\boldsymbol{I}&:\boldsymbol{\nabla}\boldsymbol{v})+[\boldsymbol{T} - \boldsymbol{\mathcal{E}}\otimes\boldsymbol{P} - (\boldsymbol{\mathcal{M}}\cdot\boldsymbol{B})\boldsymbol{I} + \boldsymbol{\mathcal{M}}\otimes\boldsymbol{B}]:\boldsymbol{\nabla}\boldsymbol{v} + \rho r\\&+\sum_i \boldsymbol{j}_i \cdot \boldsymbol{b}_i- \boldsymbol{\nabla}\cdot\bar{\boldsymbol{J}}_\mathrm{q}+ \boldsymbol{\mathcal{J}}^{\mathrm{f}}\cdot\boldsymbol{\mathcal{E}} - \dot{\boldsymbol{\mathcal{E}}}\cdot\boldsymbol{P} -\boldsymbol{\mathcal{M}}\cdot\dot{\boldsymbol{B}} - \rho s \dot{T}\bigg]~.
\end{split}
\end{equation}
Equation \eqref{entropy_balance_1} can be simplified further by evaluating $\dot{\tilde{f}}$ in terms of its natural variables. In Appendix \ref{appendix:thermo_potentials}, we argue that $\tilde{f}$ is a function of $N+3$ quantities: [$T$, $c_1, c_2, ..., c_N$, $\boldsymbol{\mathcal{E}}$, $\boldsymbol{B}$], which leads to
\begin{equation}\label{f_tilde_dependencies}
    \dot{\tilde{f}} = \frac{\partial\tilde{f}}{\partial T}\dot{T} + \sum_i \frac{\partial\tilde{f}}{\partial c_i}\dot{c}_i + \frac{\partial\tilde{f}}{\partial \boldsymbol{\mathcal{E}}}\cdot \dot{\boldsymbol{\mathcal{E}}}+ \frac{\partial\tilde{f}}{\partial\boldsymbol{B}}\cdot \dot{\boldsymbol{B}}~.
\end{equation}
Substituting Eq. \eqref{f_tilde_dependencies} into Eq. \eqref{entropy_balance_1} yields
\begin{equation}\label{entropy_balance_1_2}
    \begin{split}
    \rho \dot{s} = \frac{1}{T}\bigg[& -\dot{T}\bigg(\rho s +\frac{\partial\tilde{f}}{\partial T}\bigg) - \sum_i
    \frac{\partial\tilde{f}}{\partial c_i}\dot{c}_i - \bigg(\boldsymbol{P}+\frac{\partial\tilde{f}}{\partial \boldsymbol{\mathcal{E}}}\bigg)\cdot\dot{\boldsymbol{\mathcal{E}}}-\bigg(\boldsymbol{\mathcal{M}}+\frac{\partial\tilde{f}}{\partial\boldsymbol{B}}\bigg)\cdot\dot{\boldsymbol{B}}-\tilde{f}(\boldsymbol{I}:\boldsymbol{\nabla}\boldsymbol{v})\\&+[\boldsymbol{T} - \boldsymbol{\mathcal{E}}\otimes\boldsymbol{P} - (\boldsymbol{\mathcal{M}}\cdot\boldsymbol{B})\boldsymbol{I} + \boldsymbol{\mathcal{M}}\otimes\boldsymbol{B}]:\boldsymbol{\nabla}\boldsymbol{v} + \rho r+\sum_i \boldsymbol{j}_i \cdot \boldsymbol{b}_i- \boldsymbol{\nabla}\cdot\bar{\boldsymbol{J}}_\mathrm{q}+ \boldsymbol{\mathcal{J}}^{\mathrm{f}}\cdot\boldsymbol{\mathcal{E}} \bigg]~.
\end{split}
\end{equation}

We now invoke the local equilibrium assumption, 
\begin{equation}
    \rho s = -\bigg(\frac{\partial\tilde{f}}{\partial T}\bigg)_{c_1, c_2, ..., c_N, \boldsymbol{\mathcal{E}},\boldsymbol{B}}~,
\end{equation}
and define the chemical potential of species $i$ as
\begin{equation}
    \mu_i \coloneqq \bigg(\frac{\partial\tilde{f}}{\partial c_i}\bigg)_{T,c_{j\neq i}, \boldsymbol{\mathcal{E}},\boldsymbol{B}}~.
\end{equation}
These definitions along with the mass balance (Eq. \eqref{mass_balance_conc}) allow us to rewrite Eq. \eqref{entropy_balance_1_2} as
\begin{equation}\label{entropy_balance_1_3}
\begin{split}
    \rho \dot{s} = \frac{\boldsymbol{\tau}:\boldsymbol{\nabla}\boldsymbol{v} }{T}+ \frac{\rho r}{T} +\frac{\sum_i \boldsymbol{j}_i \cdot \boldsymbol{b}_i}{T}+& \frac{1}{T}\sum_i\mu_i \boldsymbol{\nabla}\cdot\boldsymbol{J}_i - \frac{\boldsymbol{\nabla}\cdot\bar{\boldsymbol{J}}_\mathrm{q}}{T} +\frac{1}{T}\boldsymbol{\mathcal{J}}^{\mathrm{f}}\cdot\boldsymbol{\mathcal{E}}\\&- \frac{1}{T}\bigg(\boldsymbol{P}+\frac{\partial\tilde{f}}{\partial \boldsymbol{\mathcal{E}}}\bigg)\cdot\dot{\boldsymbol{\mathcal{E}}}-\frac{1}{T}\bigg(\boldsymbol{\mathcal{M}}+\frac{\partial\tilde{f}}{\partial\boldsymbol{B}}\bigg)\cdot\dot{\boldsymbol{B}}~,
\end{split}
\end{equation}
where $\boldsymbol{\tau}$ is defined as 
\begin{equation}\label{tau_1}
    \boldsymbol{\tau} = \boldsymbol{T}+\bigg(\sum_i \mu_i c_i - \tilde{f}\bigg)\boldsymbol{I} - \boldsymbol{\mathcal{E}}\otimes\boldsymbol{P} - (\boldsymbol{\mathcal{M}}\cdot\boldsymbol{B})\boldsymbol{I} + \boldsymbol{\mathcal{M}}\otimes\boldsymbol{B}~,
\end{equation}
which can also be expressed in terms of $\bar{\boldsymbol{T}}$ as
\begin{equation}\label{tau_2}
\begin{split}
    \boldsymbol{\tau}= \bar{\boldsymbol{T}} -\epsilon_0(\boldsymbol{E}\times\boldsymbol{B})\otimes\boldsymbol{v}-\epsilon_0\boldsymbol{E}\otimes\boldsymbol{E} - \frac{1}{\upmu_0}\boldsymbol{B}\otimes\boldsymbol{B}+\frac{1}{2}\bigg[\epsilon_0 E^2 + \frac{1}{\upmu_0} B^2\bigg]\boldsymbol{I}\\+ \bigg(\sum_i \mu_i c_i - \tilde{f}\bigg)\boldsymbol{I} - \boldsymbol{\mathcal{E}}\otimes\boldsymbol{P} - (\boldsymbol{\mathcal{M}}\cdot\boldsymbol{B})\boldsymbol{I} + \boldsymbol{\mathcal{M}}\otimes\boldsymbol{B}~.
\end{split}
\end{equation}
Rearranging, Eq. \eqref{entropy_balance_1_3} becomes
\kf{\begin{equation}\label{entropy_balance_1_4}
\begin{split}
    \rho \dot{s} = - \boldsymbol{\nabla}\cdot\bigg(\frac{\bar{\boldsymbol{J}}_\mathrm{q}-\sum_i\mu_i \boldsymbol{J}_i}{T}\bigg)+\frac{\rho r}{T}+\frac{\boldsymbol{\tau}:\boldsymbol{\nabla}\boldsymbol{v}}{T}- \frac{1}{T}\bigg(\boldsymbol{P}+\frac{\partial\tilde{f}}{\partial \boldsymbol{\mathcal{E}}}\bigg)\cdot\dot{\boldsymbol{\mathcal{E}}}-\frac{1}{T}\bigg(\boldsymbol{\mathcal{M}}+\frac{\partial\tilde{f}}{\partial\boldsymbol{B}}\bigg)\cdot \dot{\boldsymbol{B}}&\\ -\bar{\boldsymbol{J}}_\mathrm{q}\cdot\frac{\boldsymbol{\nabla}T}{T^2}-\sum_i \bigg[\boldsymbol{\nabla}\bigg(\frac{\mu_i}{T}\bigg) -\frac{M_i \boldsymbol{b}_i}{T}\bigg]\cdot\boldsymbol{J}_i+\frac{\boldsymbol{\mathcal{J}}^{\mathrm{f}}\cdot\boldsymbol{\mathcal{E}}}{T}&~.
\end{split}
\end{equation}}
To convert the entropy balance in Eq. \eqref{entropy_balance_1_4} into the form of Eq. \eqref{entropy_balance_form}, we express $\boldsymbol{\mathcal{J}}^{\mathrm{f}}$ in terms of the fluxes of ionic species:
\begin{equation}\label{J_free_microscopic}
    \boldsymbol{\mathcal{J}}^{\mathrm{f}}=\tilde{\boldsymbol{j}}^{\mathrm{f}}-q^{\mathrm{f}}\boldsymbol{v} = \sum_i z_i c_i F(\boldsymbol{v}_i - \boldsymbol{v}) = \sum_i z_i F \boldsymbol{J}_i~,
\end{equation}
where $F$ is Faraday's constant. Note that while the sum over $i$ in this expression includes all types of species in the system, the factor of $z_i$ (the charge valency of species $i$) means that net neutral species such as solvent do not contribute to the free charge conduction current density. In contrast, quantities such as $\sum_i\mu_i \boldsymbol{J}_i$ are influenced by both charged and neutral species. Substitution of Eq. \eqref{J_free_microscopic} into Eq. \eqref{entropy_balance_1_4} yields
\kf{\begin{equation}\label{entropy_balance_1_5}
    \begin{split}
    \rho \dot{s} = - \boldsymbol{\nabla}\cdot\bigg(\frac{\bar{\boldsymbol{J}}_\mathrm{q}-\sum_i\mu_i \boldsymbol{J}_i}{T}\bigg)+\frac{\rho r}{T}+\frac{\boldsymbol{\tau}:\boldsymbol{\nabla}\boldsymbol{v}}{T}- \frac{1}{T}\bigg(\boldsymbol{P}+\frac{\partial\tilde{f}}{\partial \boldsymbol{\mathcal{E}}}\bigg)\cdot\dot{\boldsymbol{\mathcal{E}}}-\frac{1}{T}\bigg(\boldsymbol{\mathcal{M}}+\frac{\partial\tilde{f}}{\partial\boldsymbol{B}}\bigg)\cdot \dot{\boldsymbol{B}} &\\-\bar{\boldsymbol{J}}_\mathrm{q}\cdot\frac{\boldsymbol{\nabla}T}{T^2}-\sum_i \bigg[\boldsymbol{\nabla}\bigg(\frac{\mu_i}{T}\bigg) -\frac{z_i F  \boldsymbol{\mathcal{E}}}{T}-\frac{M_i \boldsymbol{b}_i}{T}\bigg]\cdot\boldsymbol{J}_i&~.
\end{split}
\end{equation}}
Eq. \eqref{entropy_balance_1_5} is the most general form of the entropy balance for a mixture subject to an electromagnetic field. Comparing to Eq. \eqref{entropy_balance_form}, we can deduce that the entropy flux is
\begin{equation}
    \boldsymbol{J}_{\mathrm{s}} = \frac{\bar{\boldsymbol{J}}_\mathrm{q}-\sum_i\mu_i \boldsymbol{J}_i}{T}~,
\end{equation}
the external entropy production is
\begin{equation}
    \rho\sigma_{\mathrm{e}} = \frac{\rho r}{T}~,
\end{equation}
and the internal entropy production is
\kf{\begin{equation}\label{entropy_production_1}
\begin{split}
    \rho\sigma_{\mathrm{i}} = \frac{\boldsymbol{\tau}:\boldsymbol{\nabla}\boldsymbol{v}}{T}- \frac{1}{T}\bigg(\boldsymbol{P}+\frac{\partial\tilde{f}}{\partial \boldsymbol{\mathcal{E}}}\bigg)\cdot\dot{\boldsymbol{\mathcal{E}}}-\frac{1}{T}\bigg(\boldsymbol{\mathcal{M}}+\frac{\partial\tilde{f}}{\partial\boldsymbol{B}}\bigg)\cdot \dot{\boldsymbol{B}} -\bar{\boldsymbol{J}}_\mathrm{q}\cdot\frac{\boldsymbol{\nabla}T}{T^2}\\-\sum_i \bigg[\boldsymbol{\nabla}\bigg(\frac{\mu_i}{T}\bigg) -\frac{z_i F  \boldsymbol{\mathcal{E}}}{T}-\frac{M_i \boldsymbol{b}_i}{T}\bigg]\cdot\boldsymbol{J}_i \geq 0~.
\end{split}
\end{equation}}
If we assume there are no dissipation processes associated with polarization or magnetization, we may define
\begin{equation}\label{p_m}
    \begin{gathered}
        \boldsymbol{P} = -\bigg(\frac{\partial\tilde{f}}{\partial \boldsymbol{\mathcal{E}}}\bigg)_{T, c_1, c_2, ..., c_N,\boldsymbol{B}}~,\\
        \boldsymbol{\mathcal{M}} = -\bigg(\frac{\partial\tilde{f}}{\partial \boldsymbol{B}}\bigg)_{T, c_1, c_2, ..., c_N,\boldsymbol{\mathcal{E}}}~.
    \end{gathered}
\end{equation}
This brings the second and third terms of Eq. \eqref{entropy_production_1} to zero, giving
\kf{\begin{equation}
    \rho\sigma_{\mathrm{i}} = \frac{\boldsymbol{\tau}:\boldsymbol{\nabla}\boldsymbol{v}}{T} -\bar{\boldsymbol{J}}_\mathrm{q}\cdot\frac{\boldsymbol{\nabla}T}{T^2}-\sum_i \bigg[\boldsymbol{\nabla}\bigg(\frac{\mu_i}{T}\bigg) -\frac{z_i F  \boldsymbol{\mathcal{E}}}{T}-\frac{M_i \boldsymbol{b}_i}{T}\bigg]\cdot\boldsymbol{J}_i \geq 0~.
\end{equation}}

$\\$\textbf{Specialization to electrolytes in the absence of a magnetic field}. When $\boldsymbol{B} = 0$, $\boldsymbol{\mathcal{E}} = \boldsymbol{E} = -\boldsymbol{\nabla}\phi$ and Eq. \eqref{entropy_balance_1_5} reduces to
\kf{\begin{equation}\label{entropy_balance_no_magnetic}
    \begin{split}
    \rho \dot{s} = - \boldsymbol{\nabla}\cdot\bigg(\frac{\bar{\boldsymbol{J}}_\mathrm{q}-\sum_i\mu_i \boldsymbol{J}_i}{T}\bigg)&+\frac{\rho r}{T}+\frac{\boldsymbol{\tau}:\boldsymbol{\nabla}\boldsymbol{v}}{T}-\bar{\boldsymbol{J}}_\mathrm{q}\cdot\frac{\boldsymbol{\nabla}T}{T^2}\\&-\sum_i \bigg[\boldsymbol{\nabla}\bigg(\frac{\mu_i}{T}\bigg) +\frac{z_i F  \boldsymbol{\nabla}\phi}{T}-\frac{M_i \boldsymbol{b}_i}{T}\bigg]\cdot\boldsymbol{J}_i~,
\end{split}
\end{equation}}
where $\boldsymbol{\tau}$ is now
\begin{equation}\label{tau_no_magnetic}
\begin{split}
     \boldsymbol{\tau} =& \boldsymbol{T}+\bigg(\sum_i \mu_i c_i - \tilde{f}\bigg)\boldsymbol{I} - \boldsymbol{E}\otimes\boldsymbol{P} \\
      =&\bar{\boldsymbol{T}} -\epsilon_0\boldsymbol{E}\otimes\boldsymbol{E} +\frac{1}{2}\epsilon_0 E^2 \boldsymbol{I}+ \bigg(\sum_i \mu_i c_i - \tilde{f}\bigg)\boldsymbol{I} - \boldsymbol{E}\otimes\boldsymbol{P} ~.
\end{split}
\end{equation}
In this case, internal entropy production is
\kf{\begin{equation}\label{internal_entropy_1}
    \rho\sigma_{\mathrm{i}} = \frac{\boldsymbol{\tau}:\boldsymbol{\nabla}\boldsymbol{v}}{T} -\bar{\boldsymbol{J}}_\mathrm{q}\cdot\frac{\boldsymbol{\nabla}T}{T^2}-\sum_i \bigg[\boldsymbol{\nabla}\bigg(\frac{\mu_i}{T}\bigg) +\frac{z_i F  \boldsymbol{\nabla}\phi}{T}-\frac{M_i \boldsymbol{b}_i}{T}\bigg]\cdot\boldsymbol{J}_i \geq 0
\end{equation}}
We can further simplify this expression by assuming that thermodynamic forces and fluxes of different tensorial characters do not couple with each other in isotropic systems, also called the Curie principle\cite{Curie,DeGroot1969Non-EquilibriumThermodynamics}. Thus, we can split our entropy production inequality as follows:
\kf{\begin{equation}\label{internal_entropy_split}
\begin{gathered}
    \frac{\boldsymbol{\tau}:\boldsymbol{\nabla}\boldsymbol{v}}{T} \geq 0~,\\
    -\bar{\boldsymbol{J}}_\mathrm{q}\cdot\frac{\boldsymbol{\nabla}T}{T^2}-\sum_i \bigg[\boldsymbol{\nabla}\bigg(\frac{\mu_i}{T}\bigg) + \frac{z_i F\boldsymbol{\nabla} \phi}{T} -\frac{M_i \boldsymbol{b}_i}{T}\bigg]\cdot\boldsymbol{J}_i \geq 0~.
\end{gathered}
\end{equation}}
Before proceeding, we must modify the second equality of Eq. \eqref{internal_entropy_split} to account for the fact that a system of $n$ components contains only $n-1$ independent fluxes due to the constraint that $\sum_i \boldsymbol{j}_i = \sum_i M_i \boldsymbol{J}_i = 0$, i.e., we can express the solvent flux, $\boldsymbol{J}_0$, as $\boldsymbol{J}_0 = -\sum_{i\neq 0} \frac{M_i}{M_0} \boldsymbol{J}_i$. Inclusion of this constraint into the second equality of Eq. \eqref{internal_entropy_split} yields an expression of the form
\kf{\begin{equation}\label{entropy_prod_with_constraint}
    -\bar{\boldsymbol{J}}_\mathrm{q}\cdot\frac{\boldsymbol{\nabla}T}{T^2}-\sum_{i\neq 0} \bigg[\boldsymbol{\nabla}\bigg(\frac{\mu_i- \frac{M_i}{M_0}\mu_0}{T}\bigg) + \frac{(z_i - \frac{M_i}{M_0} z_0) F\boldsymbol{\nabla} \phi}{T} -\frac{M_i (\boldsymbol{b}_i-\boldsymbol{b}_0)}{T}\bigg]\cdot\boldsymbol{J}_i \geq 0~.
\end{equation}}
Note that in most cases the solvent charge valency $z_0$ will be equal to zero.

For the remainder of our analysis we will consider the case of an isothermal system with no additional body forces $\boldsymbol{b}_i$, in which case Eq. \eqref{entropy_prod_with_constraint} reduces to
\begin{equation}\label{entropy_inequality}
    -\frac{1}{T}\sum_{i\neq 0}\bigg(\boldsymbol{\nabla}\overline{\mu}_i - \frac{M_i}{M_0}\boldsymbol{\nabla}\overline{\mu}_0\bigg)\cdot \boldsymbol{J}_i \geq 0~.
\end{equation}
In this final expression we have combined the chemical potential and the body force from the external electric field into a single term, the electrochemical potential: $\overline{\mu}_i \coloneqq \mu_i + z_i F \phi$. Note that it is only possible to directly combine these terms after assuming that temperature is constant.

In general, we expect to be able to write internal entropy production as the sum of thermodynamic driving forces, $\boldsymbol{X}_i$, and fluxes, $\boldsymbol{J}_i$\cite{Prigogine1967IntroductionProcesses,DeGroot1969Non-EquilibriumThermodynamics}:
\begin{equation}\label{eq:force_flux}
    \sigma_\mathrm{i} = \sum_i \boldsymbol{J}_i \cdot \boldsymbol{X}_i \geq 0~.
\end{equation}
It is clear from Eqs. \eqref{entropy_inequality} and \eqref{eq:force_flux} that for this system we can choose
\begin{equation}\label{x_i}
    \hat{\boldsymbol{X}}_i = -(\boldsymbol{\nabla}\overline{\mu}_i - \frac{M_i}{M_0}\boldsymbol{\nabla}\overline{\mu}_0)
\end{equation}
and
\begin{equation}\label{j_i}
    \boldsymbol{J}_i = c_i(\boldsymbol{v}_i-\boldsymbol{v})~,
\end{equation}
where $i \neq 0$. In Sec. \ref{sec:transport_coeff}, we will use these definitions to define transport coefficients.

In this section, we derived an expression for internal entropy production using the Helmholtz free energy. The entropy balance is more conventionally derived, however, using the local equilibrium hypothesis and the Gibbs equation\cite{Prigogine1967IntroductionProcesses,DeGroot1969Non-EquilibriumThermodynamics}. For mixtures subject to an electromagnetic field, we do not a priori know the form of the Gibbs equation and therefore could not begin with this approach. We can, however, use our final expressions for the energy and entropy balances to derive the Gibbs equation for these systems (see Appendix \ref{appendix:local_equilibrium}). Equation \eqref{gibbs_eq_dt} could be used as the starting point for deriving internal entropy production in a manner consistent with that presented in this section.

\section{Linear constitutive relations and linear irreversible thermodynamics}\label{sec:linear_irr_thermo}
In what follows, we consider the simplification of the momentum, energy, and entropy balances after proposing linear constitutive relations for the polarization and shear stress. In defining these linear relations, we will restrict our discussion to isotropic materials in the absence of a magnetic field. The assumption of isotropy allows us to make use of the representation theorem saying that any $n$-dimensional isotropic tensor can be generated using the Kronecker delta tensor $\delta_{ij}$ and the $n$-dimensional Levi-Civita tensor $\varepsilon_{i_1, i_2, ... i_n}$\cite{epstein2020time}.

\subsection{Linear isotropic dielectrics}\label{sec:linear_dielectric}
In a linear isotropic dielectric with no dissipation effects, the polarization $\boldsymbol{P}$ is directly proportional to the electric field $\boldsymbol{E}$. The most general such linear relationship is given by a rank-2 isotropic tensor, which by the aforementioned representation theorem must be proportional to the Kronecker delta. This leads to
\begin{equation}\label{linear_relation_p_e}
    \boldsymbol{P} = (\epsilon-\epsilon_0)\boldsymbol{E}~,
\end{equation}
where $\epsilon$ is the dielectric constant of the medium. By Eq. \eqref{displacement_polarization} we also see that 
\begin{equation}\label{d_eps_e}
    \boldsymbol{D}^{\mathrm{f}} = \epsilon \boldsymbol{E}~.
\end{equation}
The assumption of a linear dielectric allows us to integrate the first equality of Eq. \eqref{p_m} to obtain
\begin{equation}\label{f_minus_f0}
    \tilde{f} - \tilde{f}_0 = -\int (\epsilon-\epsilon_0)\boldsymbol{E}\cdot d\boldsymbol{E} = -\frac{1}{2} (\epsilon-\epsilon_0) E^2~,
\end{equation}
where $\tilde{f}_0 (T, c_1, c_2, ..., c_N)$ is the Helmholtz free energy per volume in the absence of an electric field. Equation \eqref{f_minus_f0} can be used to evaluate the electromagnetic contribution to the pressure, $p$. Recall that the pressure is conventionally defined in terms of the total Helmholtz free energy $\mathcal{F} = \tilde{f} V$ as $p := -\frac{\partial \mathcal{F}}{\partial V}\bigg|_{T,n_1,...,n_N,\boldsymbol{\mathcal{E}},\boldsymbol{B}}$, where $V$ is volume. The pressure can be expressed in terms of the free energy per volume as 
\begin{equation}
    p = -\frac{\partial}{\partial V}(\tilde{f}V) = -\tilde{f} - V\frac{\partial\tilde{f}}{\partial V} = -\tilde{f} - V\sum_i\frac{\partial\tilde{f}}{\partial c_i}\frac{\partial c_i}{\partial V}~,
\end{equation}
or
\begin{equation}\label{f_tilde_2}
    \tilde{f} = -p + \sum_i \mu_i c_i~.
\end{equation}
This is identical to the result derived using extensivity arguments in Appendix \ref{appendix:thermo_potentials} (Eq. \eqref{f_tilde}). 
Incorporating Eq. \eqref{f_minus_f0}, we can alternatively write
\begin{equation}\label{p_em_1}
\begin{split}
    p =& -\tilde{f} - V\frac{\partial(\tilde{f_0}-\frac{1}{2} (\epsilon-\epsilon_0) E^2)}{\partial V} \\=& -\tilde{f}_0 +\frac{1}{2} (\epsilon-\epsilon_0) E^2- V\sum_i\frac{\partial\tilde{f_0}}{\partial c_i}\frac{\partial c_i}{\partial V} + V\frac{1}{2} E^2 \frac{\partial\epsilon}{\partial V}~.
\end{split}
\end{equation}
Defining $\mu_{i,0} \coloneqq \frac{\partial\tilde{f_0}}{\partial c_i}$ and $p_0 \coloneqq \sum_i \mu_{i,0} c_i - \tilde{f}_0$ to be the chemical potential of species $i$ and the pressure in the absence of an electric field, respectively, allows us to rewrite Eq. \eqref{p_em_1} as
\begin{equation}\label{pressure}
    p = p_0 + \frac{1}{2}\bigg[\epsilon-\epsilon_0 - \rho \frac{\partial\epsilon}{\partial \rho} \bigg]E^2~.
\end{equation}

\subsection{Newtonian fluids}
In a Newtonian fluid, the shear stress $\boldsymbol{\tau}$ is directly proportional to the velocity gradient: $\boldsymbol{\tau} = \boldsymbol{\eta}^{(4)}\boldsymbol{\nabla}\boldsymbol{v}$, where $\boldsymbol{\eta}^{(4)}$ is the fourth order viscosity tensor. A general fourth order tensor in three dimensions can be written as $\eta_{ijkl} = \eta_1 \delta_{ij}\delta_{kl} + \eta_2 \delta_{ik}\delta_{jl}+\eta_3 \delta_{il}\delta_{jk}$, with three independent parameters. Imposing the symmetry of the stress tensor derived from the angular momentum balance (Sec. \ref{sec:angular_momentum}) eliminates one of these parameters and reduces $\boldsymbol{\tau} = \boldsymbol{\eta}^{(4)}\boldsymbol{\nabla}\boldsymbol{v}$ to
\begin{equation}\label{Newtons_law_viscosity}
    \boldsymbol{\tau} = 2\eta\boldsymbol{d} + \lambda (\boldsymbol{d}:\boldsymbol{I})\boldsymbol{I}~,
\end{equation}
where $\eta$ and $\lambda$ are the two coefficients of viscosity and $\boldsymbol{d}$ is the symmetric part of the velocity gradient tensor, $\boldsymbol{d} = \frac{\boldsymbol{\nabla}\boldsymbol{v} + (\boldsymbol{\nabla}\boldsymbol{v})^T}{2}$. 

The expression for $\tilde{f}$ in Eq. \eqref{f_tilde_2} can be combined with the linear constitutive relation for shear stress to directly evaluate $\boldsymbol{T}$ and $\bar{\boldsymbol{T}}$ and thus write the momentum balances in more useful forms. Substituting Newton's law of viscosity (Eq. \eqref{Newtons_law_viscosity}) and Eqs. \eqref{f_tilde_2} and \eqref{pressure} into our expressions for $\boldsymbol{\tau}$ (Eqs. \eqref{tau_1} and \eqref{tau_2}) in the case of no magnetic field, we see that 
\begin{equation}\label{T_visocsity}
\begin{split}
    \boldsymbol{T} &=  2\eta\boldsymbol{d}+ \lambda (\boldsymbol{d}:\boldsymbol{I})\boldsymbol{I}-p\boldsymbol{I} + \boldsymbol{E}\otimes\boldsymbol{P} \\&= 2\eta\boldsymbol{d}+ \lambda (\boldsymbol{d}:\boldsymbol{I})\boldsymbol{I}-p_0\boldsymbol{I} - \frac{1}{2} \bigg(\epsilon-\epsilon_0-\rho \frac{\partial\epsilon}{\partial \rho}\bigg) E^2\boldsymbol{I} + (\epsilon-\epsilon_0)\boldsymbol{E}\otimes\boldsymbol{E}
\end{split}
\end{equation}
and
\begin{equation}\label{T_bar_viscosity}
\begin{split}
     \bar{\boldsymbol{T}}&=  2\eta\boldsymbol{d}+ \lambda (\boldsymbol{d}:\boldsymbol{I})\boldsymbol{I}+\epsilon_0\boldsymbol{E}\otimes\boldsymbol{E}-\frac{1}{2}\epsilon_0 E^2\boldsymbol{I}-p\boldsymbol{I} + \boldsymbol{E}\otimes\boldsymbol{P} \\&= 2\eta\boldsymbol{d}+ \lambda (\boldsymbol{d}:\boldsymbol{I})\boldsymbol{I}-p_0\boldsymbol{I}+\epsilon\boldsymbol{E}\otimes\boldsymbol{E}-\frac{1}{2}\bigg[\epsilon-\rho \frac{\partial\epsilon}{\partial \rho}\bigg] E^2\boldsymbol{I}~.
\end{split}
\end{equation}
We can substitute these expressions directly into the local momentum balances. Incorporation of Eq. \eqref{T_visocsity} into Eq. \eqref{local_momentum_free_charge} (when $\boldsymbol{B} = 0$ and dielectric constant does not vary with density) gives
\begin{equation}
    \rho \dot{\boldsymbol{v}} = -\boldsymbol{\nabla}p_0 + \eta \nabla^2\boldsymbol{v} + (\eta + \lambda) \boldsymbol{\nabla} (\boldsymbol{\nabla}\cdot\boldsymbol{v}) +\sum_i \rho_i \boldsymbol{b}_i + q^f \boldsymbol{E}~.
\end{equation}
This expression is simply the Navier-Stokes equations with an additional body force acting on the free charges in the system. For an electroneutral system ($q^f = 0$) in a linear dielectric, it is therefore appropriate to use the standard Navier-Stokes equations to analyze momentum transport in an electrolyte.

Analogously, substituting Eq. \eqref{T_bar_viscosity} into the momentum balance in the form of Eq. \eqref{local_momentum_2} yields
\begin{equation}
    \rho \dot{\boldsymbol{v}}  = -\boldsymbol{\nabla}p_0 + \eta \nabla^2\boldsymbol{v} + (\eta + \lambda) \boldsymbol{\nabla} (\boldsymbol{\nabla}\cdot\boldsymbol{v})+\boldsymbol{\nabla}\cdot \bigg[\epsilon\boldsymbol{E}\otimes\boldsymbol{E}-\frac{1}{2}\epsilon E^2\boldsymbol{I}\bigg] + \sum_i \rho_i \boldsymbol{b}_i~.
\end{equation}
We have used the fact that for the case of no magnetic field, $\dot{\boldsymbol{g}} = \dot{\boldsymbol{v}}$. Whichever form of the momentum balance is used, the boundary conditions for momentum transport are always obtained from $\bar{\boldsymbol{T}}\boldsymbol{n}$ (Eq. \eqref{T_bar_viscosity}), not from $\boldsymbol{T}\boldsymbol{n}$ (Eq. \eqref{T_visocsity}).

\subsection{Diffusive transport coefficients}\label{sec:transport_coeff}
Recall from Sec. \ref{sec:entropy_bal} that internal entropy production can be written as a bilinear form relating thermodynamic driving forces and fluxes, i.e., $ \sigma_\mathrm{i} = \sum_i \boldsymbol{J}_i \cdot \boldsymbol{X}_i$. We now postulate linear relations between these forces and fluxes of the following form:
\begin{equation}\label{lij_general}
    \boldsymbol{J}_i = \sum_j \boldsymbol{L}^{ij}\boldsymbol{X}_j
\end{equation}
and
\begin{equation}
    \boldsymbol{X}_j = \sum_i \boldsymbol{M}^{ij}\boldsymbol{J}_i~.
\end{equation}
Each transport coefficient $\boldsymbol{L}^{ij}$ or $\boldsymbol{M}^{ij}$  is a second order tensor in three dimensions, which for an isotropic system may be expressed as $\boldsymbol{L}^{ij} = L^{ij}\boldsymbol{I}$ and $\boldsymbol{M}^{ij} = M^{ij}\boldsymbol{I}$. For the subsequent analysis we consider only the scalar transport coefficients $L^{ij}$ and $M^{ij}$. Additionally, note that $L^{ij}=L^{ji}$ and $M^{ij}=M^{ji}$ by the Onsager reciprocal relations\cite{onsager_reciprocal_1,onsager_reciprocal_2}, as will be apparent from the Green-Kubo relations derived in Sec. \ref{sec:GK_derivation}. 

The second law dictates that
\begin{equation}
    \sigma_\mathrm{i} = \sum_i \boldsymbol{J}_i \cdot \boldsymbol{X}_i = \sum_i \sum_j L^{ij} \boldsymbol{X}_j \cdot \boldsymbol{X}_i \geq 0~.
\end{equation}
Thus the matrix $\boldsymbol{L}$ composed of each of the $L^{ij}$ coefficients is positive semi-definite. This provides some information on the possible values for each $L^{ij}$, for example that the diagonal elements $L^{ii}$ must be greater than or equal to zero and that $\sum_i \sum_j L^{ij} \geq 0$. Furthermore, the condition that the eigenvalues of $\boldsymbol{L}$ must be real and greater than or equal to zero tells us that the principal invariants of $\boldsymbol{L}$ are positive. Thus, the determinant of $\boldsymbol{L}$ is positive, for example $L^{++}L^{--}-L^{+-2} \geq 0$ for a binary electrolyte of a single type of cation ($+$) and anion ($-$).

The choices of force and flux defined in Eqs. \eqref{x_i} and \eqref{j_i} yield the following relations:\footnote{The linear laws can easily be generalized to the case of non-isothermal systems, where we could have \begin{equation}\label{eq:temp_coupling}
    \boldsymbol{J}_i =  c_i(\boldsymbol{v}_i - \boldsymbol{v}) = - \frac{L^{iT}}{T^2}\boldsymbol{\nabla}T -\sum_{j\neq 0} L^{ij} \bigg(\boldsymbol{\nabla}\bigg(\frac{\mu_j}{T}\bigg) + \frac{z_i F \boldsymbol{\nabla}\phi}{T} - \frac{M_j}{M_0}\bigg[\boldsymbol{\nabla}\bigg(\frac{\mu_0}{T}\bigg) + \frac{z_0 F \boldsymbol{\nabla}\phi}{T}\bigg]\bigg)~,
\end{equation}
which captures cross-coupling effects between temperature gradients and species flux, i.e., the Soret effect.}
\begin{equation}\label{eq:Lij_hat}
    c_i(\boldsymbol{v}_i - \boldsymbol{v}) = -\sum_{j\neq 0} L^{ij} \bigg(\boldsymbol{\nabla}\overline{\mu}_j - \frac{M_j}{M_0}\boldsymbol{\nabla}\overline{\mu}_0\bigg)
\end{equation}
and
\begin{equation}\label{eq:Mij_hat}
     - \bigg(\boldsymbol{\nabla}\overline{\mu}_i - \frac{M_i}{M_0}\boldsymbol{\nabla}\overline{\mu}_0\bigg) = \sum_{j\neq 0} M^{ij}c_j(\boldsymbol{v}_j - \boldsymbol{v})~.
\end{equation}
Note that based on this formulation, the transport coefficients $L^{ij}$ and $M^{ij}$ are not defined for $i$ or $j$ equal to the solvent, species $0$. To reformulate Eqs. \eqref{eq:Lij_hat} and \eqref{eq:Mij_hat} when the solvent is also included as one of the species, one can define
\begin{equation}\label{eq:L_solvent}
    L^{i0} = L^{0i} = - \sum_{j\neq0}\frac{M_j}{M_0}L^{ij}~,
\end{equation}
which yields a simpler, more convenient equation:
\begin{equation}\label{Lij}
    c_i(\boldsymbol{v}_i - \boldsymbol{v}) = -\sum_{j} L^{ij} \boldsymbol{\nabla}\overline{\mu}_j~,
 \end{equation}
where the summation is now over all species. Note that isotropy, the Onsager reciprocal relations, and the constraint that $\sum_i M_i L^{ij} = 0$ (Eq. \eqref{eq:L_solvent}) implies that an $n$-component electrolyte has $n(n-1)/2$ independent transport coefficients.

\section{\label{sec:statMech}Statistical mechanics of transport phenomena}
\subsection{\label{sec:GK_derivation}Green-Kubo relations}
The balance laws presented in the previous sections allow us to describe macroscopic, boundary-driven transport phenomena in systems out of equilibrium. In this section, we will derive Green-Kubo relations\cite{green1954markoff,kubo1957statistical} to relate the diffusive transport coefficients $L^{ij}$ to the decay of fluctuations at equilibrium. This connection between the deterministic, continuum level transport theory and the stochastic behavior observed at the molecular level is enabled by the Onsager regression hypothesis, one of the most important developments of nonequilibrium statistical mechanics\cite{onsager_reciprocal_1,onsager_reciprocal_2}. This hypothesis states that the relaxation, or regression, of spontaneous fluctuations in an aged system in equilibrium is governed by the same laws which describe the response to macroscopic perturbations away from equilibrium. The regression hypothesis was used by Kubo \cite{kubo1957statistical} to derive the Green-Kubo relations, which enable facile computation of transport coefficients from molecular dynamics simulations. This offers a means of rigorously studying transport in systems where experimental characterization may be challenging or impractical, for example in screening new electrolyte chemistries or studying systems with more than two types of ionic species.

Consider a system at equilibrium in which each species $k$ has a mean concentration $\overline{c}_k$. Thermal fluctuations at equilibrium will induce small fluctuations in concentration, $\delta c_k$, about this mean value. The concentration of species $k$ at any instant $c_k(\boldsymbol{x},t)$ may thus be expressed as $c_k(\boldsymbol{x},t) = \overline{c}_k + \delta c_k(\boldsymbol{x},t)$. The species mass balance (Eq. \eqref{mass_balance_conc}) may be modified at equilibrium to be
\begin{equation}\label{mass_bal_fluctuations}
    \frac{\partial (\delta c_i)}{\partial t} = -\boldsymbol{\nabla}\cdot \boldsymbol{J}_i~.
\end{equation}
In writing Eq. \eqref{mass_bal_fluctuations} we have used the fact that $\boldsymbol{v}=0$ for a system at equilibrium. Substituting the constitutive relation in Eq. \eqref{Lij} into Eq. \eqref{mass_bal_fluctuations} yields
\begin{equation} \label{mass_bal_withLij}
    \frac{\partial (\delta c_i)}{\partial t} = \boldsymbol{\nabla} \cdot \bigg[\sum_j L^{ij} \boldsymbol{\nabla} \overline{\mu}_j\bigg]~.
\end{equation}
Using Eq. \eqref{df_tilde} with constant temperature and the magnetic field $\boldsymbol{B} = 0$, the quantity $\boldsymbol{\nabla} \overline{\mu}_j$ can be written in terms of the electric field and concentration as
\begin{equation}
    \boldsymbol{\nabla} \overline{\mu}_j = \boldsymbol{\nabla}\mu_j + z_j F \boldsymbol{\nabla}\phi= - \bigg(\frac{\partial \boldsymbol{P}}{\partial c_j}\bigg)_{T,c_{l\neq j},\boldsymbol{E}}\cdot\boldsymbol{\nabla}\boldsymbol{E} + \sum_{k}\bigg(\frac{\partial\mu_j}{\partial c_k}\bigg)_{T,c_{l\neq k}, \boldsymbol{E}}\boldsymbol{\nabla} c_k+ z_j F \boldsymbol{\nabla}\phi~,
\end{equation}
where as before $\boldsymbol{E} = -\boldsymbol{\nabla}\phi$.
Using Eq. \eqref{linear_relation_p_e}, we can write $\big(\frac{\partial \boldsymbol{P}}{\partial c_j}\big)_{T,c_{l\neq j},\boldsymbol{E}}\cdot\boldsymbol{\nabla}\boldsymbol{E} = - \big(\frac{\partial (\epsilon-\epsilon_0)\boldsymbol{E}}{\partial c_j}\big)_{T,c_{l\neq j},\boldsymbol{E}}\cdot\boldsymbol{\nabla}\boldsymbol{E}$. Assuming changes in dielectric constant with respect to concentration are negligible, this term can be eliminated. Thus, the gradient in electrochemical potential is
\begin{equation}
    \boldsymbol{\nabla}\overline{\mu}_j = \sum_{k}\bigg(\frac{\partial\mu_j}{\partial c_k}\bigg)_{T,c_{l\neq k}, \boldsymbol{E}}\boldsymbol{\nabla} c_k + z_j F \boldsymbol{\nabla}\phi~.
\end{equation}
Now, Eq. \eqref{mass_bal_withLij} becomes
\begin{equation} 
    \frac{\partial (\delta c_i)}{\partial t}  = \boldsymbol{\nabla} \cdot \bigg[\sum_j L^{ij} \bigg(\sum_k \frac{\partial\mu_j}{\partial c_k}\bigg|_{T,c_{l\neq k},\boldsymbol{E}}\boldsymbol{\nabla} c_k + z_j F \boldsymbol{\nabla}\phi\bigg)\bigg]~.
\end{equation}
The term $\frac{\partial\mu_j}{\partial c_k}$ can be rewritten in terms of $\delta c_i$ at equilibrium (see Appendix \ref{appendix:concFluctuations}):
\begin{equation}
	\frac{\partial\mu_j}{\partial c_k} = \frac{1}{\beta V}(\boldsymbol{K}_{\mathrm{CC}}^{-1})^{kj}~,
\end{equation}
where $\boldsymbol{K}_{\mathrm{CC}}$ is the covariance matrix with elements $\big<\delta c_i \delta c_j\big>$ and $\beta = (k_\mathrm{B}T)^{-1}$, where $k_\mathrm{B}$ is the Boltzmann constant. We thus have
\begin{equation} 
    \frac{\partial (\delta c_i)}{\partial t}  = \boldsymbol{\nabla} \cdot \bigg[\sum_j L^{ij} \bigg(\sum_k \frac{1}{\beta V}(\boldsymbol{K}_{\mathrm{CC}}^{-1})^{kj}\boldsymbol{\nabla} c_k + z_j F \boldsymbol{\nabla}\phi\bigg)\bigg]~.
\end{equation}
Expanding the right hand side yields
\kf{\begin{equation}\label{mass_bal_expanded}
\begin{split}
    \frac{\partial (\delta c_i)}{\partial t}  = \sum_j L^{ij}\bigg[ \sum_k \frac{1}{\beta V}(\boldsymbol{K}_{\mathrm{CC}}^{-1})^{kj} &\boldsymbol{\nabla}^2 c_k- z_j F \boldsymbol{\nabla}\cdot \boldsymbol{E}\bigg] \\&+ \sum_j \boldsymbol{\nabla} L^{ij} \cdot \bigg[ \sum_k \frac{1}{\beta V}(\boldsymbol{K}_{\mathrm{CC}}^{-1})^{kj} \boldsymbol{\nabla} c_k - z_j F \boldsymbol{E}\bigg]~.
\end{split}
\end{equation}}
By Eqs. \eqref{gauss_medium} and \eqref{d_eps_e} for a system with uniform dielectric constant, $\boldsymbol{\nabla}\cdot \boldsymbol{E} = q^{\mathrm{f}}/\epsilon$. Over length scales shorter than the Debye length, electroneutrality may be violated, yielding a nonzero value of $q^{\mathrm{f}}$. In Green-Kubo relations, however, we are only interested in describing long wavelength fluctuations at equilibrium; thus, the quantity $\boldsymbol{\nabla}\cdot \boldsymbol{E}$ may be neglected. 

The transport coefficients $L^{ij}$ depend on concentration. Given that the concentration fluctuations at equilibrium are small, however, we can linearize $L^{ij}$ around the mean solution concentration to obtain
\begin{equation}\label{linearization}
    L^{ij} = L^{ij}\bigg|_{\overline{c}_m} + \sum_l\frac{\partial L^{ij}}{\partial c_l}\bigg|_{\overline{c}_{m}} (\delta c_l) ~.
\end{equation}
Using Eq. \eqref{linearization} and rewriting all terms in terms of concentration fluctuations, Eq. \eqref{mass_bal_expanded} becomes
\kf{\begin{equation}\label{linearization_1}
\begin{split}
     \frac{\partial (\delta c_i)}{\partial t} = \sum_j \bigg(L^{ij}\bigg|_{\overline{c}_m} +& \sum_l\frac{\partial L^{ij}}{\partial c_l}\bigg|_{\overline{c}_{m}} (\delta c_l) \bigg)\bigg(\sum_k\frac{(\boldsymbol{K}_{\mathrm{CC}}^{-1})^{kj}}{\beta V}\boldsymbol{\nabla}^2 \delta c_k\bigg) \\&+ \sum_j \sum_l\frac{\partial L^{ij}}{\partial c_l}\bigg|_{\overline{c}_{m}}(\boldsymbol{\nabla} \delta c_l)\bigg( \sum_k\frac{(\boldsymbol{K}_{\mathrm{CC}}^{-1})^{kj}}{\beta V}\boldsymbol{\nabla} c_k - z_j F \boldsymbol{E}\bigg)~.
\end{split}
\end{equation}}
\kf{Eliminating the terms in Eq. \eqref{linearization_1} which are negligibly small leads to an evolution equation for the concentration:}
\begin{equation} \label{eq:linearized_Lij_MB}
    \frac{\partial (\delta c_i)}{\partial t} = \sum_j \sum_k\frac{(\boldsymbol{K}_{\mathrm{CC}}^{-1})^{kj}}{\beta V} L^{ij} \boldsymbol{\nabla}^2 (\delta c_k) ~,
\end{equation}
where we have removed the subscript on $L^{ij}\big|_{\overline{c}_m}$ for simplicity.

Let us express the concentration as a Fourier series in terms of the wavevector $\boldsymbol{k}$:
\begin{equation}
    \delta c_j = \sum_{\boldsymbol{k}} \delta c_j (\boldsymbol{k},t) e^{i\boldsymbol{k}\cdot \boldsymbol{x}}~,
\end{equation}
which leads to
\begin{equation}
    \boldsymbol{\nabla}^2 (\delta c_j) = \sum_{\boldsymbol{k}} -k^2 \delta c_j (\boldsymbol{k},t) e^{i\boldsymbol{k}\cdot \boldsymbol{x}}~.
\end{equation}
Equation \eqref{mass_bal_fluctuations} may be thus be written as
\begin{equation}\label{mass_bal_fluctuations_fourier}
    \delta\dot{c}_i(\boldsymbol{k},t) = -i \boldsymbol{k}\cdot \boldsymbol{J}_i(\boldsymbol{k},t)~,
\end{equation}
where we have used $\boldsymbol{J}_i = \sum_{\boldsymbol{k}} \boldsymbol{J}_i(\boldsymbol{k},t) e^{i\boldsymbol{k}\cdot \boldsymbol{x}}$. Analogously, Eq. \eqref{eq:linearized_Lij_MB} is transformed into
\begin{equation}\label{after_fourier}
    \delta \dot{c}_i (\boldsymbol{k},t) = -k^2 \frac{k_{\mathrm{B}} T }{V} \bigg[\sum_j \sum_m (\boldsymbol{K}_{\mathrm{CC}}^{-1})^{mj} L^{ij}\delta c_m(\boldsymbol{k},t)\bigg] ~.
\end{equation}
We now multiply both sides of Eq. \eqref{after_fourier} by $\delta c_l(\boldsymbol{-k}, 0)$ and take an ensemble average, giving
\begin{equation}\label{after_fourier_2}
\begin{split}
    \big<\delta \dot{c}_i(\boldsymbol{k},t) \delta c_l(-\boldsymbol{k}, 0)\big>  = -k^2 \frac{k_{\mathrm{B}} T }{V} \bigg[\sum_j \sum_m (\boldsymbol{K}_{\mathrm{CC}}^{-1})^{mj} L^{ij}\big<\delta c_m(\boldsymbol{k}, t)\delta c_l(-\boldsymbol{k}, 0)\big> \bigg]~.
\end{split}
\end{equation}
Defining the correlation function $A^{ij} = \big<\delta c_i(\boldsymbol{k}, t)\delta c_j(-\boldsymbol{k}, 0) \big>$, Eq. \eqref{after_fourier_2} can be rewritten as
\begin{equation}\label{dAdt}
    \frac{dA^{il}}{dt} = -k^2 \frac{k_{\mathrm{B}} T }{V} \bigg[\sum_j \sum_m (\boldsymbol{K}_{\mathrm{CC}}^{-1})^{mj} L^{ij}A^{ml}(\boldsymbol{k},t) \bigg]~.
\end{equation}

\kf{We proceed by taking a Laplace transform of Eq. \eqref{dAdt}, defining the Laplace transform $\widetilde{A}(s)$ of a quantity $A(t)$ to be $\widetilde{A}(s) = \int_0^\infty dt e^{-st}A(t)$, where $s$ is a complex frequency parameter.} Using integration by parts, the left hand side of Eq. \eqref{dAdt} becomes
\begin{equation}\label{laplace_lhs}
\begin{split}
    \int_0^\infty dt e^{-st} \frac{dA^{il}}{dt} = e^{-st} A^{il}\bigg|_0^\infty - \int_0^\infty dt e^{-st}(-s)A^{il}  = s\widetilde{A}^{il}(\boldsymbol{k},s) - A^{il}(\boldsymbol{k},0)~,
\end{split}
\end{equation}
 and the right side of the equation is
\begin{equation}\label{laplace_rhs}
\begin{split}
    \int_0^\infty dt e^{-st} \bigg{(}-k^2 \frac{k_{\mathrm{B}} T }{V} \bigg[\sum_j \sum_m& (\boldsymbol{K}_{\mathrm{CC}}^{-1})^{mj} L^{ij}A^{ml}(\boldsymbol{k},t) \bigg] \bigg{)}\\&= -k^2  \frac{k_{\mathrm{B}} T }{V} \sum_j \sum_m (\boldsymbol{K}_{\mathrm{CC}}^{-1})^{mj}L^{ij}\widetilde{A}^{ml}(\boldsymbol{k},s)~.
\end{split}
\end{equation}
Combining Eqs. \eqref{laplace_lhs} and \eqref{laplace_rhs}, Eq. \eqref{dAdt} reduces to
\begin{equation}
    s\widetilde{A}^{il}(\boldsymbol{k},s) - A^{il}(\boldsymbol{k},0) = -k^2  \frac{k_{\mathrm{B}} T }{V} \sum_j \sum_m (\boldsymbol{K}_{\mathrm{CC}}^{-1})^{mj}L^{ij}\widetilde{A}^{ml}(\boldsymbol{k},s)~.
\end{equation}
Solving for the transport coefficient $L^{il}$ yields
\begin{equation}\label{L_il_1}
    L^{il} = \frac{s\widetilde{A}^{il}(\boldsymbol{k},s) - A^{il}(\boldsymbol{k},0) + k^2  \frac{k_{\mathrm{B}} T }{V} \sum_{j\neq l} \sum_m (\boldsymbol{K}_{\mathrm{CC}}^{-1})^{mj}L^{ij}\widetilde{A}^{ml}(\boldsymbol{k},s)}{-k^2  \frac{k_{\mathrm{B}} T }{V} \sum_m (\boldsymbol{K}_{\mathrm{CC}}^{-1})^{ml}\widetilde{A}^{ml}(\boldsymbol{k},s)}~.
\end{equation}

Now consider a new function, $\phi^{ij}(\boldsymbol{k},t)$, defined as
\begin{equation}
    \phi^{ij}(\boldsymbol{k},t) \coloneqq \big<\delta \dot{c}_i(\boldsymbol{k},t)  \delta \dot{c}_j(-\boldsymbol{k},0) \big> = -\frac{d^2A^{ij}}{dt^2}~.
\end{equation}
The Laplace transform of $\phi^{ij}(\boldsymbol{k},t)$ is
\begin{equation}
    -\widetilde{\phi}^{ij}(\boldsymbol{k},s) = s^2 \widetilde{A}^{ij}(\boldsymbol{k},s)-sA^{ij}(\boldsymbol{k},0)~,
\end{equation}
where we used the relation $\dot{A}^{ij}(0)=0$. Substituting $\phi^{ij}(\boldsymbol{k},t)$ into Eq. \eqref{L_il_1} yields
\begin{equation}\label{L_il_2}
    L^{il} = \frac{-\widetilde{\phi}^{il}(\boldsymbol{k}, s)\big/s+k^2 \frac{k_{\mathrm{B}} T }{V} \sum_{j\neq l} \sum_m (\boldsymbol{K}_{\mathrm{CC}}^{-1})^{mj}L^{ij}
    \big(-\widetilde{\phi}^{ml}(\boldsymbol{k}, s)\big/s^2 + A^{ml}(\boldsymbol{k},0)\big/s\big)}
    {-k^2\frac{k_{\mathrm{B}} T }{V}  \sum_m (\boldsymbol{K}_{\mathrm{CC}}^{-1})^{ml}\big(-\widetilde{\phi}^{ml}(\boldsymbol{k}, s)\big/s^2 + A^{ml}(\boldsymbol{k},0)\big/s\big)}~.
\end{equation}

Let us consider large wavelength fluctuations, corresponding to the limit of $\boldsymbol{k}$ tending to zero. Under this limit, Equation \eqref{L_il_2} simplifies to:
\begin{equation}
\begin{split}
    L^{il} &= \lim_{\boldsymbol{k}\to0} \frac{\widetilde{\phi}^{il}(\boldsymbol{k},s)}{k^2 \frac{k_{\mathrm{B}} T }{V}\sum_m (\boldsymbol{K}_{\mathrm{CC}}^{-1})^{ml}(-\widetilde{\phi}^{ml}(\boldsymbol{k}, s)\big/s + A^{ml}(\boldsymbol{k},0)\big)}
    \\&= \lim_{\boldsymbol{k}\to0}
    \frac{\int_0^\infty dt e^{-st}\big<\delta \dot{c}_i(\boldsymbol{k},t) \delta \dot{c}_l(-\boldsymbol{k},0) \big>}
    {k^2 \frac{k_{\mathrm{B}} T}{V}\sum_m (\boldsymbol{K}_{\mathrm{CC}}^{-1})^{ml}(-\int_0^\infty dt e^{-st}/s\big<\delta \dot{c}_m(\boldsymbol{k},t) \delta \dot{c}_l(-\boldsymbol{k},0) \big> + A^{ml}(\boldsymbol{k},0))}~.
\end{split}
\end{equation}
Further simplification by substituting Eq. \eqref{mass_bal_fluctuations_fourier} leads to
\begin{equation}
    L^{il} = \lim_{\boldsymbol{k}\to0}
      \frac{\frac{V}{k_{\mathrm{B}} T }\int_0^\infty dt e^{-st}\big<\big[(-i\boldsymbol{k})\cdot \boldsymbol{J}_i(\boldsymbol{k},t)\big]\big[(i\boldsymbol{k})\cdot \boldsymbol{J}_l(-\boldsymbol{k},0)\big]\big>}{k^2 \sum_m (\boldsymbol{K}_{\mathrm{CC}}^{-1})^{ml}(-\int_0^\infty dt e^{-st}/s\big<\big[(-i\boldsymbol{k})\cdot \boldsymbol{J}_m(\boldsymbol{k},t)\big]\big[(i\boldsymbol{k})\cdot \boldsymbol{J}_l(-\boldsymbol{k},0)\big]\big>+ A^{ml}(\boldsymbol{k},0))}~.
\end{equation}
We now invoke the assumption that the system is isotropic and continue by using $\boldsymbol{k} = k \boldsymbol{e}_x$,  which yields
\begin{equation}
\begin{split}
    L^{il} &= \lim_{\boldsymbol{k}\to0}
      \frac{k^2\frac{V}{k_{\mathrm{B}} T }\int_0^\infty dt e^{-st}\big<J_{i,x}(\boldsymbol{k},t) J_{l,x}(-\boldsymbol{k},0)\big>}{k^2 \sum_m (\boldsymbol{K}_{\mathrm{CC}}^{-1})^{ml}\big(k^2\int_0^\infty dt e^{-st}/s\big<J_{m,x}(\boldsymbol{k},t)J_{l,x}(-\boldsymbol{k},0)\big>+ A^{ml}(\boldsymbol{k},0)\big)}
      \\&= \lim_{\boldsymbol{k}\to0}
      \frac{\frac{V}{k_{\mathrm{B}} T }\int_0^\infty dt e^{-st}\big<J_{i,x}(\boldsymbol{k},t) J_{l,x}(-\boldsymbol{k},0)\big>}{\sum_m (\boldsymbol{K}_{\mathrm{CC}}^{-1})^{ml} A^{ml}(\boldsymbol{k},0)}~.
\end{split}
\end{equation}
Note that $A^{ml}(0,0) = \boldsymbol{K}_{\mathrm{CC}}^{ml}$; thus the quantity $\sum_m (\boldsymbol{K}_{\mathrm{CC}}^{-1})^{ml} A^{ml}(0,0) = 1$. Further, taking the limit as $s$ tends to zero (corresponding to the long-time limit of equilibrium processes) yields the Green-Kubo expression
\begin{equation}
    L^{il} = \frac{V}{k_{\mathrm{B}}T}\int_0^\infty dt \big<J_{i,x}(0,t)\cdot J_{l,x}(0,0) \big>~.
\end{equation}
Equivalent expressions can be obtained using the $y$- or $z$-components of $\boldsymbol{J}_i$ as well. We can thus average over all three spatial dimensions to obtain (after a change of indices) the final form of the Green-Kubo relations for transport coeffieints $L^{ij}$ as 
\begin{equation}\label{eq:GK_1}
    L^{ij} = \frac{V}{3k_{\mathrm{B}}T}\int_0^\infty dt \big<\boldsymbol{J}_i(t)\cdot \boldsymbol{J}_j(0) \big>~.
\end{equation}

Our definition for $L^{i0}$ in Eq. \eqref{eq:L_solvent} is automatically satisfied by Eq. \eqref{eq:GK_1}:
\begin{equation}
    L^{i0} = \frac{V}{3k_{\mathrm{B}}T}\int_0^\infty dt \big<\boldsymbol{J}_i(t)\cdot \boldsymbol{J}_0(0) \big>~.
\end{equation}
Incorporating the constraint that all fluxes sum to zero yields
\begin{equation}
\begin{split}
    L^{i0} &= \frac{V}{3k_{\mathrm{B}}T}\int_0^\infty dt \bigg<\boldsymbol{J}_i(t)\cdot \bigg(-\sum_{j\neq0} \frac{M_j}{M_0}\boldsymbol{J}_j(0)\bigg) \bigg> \\&= -\frac{V}{3k_{\mathrm{B}}T}\sum_{j\neq0}\frac{M_j}{M_0}\int_0^\infty dt \big<\boldsymbol{J}_i(t)\cdot \boldsymbol{J}_j(0) \big> =-\sum_{j\neq0}\frac{M_j}{M_0}L^{ij} ~.
\end{split}
\end{equation}

In summary, we note that the derivation presented in this section has made use of the following assumptions: the system is an isotropic, isothermal, linear dielectric; there is no applied magnetic field; and changes in the dielectric constant with concentration are negligible. Furthermore, the final Green-Kubo relations capture only long wavelength fluctuations, i.e., they are valid on larger length scales for which we may assume electroneutrality.

The Green-Kubo relations for $L^{ij}$ (Eq. \eqref{eq:GK_1}) provide direct insight into the physical meaning of the Onsager transport coefficients on a molecular level. As these expressions consist of correlation functions between fluxes, it is clear that $L^{ij}$ captures the extent of correlation between the motion of species $i$ and $j$. For a binary electrolyte, the quantity $L^{+-}$ captures correlations between cations and anions, i.e. a positive value of $L^{+-}$ suggests that cations and anions are moving in a concerted manner, for example as ion pairs or aggregates. The diagonal terms $L^{++}$ and $L^{--}$ capture both self-diffusion of individual cations or anions, respectively, as well as correlations between distinct ions of the same type. In Sec. \ref{sec:MD}, we will demonstrate the types of physical insight one can obtain from knowing $L^{ij}$ using molecular simulations of a model electrolyte.

\subsection{Linear response theory}\label{sec:linearReponse}
For systems which can be described with a Hamiltonian, the Green-Kubo relations for the transport coefficients $L^{ij}$ can also be derived through linear response theory, where we couple the system to a weak external perturbation and observe the resulting response. This derivation parallels that of Evans and Morriss\cite{MorrissP.2007StatisticalLiquids} as well as Wheeler and Newman\cite{Wheeler2004MolecularMethod}. 

Let a system in equilibrium be described by a Hamiltonian $H = H_0(\{\boldsymbol{r}^\alpha,\boldsymbol{p}^\alpha\})$, where $\{\boldsymbol{r}^\alpha,\boldsymbol{p}^\alpha\}$ is the set of all particle positions and momenta. For a conservative system, $H_0$ gives the sum of the kinetic and potential energy of the system. Time evolution of the system is governed by Hamilton's equations of motion, $\dot{\boldsymbol{r}}^\alpha =  \frac{\partial H_0}{\partial \boldsymbol{p}^\alpha}$ and $\dot{\boldsymbol{p}}^\alpha =  -\frac{\partial H_0}{\partial \boldsymbol{r}^\alpha}$.

We now introduce a small, constant external force on the equilibrium ensemble. The Hamiltonian for this perturbed system $H$ is
\begin{equation}
    H = H_0 -\sum_j \boldsymbol{\mathcal{R}}_j\cdot\boldsymbol{\mathcal{F}}_j~,
\end{equation}
where $\boldsymbol{\mathcal{F}}_j$ is a force acting on species $j$ and $\boldsymbol{\mathcal{R}}_j$ is a function of the position of species $j$. For sufficiently small $\boldsymbol{\mathcal{F}}_j$, the expectation value of any observable $\boldsymbol{\mathcal{B}}$ in this perturbed system (derived in Evans and Morriss\cite{MorrissP.2007StatisticalLiquids}) is:
\begin{equation}\label{eq:B}
    \big<\boldsymbol{\mathcal{B}}(t)\big> =\big<\boldsymbol{\mathcal{B}}(0)\big>_0 + \beta\int_0^t ds \big<\boldsymbol{\mathcal{B}}(s)\sum_j\dot{\boldsymbol{\mathcal{R}}}_j(0)\cdot\boldsymbol{\mathcal{F}}_j\big>_0~,
\end{equation}
where the notation $<\;>_0$ denotes an average over the equilibrium ensemble corresponding to $H_0$. 

For electrolyte solutions, we choose $\boldsymbol{\mathcal{F}}_j = \boldsymbol{X}_j c_j V$, where once again $\boldsymbol{X}_j = -\boldsymbol{\nabla} \overline{\mu}_j$, and $\boldsymbol{\mathcal{R}}_j = 
\overline{\boldsymbol{r}}_j - \overline{\boldsymbol{r}}$. The quantity $\overline{\boldsymbol{r}}_j$ is the average position of species $j$, i.e., $\overline{\boldsymbol{r}}_j = \frac{1}{N_j}\sum_{\alpha} \boldsymbol{r}_j^\alpha$, where the notation $\boldsymbol{r}_j^\alpha$ refers to an atom/molecule $\alpha$ of type $j$ and $N_j$ is the number of particles of species $j$. The quantity $\overline{\boldsymbol{r}}$ is the center-of-mass position of the system, defined using the mass of each particle $\alpha$, $m^\alpha$, as $\overline{\boldsymbol{r}} = (\sum_\alpha m^\alpha \boldsymbol{r}^\alpha)/\sum_\beta m^\beta$. This choice of $\boldsymbol{\mathcal{F}}_j$ and $\boldsymbol{\mathcal{R}}_j$ corresponds to a perturbed Hamiltonian of
\begin{equation}
    H = H_0 -\sum_j (\overline{\boldsymbol{r}}_j - \overline{\boldsymbol{r}})\cdot \boldsymbol{X}_j c_j V~.
\end{equation}
This perturbation to the Hamiltonian modifies both the energy of the system as well as the equations of motion, which can now be written as $\dot{\boldsymbol{r}}^\alpha =  \frac{\partial H_0}{\partial \boldsymbol{p}^\alpha}$ and $\dot{\boldsymbol{p}}^\alpha =  -\frac{\partial H_0}{\partial \boldsymbol{r}^\alpha}-\boldsymbol{\nabla}\overline{\mu}_i$, where the additional force $-\boldsymbol{\nabla}\overline{\mu}_i$ is only applied to atoms $\alpha$ corresponding to type $i$. It is clear how the presence of $\boldsymbol{\nabla}\overline{\mu}_i$ results in an additional force driving the acceleration of particle $\alpha$ down its electrochemical potential gradient. In the absence of chemical potential gradients, the second equation of motion simply reduces to $\dot{\boldsymbol{p}}^\alpha =  -\frac{\partial H_0}{\partial \boldsymbol{r}^\alpha}+\hat{q}^\alpha\boldsymbol{E}$.

Noting that $\dot{\boldsymbol{\mathcal{R}}}_j = \boldsymbol{v}_j - \boldsymbol{v}$, the quantity $\sum_j\dot{\boldsymbol{\mathcal{R}}}_j(0)\cdot\boldsymbol{\mathcal{F}}_j$ appearing in Eq. \eqref{eq:B} is
\begin{equation} \label{rDotF}
    \sum_j\dot{\boldsymbol{\mathcal{R}}}_j(0)\cdot\boldsymbol{\mathcal{F}}_j = \sum_j c_j V (\boldsymbol{v}_j(0) - \boldsymbol{v}(0))\cdot\boldsymbol{X}_j~.
\end{equation}
Recalling the definition of species flux, $\boldsymbol{J}_i = c_i(\boldsymbol{v}_i-\boldsymbol{v})$, we can rewrite Eq. \eqref{rDotF} as
\begin{equation}\label{rDotF_1}
    \sum_j\dot{\boldsymbol{\mathcal{R}}}_j(0)\cdot\boldsymbol{\mathcal{F}}_j= \sum_j \boldsymbol{J}_j(0)\cdot\boldsymbol{X}_jV~.
\end{equation}
Substituting Eq. \eqref{rDotF_1} into Eq. \eqref{eq:B} gives
\begin{equation}
     \big<\boldsymbol{\mathcal{B}}(t)\big> =  \big<\boldsymbol{\mathcal{B}}(0)\big>_0 + V\beta \int_0^t ds \big<\boldsymbol{\mathcal{B}}(s)\sum_j\boldsymbol{J}_j(0)\cdot\boldsymbol{X}_j
     \big>_0~.
\end{equation}
We proceed by choosing $\boldsymbol{\mathcal{B}} = \boldsymbol{J}_i$, yielding
\begin{equation}\label{eq:linearResponse_J}
    \big<\boldsymbol{J}_i(t)\big> =  \big<\boldsymbol{J}_i(0)\big>_0 + V\beta \int_0^t ds \big<\boldsymbol{J}_i(s)\sum_j\boldsymbol{J}_j(0)\cdot\boldsymbol{X}_j
     \big>_0~.
\end{equation}
The quantity $\big<\boldsymbol{J}_i(0)\big>_0$ is equal to zero, as there is no net flux of any species at equilibrium. Furthermore, $\boldsymbol{X}_j$ is time-independent and can be written outside the time integral. Thus, Eq. \eqref{eq:linearResponse_J} can be written as
\begin{equation}
    \big<\boldsymbol{J}_i(t)\big> = V\beta \int_0^t ds \big<\sum_j\boldsymbol{J}_i(s)\otimes\boldsymbol{J}_j(0)
     \big>_0 \boldsymbol{X}_j = \sum_j \boldsymbol{L}^{ij}\boldsymbol{X}_j~,
\end{equation}
where in the last equality we have incorporated Eq. \eqref{lij_general}. Taking the limit as $t$ approaches infinity to get the long-time behavior of the system allows us to obtain an expression for $\boldsymbol{L}^{ij}$:
\begin{equation}
    \boldsymbol{L}^{ij} =  V\beta \int_0^\infty dt \big< \boldsymbol{J}_i(t)\otimes \boldsymbol{J}_j(0)\big>_0~.
\end{equation}
Once again assuming isotropy, we reach the same Green-Kubo relations as obtained previously (Eq. \eqref{eq:GK_1}):
\begin{equation}\label{eq:GK_2}
    L^{ij} = \frac{V}{3k_{\mathrm{B}}T}\int_0^\infty dt \big<\boldsymbol{J}_i(t)\cdot \boldsymbol{J}_j(0) \big>~,
\end{equation}
where we have now omitted the subscript $0$ on the equilibrium ensemble average.

Note that the derivation presented in this section is contingent on choosing the correct form of the modified Hamiltonian. Here, we chose the positions and force ($\boldsymbol{\mathcal{R}}_j$ and $\boldsymbol{\mathcal{F}}_j$) specifically so that the final Green-Kubo relations would be equivalent to that derived in Sec. \ref{sec:GK_derivation}. These $\boldsymbol{\mathcal{R}}_j$ and $\boldsymbol{\mathcal{F}}_j$ are not known a priori, however, and therefore a fully rigorous derivation of the Green-Kubo relations should be done using the mass balance, as in Sec. \ref{sec:GK_derivation}. 

\section{\label{sec:relating_frameworks}Relating various frameworks for electrolyte transport}

The transport relations between $\boldsymbol{X}_i$ and $\boldsymbol{J}_i$ (Eq. \eqref{Lij}) derived in this work differ from other common conventions describing transport phenomena in electrolytes, and those typically analyzed in experiments\cite{Newman2004ElectrochemicalSystems,krishna1997maxwell}. In this section, we briefly describe two major conventions and give relations for interconverting between them. 

$\\$\textbf{I. Stefan-Maxwell equations.}
As described in the introduction, the most ubiquitous convention for electrolyte transport is the Stefan-Maxwell equations for multi-component diffusion,
\begin{equation}\label{StefanMaxwell}
    c_i\boldsymbol{\nabla}\overline{\mu}_i = \sum_{j\neq i} K^{ij}(\boldsymbol{v}_j - \boldsymbol{v}_i)~,
\end{equation}
which describe the force on species $i$ as linearly proportional to the relative friction between species $i$ and each of the other species in the system. Rather than describing particle motion with respect to a reference velocity such as $\boldsymbol{v}$, the Stefan-Maxwell framework is written in terms of the relative velocity of two species. Recall that $K^{ij}$ may be written in terms of the binary interaction diffusion coefficients $D^{ij}$, also called the Stefan-Maxwell diffusion coefficients, as $K^{ij} = \frac{RTc_ic_j}{c_{\mathrm{T}} D^{ij}}$.

$\\$\textbf{II. Solvent velocity reference system.}
It is also common to choose yet another convention, with $\boldsymbol{X}_i^{\mathrm{s}} = -c_i \boldsymbol{\nabla} \overline{\mu}_i$ and $\boldsymbol{J}_i^{\mathrm{s}} = (\boldsymbol{v}_i-\boldsymbol{v}_0)$, where we use the superscript $\mathrm{s}$ to denote that the flux of species $i$ ($\boldsymbol{J}_i^\mathrm{s}$) is described with respect to the solvent velocity $\boldsymbol{v}_0$\cite{Newman2004ElectrochemicalSystems, Wheeler2004MolecularMethod,katchalsky1965nonequilibrium}. The use of $\boldsymbol{v}_0$ as the reference velocity is sometimes referred to as the Hittorf reference system\cite{barthel1998physical}. The choice of $\boldsymbol{X}_i^{\mathrm{s}}$ as $-c_i \boldsymbol{\nabla} \overline{\mu}_i$ is particularly convenient given the form of the chemical potential in the dilute/ideal limit (discussed in more detail in the following section): $\mu_i = \mu_i^o + RT\ln c_i$. In this case, $-c_i\boldsymbol{\nabla}\mu_i = -RT \boldsymbol{\nabla} c_i$, and we recover the familiar result from Fick's law, in which the negative gradient of concentration is the driving force for diffusive flux. 

The expressions for $\boldsymbol{X}_i^{\mathrm{s}}$ and $\boldsymbol{J}_i^{\mathrm{s}}$ can also be motivated directly from our entropy production expression (Eq. \eqref{internal_entropy_split}) if we apply the Gibbs-Duhem equation for electroneutral systems at constant temperature, pressure, and electric field, i.e., $\sum_i c_i \boldsymbol{\nabla}\overline{\mu}_i = 0$ or $\boldsymbol{\nabla}\overline{\mu}_0 = -\sum_{i\neq 0} \frac{c_i}{c_0}\boldsymbol{\nabla}\overline{\mu}_i$ (Eq. \eqref{gibbs_duhem_2}), instead of applying the constraint that all mass fluxes must sum to zero. In this case, the entropy production in Eq. \eqref{internal_entropy_split} at constant temperature and $\boldsymbol{b}_i = 0$ becomes
\begin{equation}\label{entropy_prod_solvent_reference}
    \sigma_{\mathrm{i}} = -\sum_i\boldsymbol{\nabla}\overline{\mu}_i\cdot \boldsymbol{J}_i = -\sum_{i\neq 0}\boldsymbol{\nabla}\overline{\mu}_i\cdot(\boldsymbol{J}_i-\boldsymbol{J}_0) = -\sum_{i\neq 0}\boldsymbol{\nabla}\overline{\mu}_i\cdot[c_i(\boldsymbol{v}_i-\boldsymbol{v}_0)] ~.
\end{equation}
From Eq. \eqref{entropy_prod_solvent_reference} it is clear that both $\boldsymbol{X}_i$ and $\boldsymbol{J}_i$ as well as $\boldsymbol{X}_i^{\mathrm{s}}$ and $\boldsymbol{J}_i^{\mathrm{s}}$ yield the same total entropy production and are thus both consistent with linear irreversible thermodynamics. The choices of $\boldsymbol{X}_i^{\mathrm{s}}$ and $\boldsymbol{J}_i^{\mathrm{s}}$, and corresponding linear relations, yield the following transport coefficient equations:
\begin{equation}\label{Mij_Newman}
     c_i\boldsymbol{\nabla}\overline{\mu}_i = \sum_{j\neq 0} M^{{ij}^\mathrm{s}}(\boldsymbol{v}_j - \boldsymbol{v}_0)~,
\end{equation}
and
\begin{equation}\label{Lij_Newman}
    (\boldsymbol{v}_i - \boldsymbol{v}_0) = -\sum_{j\neq 0} L^{{ij}^\mathrm{s}} c_j\boldsymbol{\nabla}\overline{\mu}_j~.
\end{equation}
Note that, by convention, the negative sign in the thermodynamic driving force has been absorbed into $M^{{ij}^\mathrm{s}}$, and therefore $\boldsymbol{L}^\mathrm{s} = -\boldsymbol{M}^{\mathrm{s}^{-1}}$. 

Although both reference velocities give equivalent entropy production, only the mass-averaged velocity reference system can be cleanly integrated into the mass balance, which forms the basis for the regression hypothesis and derivations of the Green-Kubo relations in Sec. \ref{sec:GK_derivation}. Wheeler and Newman\cite{Wheeler2004MolecularMethod} have obtained Green-Kubo expressions for $L^{ij^\mathrm{s}}$ using the linear response approach of Sec. \ref{sec:linearReponse}; their choice of modified Hamiltonian yields expressions for $L^{ij^\mathrm{s}}$ in terms of $\boldsymbol{J}_i^{\mathrm{s}}$, the species flux with respect to the solvent velocity. This Hamiltonian may not be consistent with conservation of mass as written in Eq. \eqref{mass_balance_conc}, which is with respect to the barycentric velocity and not the solvent velocity. Indeed, the $L^{ij^\mathrm{s}}$ obtained from our $L^{ij}$ by the mapping described in the following sections (Secs. \ref{sec:onsager_SM} and \ref{sec:onsager_solvent}) may not be consistent with $L^{ij^\mathrm{s}}$ given by Wheeler and Newman's Green-Kubo relations. Their expressions thus may not correspond to true diffusive transport in the system.

\subsection{Relating the Onsager transport and Stefan-Maxwell equations}\label{sec:onsager_SM}
In this section we provide a mapping between the Onsager transport coefficients $L^{ij}$ in Eq. \eqref{Lij} and the Stefan-Maxwell transport coefficients $K^{ij}$ (Eq. \eqref{StefanMaxwell}). The methodology to obtain this mapping parallels that described by Bird \kf{for non-electrolyte multicomponent systems}\cite{bird2013multicomponent}.

We begin by rewriting Eq. \eqref{Lij} as
\begin{equation}\label{lij_over_c}
    \boldsymbol{v}_i - \boldsymbol{v} = -\sum_k \hat{L}^{ik} c_k \boldsymbol{\nabla}\overline{\mu}_k~,
\end{equation}
where, for convenience, we have defined the quantity $\hat{L}^{ik} = \frac{L^{ik}}{c_i c_k}$. Subtracting Eq. \eqref{lij_over_c} for species $i$ and $j$ and multiplying by $K^{ij}$ gives
\begin{equation}
    K^{ij}(\boldsymbol{v}_j - \boldsymbol{v}_i) = K^{ij}\sum_k\big[ \hat{L}^{ik}- \hat{L}^{jk}\big]c_k\boldsymbol{\nabla}\overline{\mu}_k~.
\end{equation}
Summing over $j\neq i$ yields
\begin{equation}\label{sm_mapping_0}
    \sum_{j\neq i}K^{ij}(\boldsymbol{v}_j - \boldsymbol{v}_i) = \sum_k\sum_{j\neq i}  K^{ij}\big[ \hat{L}^{ik}- \hat{L}^{jk}\big]c_k\boldsymbol{\nabla}\overline{\mu}_k~.
\end{equation}
This equation takes on the form of the Stefan-Maxwell equations (Eq. \eqref{StefanMaxwell}) if
\begin{equation}
    \sum_k\sum_{j\neq i}  K^{ij}\big[ \hat{L}^{ik}- \hat{L}^{jk}\big]c_k\boldsymbol{\nabla}\overline{\mu}_k = c_i\boldsymbol{\nabla}\overline{\mu}_i~,
\end{equation}
subject to the additional constraint that $\sum_i M_i L^{ij} = 0$. Following Bird\cite{bird2013multicomponent}, we observe that these equations can be satisfied if we choose
\begin{equation}\label{sm_mapping_1}
    \sum_{j\neq i}  K^{ij}\big[ \hat{L}^{ik}- \hat{L}^{jk}\big] = \delta_{ik} - \omega_i~,
\end{equation}
where $\omega_i = \rho_i / \rho$ is the mass fraction of species $i$. We first verify that Eq. \eqref{sm_mapping_1} transforms Eq. \eqref{sm_mapping_0} into the Stefan-Maxwell equations:
\begin{equation}
    \sum_{j\neq i}K^{ij}(\boldsymbol{v}_j - \boldsymbol{v}_i) = \sum_k\delta_{ik} c_k\boldsymbol{\nabla}\overline{\mu}_k - \sum_k\omega_i c_k\boldsymbol{\nabla}\overline{\mu}_k = c_i\boldsymbol{\nabla}\overline{\mu}_i~,
\end{equation}
where the last equality is obtained by invoking the Gibbs-Duhem equation. While Eq. \eqref{sm_mapping_1} yields the Stefan-Maxwell equations without the inclusion of the $\omega_i$ term, we require the latter to satisfy the constraint $\sum_i M_i L^{ij} = 0$. We verify that this constraint is satisfied by multiplying Eq. \eqref{sm_mapping_1} by $M_k c_k$ and summing over $k$, resulting in
\begin{equation}
    \sum_k M_k c_k \sum_{j\neq i}  K^{ij}\big[ \hat{L}^{ik}- \hat{L}^{jk}\big] = \sum_k M_k c_k \delta_{ik} - \sum_k M_k c_k \omega_i~.
\end{equation}
Rearranging and noting that $\sum_k M_k c_k = \sum_k \rho_k = \rho$, we obtain
\begin{equation}
    \sum_{j\neq i}  K^{ij}\bigg[\frac{1}{c_i}\sum_k M_k L^{ik}-\frac{1}{c_j}\sum_k M_k L^{jk}\bigg] = \rho_i - \rho_i~,
\end{equation}
where invoking the constraint $\sum_i M_i L^{ij} = 0$ gives $0=0$, as required. 

To convert Eq. \eqref{sm_mapping_1} into a more useful form, let us define the matrix $\boldsymbol{W}_i$ with components $(\boldsymbol{W}_i)_{jk} = \hat{L}^{jk}- \hat{L}^{ik}$  ($j,k\neq i$). We can rewrite Eq. \eqref{sm_mapping_1} in terms of $\boldsymbol{W}_i$ as
\begin{equation}\label{mapping_b_1}
    \sum_{j\neq i}K^{ij} (\boldsymbol{W}_i)_{jk} = \omega_i \qquad (i\neq k)~.
\end{equation}
Multiplying by $(\boldsymbol{W}^{-1}_i)_{kl}$ and summing over $k$ gives (after a change of indices)
\begin{equation}\label{mapping_b_inverted}
    K^{ij} = \omega_i \sum_{k\neq i} (\boldsymbol{W}^{-1}_i)_{kj}\qquad (i\neq j)~.
\end{equation}

Analogously, the constraint $\sum_j M_j L^{jk} = 0$ can be rewritten as
\begin{equation}
    \sum_j M_j c_j c_k [\hat{L}^{jk}- \hat{L}^{ik}] + \sum_j M_j c_j c_k \hat{L}^{ik} = 0~,
\end{equation}
or, upon rearranging:
\begin{equation}
    \sum_{j\neq i} \omega_j (\boldsymbol{W}_i)_{jk} = -\hat{L}^{ik}\qquad (i\neq k)~.
\end{equation}
Equivalently,
\begin{equation}\label{constraint_b_inverted}
    \sum_{k\neq i} \hat{L}^{ik} (\boldsymbol{W}^{-1}_i)_{kj} = -\omega_j \qquad (i\neq j)~.
\end{equation}
Combining Eqs. \eqref{mapping_b_inverted} and \eqref{constraint_b_inverted} yields our final equation mapping $K^{ij}$ and $L^{ij}$:
\begin{equation}\label{final_sm_mapping}
    \frac{1}{K^{ij}} = -\frac{1}{\omega_i \omega_j}\frac{\sum_{k\neq i} \hat{L}^{ik}(\boldsymbol{W}^{-1}_i)_{kj} }{\sum_{k\neq i} (\boldsymbol{W}^{-1}_i)_{kj}} \qquad (i\neq j)~.
\end{equation}

For a two-component electrolyte such as an ionic liquid, the mapping in Eq. \eqref{final_sm_mapping} can be written as
\begin{equation}
    K^{+-} =  \frac{-\omega_+ \omega_-}{\hat{L}^{+-}} = \frac{\omega_-^2}{\hat{L}^{++}}=\frac{\omega_+^2}{\hat{L}^{--}}~,
\end{equation}
where in the second and third equalities we have used the constraint $\sum_i M_i L^{ij} = 0$. For a three-component system, such as an electrolyte with binary salt and solvent, we obtain
\kf{\begin{equation}\label{SM_mapping_binary}
\begin{split}
    &K^{+-} = \omega_+ \omega_-\frac{\hat{L}^{00} + \hat{L}^{+-} - \hat{L}^{+0}-\hat{L}^{-0}}{\hat{L}^{+0}\hat{L}^{-0}-\hat{L}^{+-}\hat{L}^{00}}~,\\
    &K^{+0} = \omega_+ \omega_0\frac{\hat{L}^{--} + \hat{L}^{+0} - \hat{L}^{+-}-\hat{L}^{-0}}{\hat{L}^{+-}\hat{L}^{-0}-\hat{L}^{+0}\hat{L}^{--}}~,\\
    &K^{-0} = \omega_- \omega_0\frac{\hat{L}^{++} + \hat{L}^{-0} - \hat{L}^{+-}-\hat{L}^{+0}}{\hat{L}^{+-}\hat{L}^{+0}-\hat{L}^{-0}\hat{L}^{++}}~.
\end{split}
\end{equation}}
Equation \eqref{final_sm_mapping} may be used to obtain analogous expressions for systems with an arbitrary number of ionic components.

\subsection{Relating the Onsager transport framework and the solvent reference velocity system}\label{sec:onsager_solvent}

The relationship between the Onsager transport coefficients $L^{ij}$ defined with reference to the barycentric velocity (Eq. \eqref{Lij}) and those of the solvent reference velocity system $L^{ij^\mathrm{s}}$ (Eq. \eqref{Lij_Newman}) are not straightforward. We can, however, easily map between the Stefan-Maxwell and solvent reference velocity frameworks. This mapping, in conjunction with Eq. \eqref{final_sm_mapping} relating the Stefan-Maxwell and Onsager transport coefficients, allows us to connect $L^{ij}$ and $L^{ij^\mathrm{s}}$.

The mapping between the Stefan-Maxwell coefficients $K^{ij}$ and $M^{{ij}^\mathrm{s}}$ of the solvent reference velocity conventions is well-established\cite{Newman2004ElectrochemicalSystems}:
\begin{equation} \label{relateMandK}
M^{{ij}^\mathrm{s,0}} = K^{ij} - \delta_{ij} \sum_k K^{ik}~,
\end{equation}
where the superscript $0$ indicates that the matrix $\boldsymbol{M}^{\mathrm{s,0}}$ includes components from all species, including the solvent. As discussed previously, when all species are included, the components of the transport matrix are not all independent due to the fact that there are only $n-1$ independent force/flux equations for an $n$-component system (as seen by either the Gibbs-Duhem equation or the fact that all fluxes must sum to zero). The independent components of the transport matrix are given by the submatrix eliminating the row and column corresponding to one species, typically the solvent. The components of this submatrix are the $M^{{ij}^\mathrm{s}}$ defined in Eq. \eqref{Mij_Newman}. Recall that the submatrix $\boldsymbol{M}^{\mathrm{s}}$ is related to $\boldsymbol{L}^\mathrm{s}$ via $\boldsymbol{L}^\mathrm{s} = -\boldsymbol{M}^{\mathrm{s}^{-1}}$. Thus, $L^{ij}$ may be mapped to $L^{ij^\mathrm{s}}$ via the following process: $L^{ij}$ may be related to $K^{ij}$ using Eq. \eqref{final_sm_mapping}, $K^{ij}$ may be related to $M^{{ij}^\mathrm{s,0}}$ with Eq. \eqref{relateMandK}, $\boldsymbol{M}^{\mathrm{s,0}}$ may be converted into the submatrix with components $M^{{ij}^\mathrm{s}}$, and finally $\boldsymbol{M}^{\mathrm{s}}$ may be inverted to give $L^{ij^\mathrm{s}}$.

In what follows, we demonstrate this mapping procedure for a binary electrolyte, consisting of a single cation, single anion, and solvent. The relation between the Stefan-Maxwell and Onsager transport coefficients have already been written for a binary electrolyte in Eq. \eqref{SM_mapping_binary}. All that remains is to explicitly write $L^{ij^\mathrm{s}}$ in terms of the Stefan-Maxwell coefficients. We choose to give this mapping in terms of the Stefan Maxwell diffusion coefficients, $D^{ij}$, rather than $K^{ij}$, as this will be useful in a later section. Writing out the components of $\boldsymbol{K}$ in terms of $D^{ij}$ gives
\begin{gather}
\boldsymbol{K} = \frac{RT}{c_{\mathrm{T}}}
\begin{bmatrix}
\frac{c_+^2}{D^{++}}    &    \frac{c_+c_-}{D^{+-}}    &    \frac{c_+c_0}{D^{+0}}\\
\frac{c_+c_-}{D^{+-}}    &    \frac{c_-^2}{D^{--}}    &    \frac{c_-c_0}{D^{-0}}\\
\frac{c_+c_0}{D^{+0}}    &    \frac{c_-c_0}{D^{-0}}    &    \frac{c_0^2}{D^{00}}\\
\end{bmatrix}~.
\end{gather}
Now applying the mapping of Eq. \eqref{relateMandK}, we obtain
\begin{gather}
\boldsymbol{M}^{\mathrm{s}^0} = \frac{RT}{c_{\mathrm{T}}}
\begin{bmatrix}
-(\frac{c_+c_-}{D^{+-}}+ \frac{c_+c_0}{D^{+0}})    &    \frac{c_+c_-}{D^{+-}}    &    \frac{c_+c_0}{D^{+0}}\\
\frac{c_+c_-}{D^{+-}}    &    -(\frac{c_+c_-}{D^{+-}}+ \frac{c_-c_0}{D^{-0}})     &    \frac{c_-c_0}{D^{-0}}\\
\frac{c_+c_0}{D^{+0}}    &    \frac{c_-c_0}{D^{-0}}    &    -(\frac{c_+c_0}{D^{+0}}+ \frac{c_-c_0}{D^{-0}}) \\
\end{bmatrix}~.
\end{gather}
As mentioned before, not all components of $\boldsymbol{M}^{\mathrm{s}^0}$ are independent. The independent coefficients are obtained by eliminating the row and column corresponding to the solvent, given by the submatrix
\begin{gather}
\boldsymbol{M}^{\mathrm{s}} = \frac{RT}{c_{\mathrm{T}}}
\begin{bmatrix}
-(\frac{c_+c_-}{D^{+-}}+ \frac{c_+c_0}{D^{+0}})    &    \frac{c_+c_-}{D^{+-}}    \\
\frac{c_+c_-}{D^{+-}}    &    -(\frac{c_+c_-}{D^{+-}}+ \frac{c_-c_0}{D^{-0}})     \\
\end{bmatrix}~.
\end{gather}
Inverting $\boldsymbol{M}^{\mathrm{s}}$ and simplifying, we obtain the following expression for $\boldsymbol{L}^{s}$.
\begin{gather}\label{L0}
\boldsymbol{L}^{\mathrm{s}} = \frac{c_{\mathrm{T}}}{RT\gamma}
\begin{bmatrix}
c_+c_-D^{-0}D^{+0} + c_-c_0D^{+-}D^{+0}    &    c_+c_-D^{+0}D^{-0}    \\
c_+c_-D^{+0}D^{-0}    &    c_+c_-D^{-0}D^{+0} + c_+c_0D^{+-}D^{-0}     \\
\end{bmatrix}~,
\end{gather}
where $\gamma = c_+^2c_-c_0D^{-0} + c_+c_-^2c_0D^{+0} + c_+c_-c_0^2D^{+-}$. 

We have now outlined mappings between the Onsager and the Stefan-Maxwell coefficients (Eq. \eqref{final_sm_mapping}), as well as between the Stefan-Maxwell coefficients and those of the solvent reference velocity framework (Eq. \eqref{L0}), thus providing a relation between the Onsager and solvent-reference transport coefficients as well. 

\section{\label{infinite_dilution}Behavior in the limit of infinite dilution}

Here we show how the Onsager transport equations (Eq. \eqref{Lij}) behave in the limit of infinite dilution, thereby recovering the familiar Nernst-Planck equation for transport in an ideal electrolyte solution. In the case of infinite dilution, we can rewrite our expressions for both $L^{ij}$ and $L^{ij^\mathrm{s}}$ by assuming that $c_{\mathrm{T}}\approx c_0 >> c_+, c_-$ and $\boldsymbol{v} \approx \boldsymbol{v}_0$. Using the latter expression and multiplying by $c_i$, Eq. \eqref{Lij_Newman} containing $L^{ij^\mathrm{s}}$ can be rewritten as 
\begin{equation}
    c_i(\boldsymbol{v}_i - \boldsymbol{v}) = -\sum_{j\neq 0} L^{{ij}^\mathrm{s}}_{\mathrm{dilute}} c_i c_j\boldsymbol{\nabla}\overline{\mu}_j~.
\end{equation}
Comparing to Eq. \eqref{Lij} containing $L^{ij}$, we can conclude that
\begin{equation}\label{l_ls_dilute}
    L^{ij}_{\mathrm{dilute}} = c_i c_j L^{{ij}^\mathrm{s}}_{\mathrm{dilute}}
\end{equation}
for $i, j\neq 0$.

$\\$\textbf{Relation to self-diffusion coefficients.}
As in the previous sections, for brevity we now consider only binary electrolytes. Extensions to multicomponent systems are straightforward. Simplifying Eq. \eqref{L0} under the assumption that  $c_{\mathrm{T}}\approx c_0 >> c_+, c_-$ yields
\begin{gather}\label{L0_dilute}
\boldsymbol{L}^{\mathrm{s}}_{\mathrm{dilute}} = \frac{1}{RT c_+c_-c_0}
\begin{bmatrix}
c_-c_0D^{+0}    &    c_+c_-D^{+0}D^{-0}/D^{+-}   \\
c_+c_-D^{+0}D^{-0}/D^{+-}    &   c_+c_0D^{-0}     \\
\end{bmatrix}~.
\end{gather}
We can also infer that there will no be correlations between distinct ions at infinite dilution, i.e., the cross-correlated transport coefficient $L^{+-} = 0$. In order for these off-diagonal terms of Eq. \eqref{L0_dilute} to tend to $0$, we require that $D^{+-} \rightarrow \infty$, yielding

\begin{gather} \label{Ls_dilute_final}
\boldsymbol{L}^{\mathrm{s}}_{\mathrm{dilute}} = \frac{1}{RT}
\begin{bmatrix}
\frac{D^{+0}}{c_+}   &    0    \\
0   &  \frac{ D^{-0}}{c_-}   \\
\end{bmatrix}~,
\end{gather}
or, using Eq. \eqref{l_ls_dilute},
\begin{gather} \label{L_dilute_final}
\boldsymbol{L}_{\mathrm{dilute}} = \frac{1}{RT}
\begin{bmatrix}
D^{+0}c_+   &    0    \\
0   &  D^{-0}c_-   \\
\end{bmatrix}~.
\end{gather}

Equations \eqref{l_ls_dilute}, \eqref{Ls_dilute_final}, and \eqref{L_dilute_final} provide direct relations between transport coefficients from the different frameworks, $L^{ij}$, $L^{{ij}^\mathrm{s}}$, and $D^{ij}$, in the limit of infinite dilution. Finally, these multicomponent transport coefficients at infinite dilution may also be related to the self-diffusion coefficients\footnote{In some texts, the term `self-diffusion coefficient' refers specifically to the motion of a labeled particle in a pure liquid of identical, unlabeled particles\cite{tyrrell1984diffuion}, whereas the diffusion of a labeled particle in a multicomponent system is referred to as an intradiffusion coefficient. In this text, however, we refer to both of these scenarios as self-diffusion coefficients, as both can be computed based on the translational Brownian motion of the particles\cite{hansen1990theory}.} of each individual species. To do so, we rewrite the Green-Kubo relations for $L^{ii}$:
\begin{equation}\label{lii_1}
    L^{ii} = \frac{Vc_i^2}{3k_{\mathrm{B}}T}\int_0^\infty dt \bigg<\bigg(\frac{1}{N_i}\sum_\alpha \boldsymbol{v}_{i}^{\alpha}(t)-\boldsymbol{v}(t)\bigg)\cdot \bigg(\frac{1}{N_i}\sum_\beta \boldsymbol{v}_{i}^{\beta}(0)-\boldsymbol{v}(0)\bigg) \bigg>~.
\end{equation}
In addition to substituting $\boldsymbol{J}_i = c_i (\boldsymbol{v}_i - \boldsymbol{v})$, we have decomposed $\boldsymbol{v}_i$ into $\frac{1}{N_i}\sum_\alpha \boldsymbol{v}_{i}^{\alpha}$, where the index $\alpha$ enumerates all atoms/molecules of species $i$. Simplifying Eq. \eqref{lii_1} yields
\begin{equation}\label{L_ii_2}
    L^{ii} = \frac{Vc_i^2}{3k_{\mathrm{B}}T N_i^2}\int_0^\infty dt \bigg<\sum_\alpha \sum_\beta \bigg(\big( \boldsymbol{v}_{i}^{\alpha}(t)-\boldsymbol{v}(t)\big)\cdot \big( \boldsymbol{v}_{i}^{\beta}(0)-\boldsymbol{v}(0)\big)\bigg) \bigg>~.
\end{equation}
Splitting the double sum in Eq. \eqref{L_ii_2} to distinguish between cases where $\alpha = \beta$ (the self terms) and those where $\alpha \neq \beta$ (the distinct terms) results in

\begin{equation}\label{L_ii_3}
\begin{split}
    L^{ii} = \frac{Vc_i^2}{3k_{\mathrm{B}}T N_i^2}\bigg[\int_0^\infty dt \bigg<\sum_\alpha \bigg(\big( \boldsymbol{v}_{i}^{\alpha}(t)-\boldsymbol{v}(t)\big)\cdot \big( \boldsymbol{v}_{i}^{\alpha}(0)-\boldsymbol{v}(0)\big)\bigg) \bigg> + \\ \int_0^\infty dt \bigg<\sum_{\beta}\sum_{\alpha\neq\beta} \bigg(\big( \boldsymbol{v}_{i}^{\alpha}(t)-\boldsymbol{v}(t)\big)\cdot \big( \boldsymbol{v}_{i}^{\beta}(0)-\boldsymbol{v}(0)\big)\bigg) \bigg>\bigg]~.
\end{split}
\end{equation}
The first term in this equation describes self-correlations, while the second term captures correlations between distinct particles of type $i$, which are negligible at infinite dilution. Therefore, we observe that
\begin{equation}
    L^{ii}_{\mathrm{dilute}} = \frac{Vc_i^2}{3k_{\mathrm{B}}T N_i^2}\sum_\alpha\bigg[\int_0^\infty dt \bigg< \big( \boldsymbol{v}_{i}^{\alpha}(t)-\boldsymbol{v}(t)\big)\cdot \big( \boldsymbol{v}_{i}^{\alpha}(0)-\boldsymbol{v}(0)\big)\bigg>\bigg]~.
\end{equation}
The term in the square brackets is the integral of the velocity autocorrelation function, which is simply three times the self-diffusion coefficient of species $i$, $D_i$\cite{Frenkel2001}. This yields
\begin{equation}\label{l_ii_dilute_0}
    L^{ii}_{\mathrm{dilute}} = \frac{Vc_i^2}{3k_{\mathrm{B}}T N_i^2}[3 N_i D_i]~,
\end{equation}
where the additional factor of $N_i$ comes from summing over all atoms/molecules $\alpha$ of species $i$. Incorporating the fact that $c_i = N_i/V$, Eq. \eqref{l_ii_dilute_0} becomes
\begin{equation}\label{lii_dilute}
L^{ii}_{\mathrm{dilute}} = \frac{D_i c_i}{RT}~.
\end{equation}
Equation \eqref{lii_dilute} shows the relations between $L^{ii}_{\mathrm{dilute}}$ and the self-diffusion coefficients and, with Eq. \eqref{L_dilute_final}, also implies that the Stefan-Maxwell diffusion coefficients $D^{+0}$ and $D^{-0}$ approach the self-diffusion coefficients $D_+$ and $D_-$, respectively, in the limit of infinite dilution.

$\\$\textbf{Derivation of the Nernst-Planck equation.}
The above simplifications allow facile derivation of the Nernst-Planck equation for the flux of species $i$, $\boldsymbol{N}_i\coloneqq c_i \boldsymbol{v}_i$, at infinite dilution. Simplification of Eq. \eqref{Lij} for the case where all but the diagonal terms of the transport matrix are zero gives
\begin{equation}
    c_i (\boldsymbol{v}_i - \boldsymbol{v}) = - L^{ii}_{\mathrm{dilute}}\boldsymbol{\nabla}\overline{\mu}_i~.
\end{equation}
Further simplification and incorporation of Eq. \eqref{lii_dilute} yields
\begin{equation}
    c_i \boldsymbol{v}_i = -\frac{D_i c_i}{RT}(\boldsymbol{\nabla}\mu_i + z_i F \boldsymbol{\nabla}\phi) + c_i \boldsymbol{v}~.
\end{equation}
We may now incorporate the definition of $\boldsymbol{N}_i$ as well as the definition of chemical potential for an ideal solution: $\mu_i = \mu_i^\theta + RT \ln c_i$, implying $\boldsymbol{\nabla}\mu_i = \frac{RT}{c_i}\boldsymbol{\nabla}c_i$. Thus,
\begin{equation}
    \boldsymbol{N}_i = -D_{i}\boldsymbol{\nabla}c_i - \frac{D_{i}c_i z_i F}{RT}\boldsymbol{\nabla}\phi + c_i \boldsymbol{v}~.
\end{equation}
As a final step, we apply the Einstein relation to relate the self-diffusion coefficient to the electrophoretic mobility $u_i$ ($u_i = \frac{D_i z_i F}{RT}$)\cite{Frenkel2001} to recover the Nernst-Planck equation:\footnote{In this work, the mobility is defined as $u_i  = \frac{\boldsymbol{v}_i - \boldsymbol{v}}{\boldsymbol{E}}$, describing the velocity of a species in response to an electric field. In some texts\cite{Newman2004ElectrochemicalSystems}, the mobility is instead defined as $u_i^\prime  =\frac{(\boldsymbol{v}_i - \boldsymbol{v})}{z_i F \boldsymbol{E}} =  \frac{u_i}{z_i F} $. The Nernst-Planck equation using this convention is $\boldsymbol{N}_i = -D_i \boldsymbol{\nabla}c_i-z_i F u_i^\prime c_i\boldsymbol{\nabla}\phi  + c_i \boldsymbol{v}$.}
\begin{equation}
    \boldsymbol{N}_i =  -D_i \boldsymbol{\nabla}c_i -u_i c_i\boldsymbol{\nabla}\phi+ c_i \boldsymbol{v}~.
\end{equation}

\section{Relation to quantities obtained from experiments and molecular simulations}\label{sec:experiment}
The Onsager transport coefficients $L^{ij}$ are not directly measurable from experiments. They may, however, be explicitly related to quantities which can be accessed experimentally, namely the ionic conductivity, electrophoretic mobility, transference number, and salt diffusion coefficient. In this section, we derive expressions for each of these experimentally measurable quantities in terms of $L^{ij}$. Utilizing our derived Green-Kubo relations for $L^{ij}$ (Eq. \eqref{eq:GK_1}), we also provide Green-Kubo expressions for some of these experimentally measurable quantities so that they can be calculated directly in molecular simulations. Finally, we give expressions for $L^{ij}$ in terms of the aforementioned experimentally measurable quantities for the special case of a binary solution.

\subsection{Ionic conductivity}

We begin by deriving an expression for the ionic conductivity in terms of $L^{ij}$. Consider a solution of uniform composition, i.e. with no gradients in chemical potential, such that $\boldsymbol{\nabla} \overline{\mu}_i = \boldsymbol{\nabla} \mu_i + z_i F \boldsymbol{\nabla} \phi = z_i F \boldsymbol{\nabla} \phi$. For conventional experimental conductivity measurements of electrolyte solutions, this condition of uniform composition is satisfied by applying a rapidly alternating voltage or current through the electrolyte about the open circuit voltage, the high frequency of which does not allow appreciable concentration gradients to form\cite{barsoukov2005impedance}. Rewriting the transport Eq. \eqref{Lij} under this condition leads to
\begin{equation}
    c_i(\boldsymbol{v}_i - \boldsymbol{v}) = -\sum_{j} L^{ij} z_j F\boldsymbol{\nabla}\phi~.
\end{equation}
Multiplying by $Fz_i$ and summing over all species $i$ results in
\begin{equation}\label{condDerivation}
    \sum_i Fz_ic_i\boldsymbol{v}_i-\sum_i Fz_ic_i\boldsymbol{v} = -\sum_i\sum_j L^{ij}z_iz_jF^2\boldsymbol{\nabla}\phi~.
\end{equation}
Note that the second term on the left side of the equation is zero due to electroneutrality, which dictates $\sum_i z_i c_i = 0$. Thus we will see that while $L^{ij}$ depends on the chosen reference velocity (in this case $\boldsymbol{v}$), the ionic conductivity will be independent of the reference velocity, as expected.

Recall that the free current density $\tilde{\boldsymbol{j}}^{\mathrm{f}}$ may be written as
\begin{equation}\label{CurrentDefinition}
    \tilde{\boldsymbol{j}}^{\mathrm{f}} = F \sum_i z_i c_i \boldsymbol{v}_i~.
\end{equation}
Equation \eqref{CurrentDefinition} can also be written in terms of the fluxes of ions as
\begin{equation}\label{current_fluxes}
    \tilde{\boldsymbol{j}}^{\mathrm{f}} = F\sum_i (z_i \boldsymbol{J}_i + z_i c_i \boldsymbol{v} ) = F\sum_i z_i \boldsymbol{J}_i~,
\end{equation}
where in the second equality we have invoked the condition of electroneutrality. Using Eqs. \eqref{condDerivation} and \eqref{CurrentDefinition}, we obtain
\begin{equation}
    \tilde{\boldsymbol{j}}^{\mathrm{f}} = -\sum_i\sum_j L^{ij}z_iz_jF^2\boldsymbol{\nabla}\phi~,
\end{equation}
showing a linear relationship between the current density and the electric field. Using Ohm's Law to define the ionic conductivity $\kappa$ from
\begin{equation}\label{OhmsLaw}
    \tilde{\boldsymbol{j}}^{\mathrm{f}} = -\kappa \boldsymbol{\nabla} \phi
\end{equation}
yields our final relation between ionic conductivity and the transport coefficients $L^{ij}$ as
\begin{equation}\label{conductivity}
    \kappa = F^2\sum_i \sum_j L^{ij}z_iz_j~.
\end{equation}

The Green-Kubo relation for ionic conductivity can be obtained from Eq. \eqref{eq:GK_1} as
\begin{equation}
    \kappa = \frac{1}{3k_{\mathrm{B}} T V}\int_0^\infty dt \bigg<\sum_{\alpha} \hat{q}^{\alpha} (\boldsymbol{v}^{\alpha}(t) - \boldsymbol{v}(t)) \cdot \sum_{\beta} \hat{q}^{\beta} (\boldsymbol{v}^{\beta}(0)- \boldsymbol{v}(0))\bigg>~,
\end{equation}
where the summations are over all individual ions in the system (denoted by the indices ${\alpha}$ and $\beta$), rather than over all types of ions $i$ as in Eq. \eqref{eq:GK_1}. Recall that $\hat{q}^{\alpha}$ is the electronic charge of the ion ${\alpha}$. With electroneutrality ($\sum_{\alpha}\hat{q}^{\alpha} =0$), the reference velocity vanishes and we obtain the Green-Kubo equation commonly presented in other works\cite{hansen1990theory}:
\begin{equation}
    \kappa = \frac{1}{3k_{\mathrm{B}} T V}\int_0^\infty dt \bigg<\sum_{\alpha} \hat{q}^{\alpha} \boldsymbol{v}^{\alpha}(t) \cdot \sum_{\beta} \hat{q}^{\beta} \boldsymbol{v}^{\beta}(0)\bigg>~.
\end{equation}

\subsection{Electrophoretic mobility}\label{sec:mobility}
We can also obtain expressions for the electrophoretic mobility of species $i$, $u_i$, defined as $\boldsymbol{v}_i - \boldsymbol{v} \eqqcolon u_i \boldsymbol{E}$, in terms of $L^{ij}$. This quantity can be measured experimentally using techniques such as electrophoretic Nuclear Magnetic Resonance (NMR) spectroscopy\cite{holz1994electrophoretic} or capillary electrophoresis\cite{whatley2001basic}. To this end, consider once again the case with no gradients in chemical potential. In this case, the definition of electrophoretic mobility can be compared with Eq. \eqref{Lij}, yielding
\begin{equation}\label{mobility}
u_i = \sum_j L^{ij} \frac{z_j F}{c_i} ~.
\end{equation}
Based on Eq. \eqref{conductivity}, it follows that the ionic conductivity is related to mobility as $\kappa = \sum_i F z_i c_i u_i$. In molecular simulations, $u_i$ can either be computed by separately calculating each $L^{ij}$ term or by directly using the Green-Kubo relations emerging from substituting the Green-Kubo relations for $L^{ij}$ (Eq. \eqref{eq:GK_1}) into Eq. \eqref{mobility}:
\begin{equation}
    u_i = \frac{1}{3 k_{\mathrm{B}} T}\int_0^\infty dt \bigg<\sum_{\alpha} \hat{q}^{\alpha} (\boldsymbol{v}^{\alpha}(0)-\boldsymbol{v}(0))\cdot(\boldsymbol{v}_i(t)-\boldsymbol{v}(t))\bigg>~.
\end{equation}
This result is consistent with the relation derived by D{\"u}nweg et al.\cite{dunweg2008colloidal} using linear response theory.

\subsection{Transference number}
The transference number of species $i$, $t_i$, can also be determined directly from $L^{ij}$. The transference number is defined as the fraction of current carried by species $i$ in a system with no concentration gradients. Using Eq. \eqref{current_fluxes}, it is given by
\begin{equation}\label{transference_0}
    t_i \coloneqq \frac{z_i \boldsymbol{J}_i}{\sum_j z_j \boldsymbol{J}_j}~.
\end{equation}
Equation \eqref{transference_0} can be expressed in terms of electrophoretic mobility and conductivity as
\begin{equation}\label{transference}
t_i = \frac{F z_i c_i u_i}{\kappa} = \frac{\sum_j L^{ij} z_i z_j}{\sum_k \sum_l L^{kl} z_k z_l}~.
\end{equation}
From the first equality, it is clear that the transference number may equivalently be interpreted as the fraction of conductivity attributed to species $i$. The second equality has incorporated Eqs. \eqref{conductivity} and \eqref{mobility} to give the transference number in terms of $L^{ij}$. 

Experimentally, the transference number can be measured via a number of methods. The most common method is a potentiostatic polarization experiment, where a fixed potential is applied to a symmetric cell and the ratio of the achieved steady state current to the Ohmic current is equal to the transference number of the reactive species \cite{EVANS19872324}. This method is only strictly valid in the infinite dilution limit. For concentrated electrolytes, additional information about the activity of the solution must be known in order to calculate transference numbers \cite{Balsara_2015}. Using another common method, the Hittorf method, the transference number can be directly obtained by measuring the concentration of ions throughout multiple connected chambers in a symmetric cell after passing current through the electrolyte for a known amount time \cite{BRUCE19921087}. The transference number can also be obtained by measuring the electrophoretic mobility of each ionic species by the methods mentioned in Sec. \ref{sec:mobility}.  

\subsection{Salt/Electrolyte diffusion coefficient}
The transport coefficients $L^{ij}$ can also be related to the salt or electrolyte diffusion coefficient, following the derivation by Katchalsky\cite{katchalsky1965nonequilibrium}. For this derivation, we restrict ourselves to a binary electrolyte, with a single salt. Rather than considering a system with no chemical potential gradients as with $\kappa$, $u_i$, and $t_i$, here we consider the condition of no electrical current. Under this condition, the salt diffusion coefficient $D_{\mathrm{el}}$ is defined by 
\begin{equation}\label{D_definition}
    \boldsymbol{J}_{\mathrm{el}}\coloneqq -D_{\mathrm{el}} \boldsymbol{\nabla}c~,
\end{equation}
where the subscript `$\mathrm{el}$' denotes quantities pertaining to the overall electrolyte. The salt flux $\boldsymbol{J}_{\mathrm{el}}$ is related to the fluxes of the cation and anion, $\boldsymbol{J}_+$ and $\boldsymbol{J}_-$, respectively, by $\boldsymbol{J}_{\mathrm{el}} = \frac{\boldsymbol{J}_+}{\nu_+} = \frac{\boldsymbol{J}_-}{\nu_-}$, where $\nu_+$ and $\nu_-$ are the stoichiometric coefficients of the cation and anion in the salt. The quantity $c$ is the concentration of salt, $c = \frac{c_+}{\nu_+} = \frac{c_-}{\nu_-}$. In what follows, we aim to express the salt diffusion coefficient $D_{\mathrm{el}}$ in terms of $L^{ij}$.

To satisfy the condition of no net current, the net charge flux must be zero, i.e., by Eq. \eqref{current_fluxes},
\begin{equation}\label{net_flux_zero}
    z_+ F \boldsymbol{J}_+ + z_- F \boldsymbol{J}_-= 0~.
\end{equation}
Incorporating the transport laws (Eq. \eqref{Lij}) into Eq. \eqref{net_flux_zero} yields
\begin{equation}\label{net_flux_zero_lij}
    -z_+ L^{++} \boldsymbol{\nabla}\overline{\mu}_+ -z_+ L^{+-} \boldsymbol{\nabla}\overline{\mu}_- -z_- L^{--} \boldsymbol{\nabla}\overline{\mu}_- -z_- L^{+-} \boldsymbol{\nabla}\overline{\mu}_+= 0~.
\end{equation}
It is convenient to define the chemical potential of the salt or electrolyte as $\mu_{\mathrm{el}} \coloneqq \nu_+ \overline{\mu}_+ + \nu_-\overline{\mu}_-$. This definition, along with Eq. \eqref{net_flux_zero_lij}, allows us to express the electrochemical potential of the ions in terms of $\mu_{\mathrm{el}}$ and $L^{ij}$:
\begin{equation}\label{mu_plusminus_diffusion}
\begin{split}
    \boldsymbol{\nabla}\overline{\mu}_+ = \boldsymbol{\nabla}\mu_{\mathrm{el}} \frac{z_-}{\nu_+}\bigg(  \frac{z_+ L^{+-} + z_- L^{--}}{z_+^2 L^{++} + 2 z_+ z_- L^{+-}+z_-^2L^{--}} \bigg)~,\\
    \boldsymbol{\nabla}\overline{\mu}_- = \boldsymbol{\nabla}\mu_{\mathrm{el}} \frac{z_+}{\nu_-}\bigg(  \frac{z_- L^{+-} + z_+ L^{++}}{z_+^2 L^{++} + 2 z_+ z_- L^{+-}+z_-^2L^{--}} \bigg)~.
\end{split}
\end{equation}
Combining Eqs. \eqref{Lij} and \eqref{mu_plusminus_diffusion} allows us to write the salt flux $\boldsymbol{J}_{\mathrm{el}}$ in terms of $\boldsymbol{\nabla}\mu_{\mathrm{el}}$ and $L^{ij}$ as
\begin{equation}\label{J_pm}
    \boldsymbol{J}_{\mathrm{el}} = \frac{z_+ z_-}{\nu_+ \nu_-}\bigg(  \frac{L^{--}L^{++}-L^{{+-}^2}}{z_+^2 L^{++} + 2 z_+ z_- L^{+-}+z_-^2L^{--}} \bigg)\boldsymbol{\nabla}\mu_{\mathrm{el}}~.
\end{equation}
Note that while all previous transport properties have been defined with respect to gradients in electrochemical potential (the true thermodynamic driving force), the salt diffusion coefficient is defined with respect to concentration gradients. Thus, to identify $D_{\mathrm{el}}$ from Eq. \eqref{J_pm}, we need a relation between $\boldsymbol{\nabla}\mu_{\mathrm{el}}$ and $\boldsymbol{\nabla}c$. To this end, we invoke the form of the chemical potential,
\begin{equation}
    \mu_{\mathrm{el}} = \mu_{\mathrm{el}}^o + 
    \nu RT \ln (c f_{\mathrm{el}}) + RT\ln(\nu_+^{\nu_+}\nu_-^{\nu_-})~,
\end{equation}
where $f_{\mathrm{el}} = (f_+^{\nu_+}f_-^{\nu_-})^{1/\nu}$ is the salt activity coefficient\cite{Newman2004ElectrochemicalSystems} and $\nu = \nu_+ + \nu_-$.\footnote{The quantities $\mu_{\mathrm{el}}$ and $f_{\mathrm{el}}$ are referred to in some texts as $\mu_{\pm}$ and $f_{\pm}$\cite{Newman2004ElectrochemicalSystems}.} Thus,
\begin{equation}\label{grad_mu_pm}
    \boldsymbol{\nabla}\mu_{\mathrm{el}} = \frac{\partial \mu_{\mathrm{el}}}{\partial c} \boldsymbol{\nabla}c = \frac{\nu RT}{c}\bigg[ 1 + \frac{d \ln f_{\mathrm{el}}}{d \ln c}\bigg] \boldsymbol{\nabla}c~.
\end{equation}
Combining Eqs. \eqref{D_definition}, \eqref{J_pm} and \eqref{grad_mu_pm}, the salt diffusion coefficient is given by
\begin{equation}\label{eq:D}
    D_{\mathrm{el}} = \frac{-z_+ z_-(L^{--}L^{++}-L^{{+-}^2)}}{\nu_+ \nu_-(z_+^2 L^{++} + 2 z_+ z_- L^{+-}+z_-^2L^{--})}\frac{\nu RT}{c}\bigg[ 1 + \frac{d \ln f_{\mathrm{el}}}{d \ln c}\bigg]~.
\end{equation}
Experimentally $D_{\mathrm{el}}$ is typically measured via the restricted diffusion method, where a concentration gradient is built across a symmetric cell by applying a potential or fixed current density\cite{doi:10.1111/j.1749-6632.1945.tb36171.x}. The potential or current is then stopped and the concentration gradient is monitored as it relaxes. The concentration can either be directly monitored using interferometry or other spectroscopic methods  or can be indirectly observed by monitoring the changing open circuit potential\cite{doi:10.1002/aic.690190220,Ehrl_2017}. 

\subsection{$L^{ij}$ in terms of experimental quantities}
The above relations enable us to compute conductivity, mobility, transference number, and salt diffusion coefficient from $L^{ij}$ values obtained from molecular simulation. In contrast, we can also manipulate these equations to solve for $L^{ij}$ in the case where the electrolyte has been characterized experimentally. Rearranging Eqs. \eqref{conductivity}, \eqref{transference}, and \eqref{eq:D} and solving for $L^{ij}$ in terms of the experimentally measurable quantities gives:
\begin{equation*}
    L^{++} = \frac{\nu_+^2 D_{\mathrm{el}}}{\frac{\nu RT}{c}\bigg[ 1 + \frac{d \ln f_{\mathrm{el}}}{d \ln c}\bigg]} + \kappa \bigg( \frac{t_+}{z_+ F}\bigg)^2~,
\end{equation*}
\begin{equation}
    L^{--} = \frac{\nu_-^2 D_{\mathrm{el}}}{\frac{\nu RT}{c}\bigg[ 1 + \frac{d \ln f_{\mathrm{el}}}{d \ln c}\bigg]} + \kappa \bigg( \frac{t_-}{z_- F}\bigg)^2~,
\end{equation}
\begin{equation*}
    L^{+-} = \frac{\nu_+\nu_- D_{\mathrm{el}}}{\frac{\nu RT}{c}\bigg[ 1 + \frac{d \ln f_{\mathrm{el}}}{d \ln c}\bigg]} + \frac{\kappa t_+ t_-}{z_+ z_- F^2}~.
\end{equation*}

In summary, we have derived equations to enable inter-conversion between $L^{ij}$ and experimentally relevant, macroscopic electrolyte transport quantities. This provides experimentalists with a means to quantitatively evaluate the extent of correlation between each of the ionic species in solution.

\section{\label{sec:MD}Applications: Molecular simulations and experimental characterization of LiCl in DMSO}
\begin{figure}[H]
    \centering
    \includegraphics[width=\textwidth]{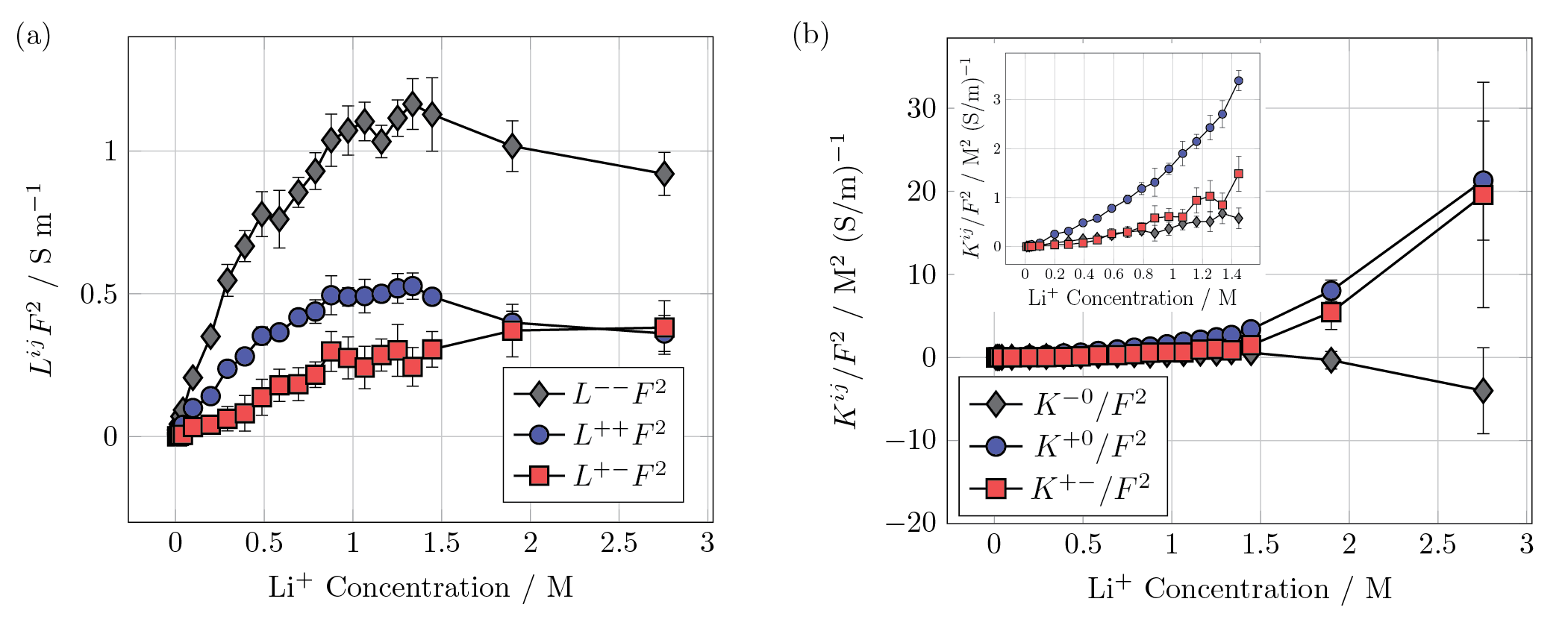}
    \caption{\kf{Transport coefficients for LiCl in DMSO solutions. (a) Onsager transport coefficients ($L^{ij}F^2$) and (b) Stefan-Maxwell transport coefficients ($K^{ij}/F^2$) of each pair of ionic species versus concentration. The inset in (b) shows the data at low concentrations. Transport coefficients are divided or multiplied by a factor of $F^2$ (where $F$ is Faraday's constant) such that the units are related to those commonly used for ionic conductivity.}}
\label{fig:Lij}
\end{figure}
In what follows, we use the Green-Kubo relations (Eq. \eqref{eq:GK_1}) derived herein to compute the transport coefficients $L^{ij}$ in a model electrolyte system using classical molecular dynamics (MD) simulations (methods are described in Appendix \ref{appendix:Sim_Methods}). Our model system consists of LiCl salt, chosen for its structural simplicity, in dimethyl sulfoxide (DMSO) solvent. DMSO was chosen over an aqueous solution to avoid the complications associated with the self-ionization of water, which introduces additional charge carrying species into solution. As the dielectric constant and donor number of DMSO are relatively high, this solvent is commonly used for its effectiveness in dissolving and dissociating salts\cite{barthel1998physical}.

Figure \ref{fig:Lij}a shows the transport coefficients $L^{ij}$ from the Green-Kubo relations as a function of salt concentration. \kf{The Stefan-Maxwell coefficients $K^{ij}$, obtained from $L^{ij}$ using the mapping in Eq. \eqref{SM_mapping_binary}, are also given for comparison in Figure \ref{fig:Lij}b.} The error bars reported here are the standard deviation of ten independent replicate simulations, although we note that the true error based on Zwanzig and Ailawadi theory\cite{Zwanzig} and its extension\cite{jones2012adaptive} is likely smaller. We observe that the anionic term $L^{--}$ is consistently the largest of the three transport coefficients. This more facile motion of the anion relative to the cation is consistent with the lithium ion's bulky solvation shell\cite{gering2017prediction}. Based on the fact that $L^{ij}$ is directly proportional to $c_i c_j$ (Eq. \eqref{eq:GK_1}), one might expect a monotonic increase in each of the transport coefficients with concentration. The non-monotonic trends in $L^{++}$ and $L^{--}$ reflect an increase in inter-ionic friction as concentration increases as a result of electrophoretic and relaxation effects\cite{onsager1932irreversible}. These effects, along with ion pairing or aggregation, also contribute to correlated cation-anion motion, captured by $L^{+-}$. We note that the magnitude of $L^{+-}$ increases monotonically with concentration, which is consistent with an increase in the number of ion pairs in solution.

\begin{figure}
    \centering
    \includegraphics[width=\textwidth]{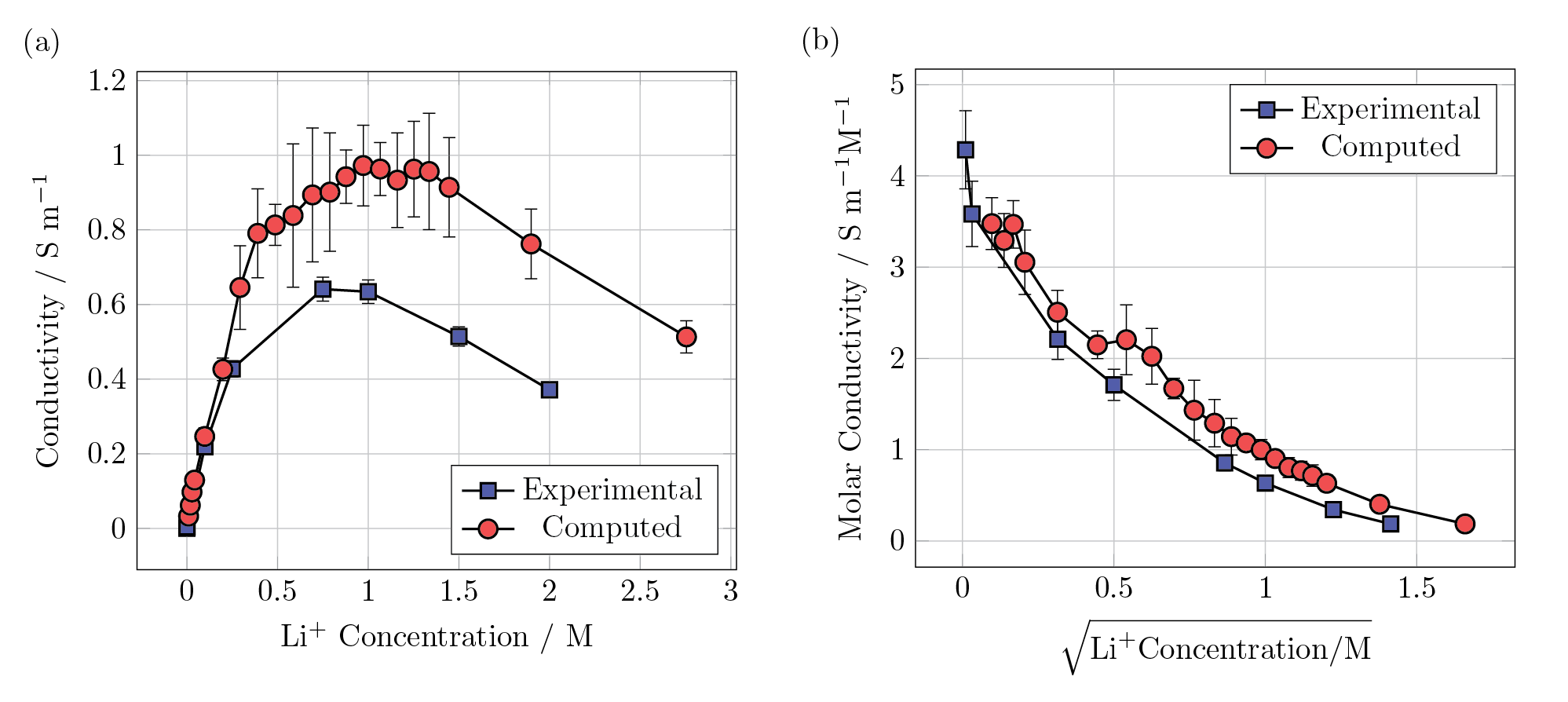}
    \caption{(a) Ionic conductivity versus concentration, with comparison to experimental data obtained from AC impedance measurements (see Appendix \ref{appendix:exp_Methods}). (b) Molar conductivity versus the square root of concentration, with comparison to experimental data.}
\label{fig:conductivity}
\end{figure}

\begin{figure}
    \centering
    \includegraphics[width=\textwidth]{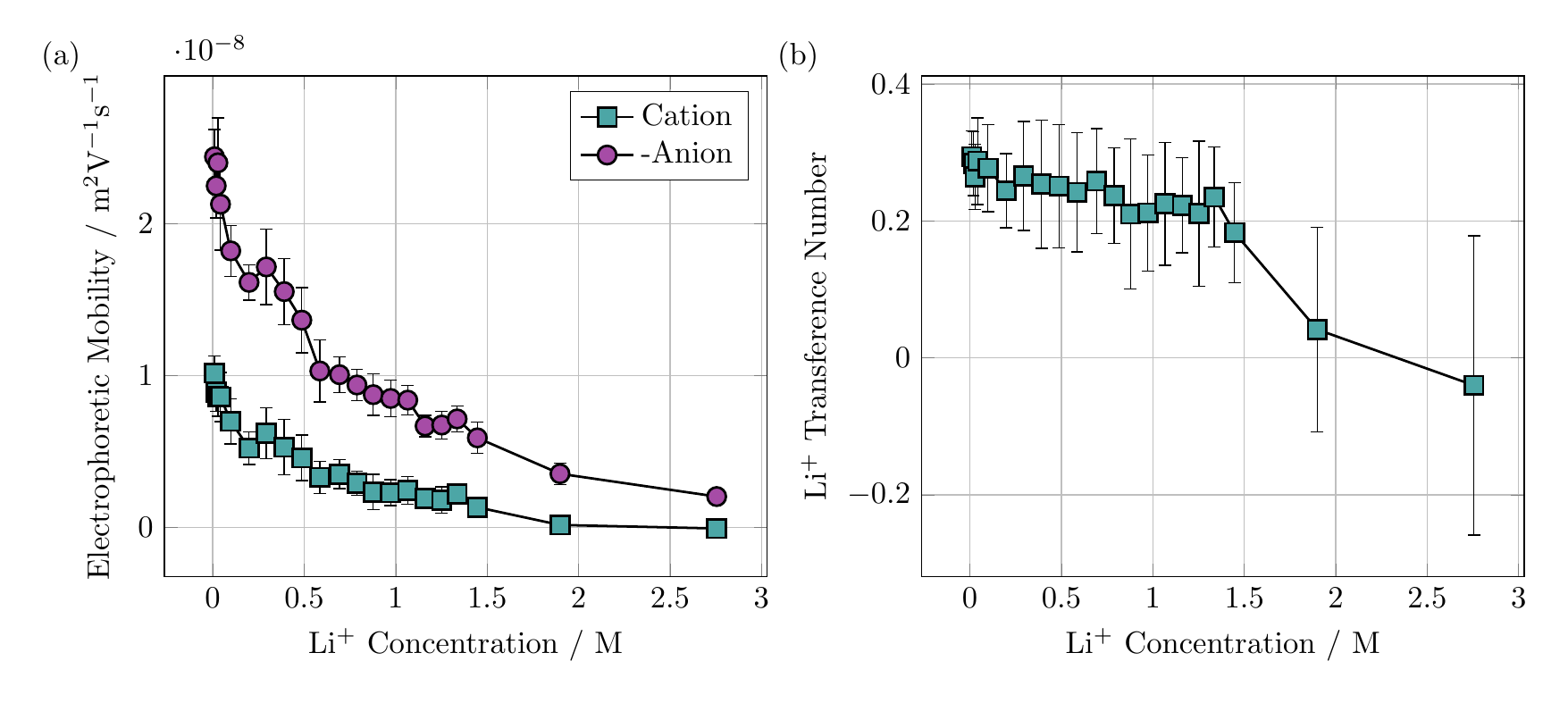}
    \caption{(a) Electrophoretic mobility $u$ of each ion. As the mobility of the anion is negative, for this species we show $-u$. (b) Cation transference number as a function of concentration.}
\label{fig:mobility_transference}
\end{figure}

In Figures \ref{fig:conductivity} and \ref{fig:mobility_transference}, we demonstrate how the computed transport coefficients $L^{ij}$ can be combined to yield experimentally relevant properties. The total ionic conductivity computed from Eq. \eqref{conductivity} is shown in Figure \ref{fig:conductivity}a. We also measure the conductivity experimentally using AC impedance spectroscopy (see Appendix \ref{appendix:exp_Methods} for a detailed description of methods). The computed values and the experimental data show reasonable agreement both qualitatively and quantitatively, showing the molecular model to be reasonable for studying transport phenomena of LiCl in DMSO. Figure \ref{fig:conductivity}b shows the same experimental and computed conductivity data as in Figure \ref{fig:conductivity}a but is plotted as molar conductivity $\Lambda$ (concentration-normalized conductivity) versus the square root of concentration. The Debye-H\"uckel-Onsager theory\cite{onsager1932irreversible} predicts that for low concentrations $\Lambda = \Lambda^0 - \xi\sqrt{c}$, where $\Lambda^0$ is the limiting molar conductivity and $
\xi$ is a constant accounting for electrophoretic and relaxation effects. As can be seen in Figure \ref{fig:conductivity}b, the linear dependence of molar conductivity on $\sqrt{c}$ is only approximately followed at the most dilute concentrations; deviations from the predicted trend can likely be attributed to incomplete dissociation of the salt\cite{wright2007introduction}. As the predicted $\sqrt{c}$ dependence is built upon Debye-H\"uckel theory, which is only valid for very dilute electrolytes, it is no surprise that we observe the molar conductivity to deviate substantially from the Debye-H\"uckel-Onsager equation at higher concentrations.

We also present the electrophoretic mobility and cation transference number in Figure \ref{fig:mobility_transference}a and b, respectively. In Figure \ref{fig:mobility_transference}a, we observe that the mobility of both the cation and anion decrease monotonically, once again due to more inter-ionic friction at higher concentrations. The magnitude of the anion mobility is larger than that of the cation, as required by the fact that $L^{--} > L^{++}$ at all concentrations (Figure \ref{fig:Lij}). The transference number in Figure \ref{fig:mobility_transference}b is observed to remain relatively constant with respect to concentration until approximately 1.5 M, after which it decreases sharply. The onset of the decrease in the lithium transference number roughly coincides with that of the ionic conductivity in Figure \ref{fig:conductivity}a. This shift likely reflects a change in the solvation environment of the lithium ions, i.e., a change in the most common ion aggregates present in solution. Note that although the transference number is defined as the fraction of conductivity attributed to a given species, it is not necessarily bounded between zero and one. Indeed, in this LiCl in DMSO system, the lithium transference number may become negative above 2.5 M. This phenomena, which has been observed in systems such as solid polymer electrolytes\cite{pesko2017negative} and ionic liquids\cite{gouverneur2018negative}, corresponds to lithium moving towards more positive potential under an applied electric field and may be due to the presence of negatively charged aggregates in solution. It is important to note that the transference number is challenging to measure experimentally and is rarely reported without the use of ideal solution approximations\cite{diederichsen2017promising}. In contrast, the Onsager framework and associated Green-Kubo relations allow for facile computation of the transference number at a computational cost equivalent to that of conventional MD calculations such as the diffusion coefficient or total ionic conductivity.

\section{Conclusions}
In this work, we present the integration of continuum mechanics, non-equilibrium thermodynamics, and electrodynamics for electrolyte solutions. This allows us to systematically obtain mass, charge, momentum, energy, and entropy balances and unambiguously derive the rate of internal entropy production in these systems. We thereby identify the relevant thermodynamic driving forces and associated fluxes governing dissipation in electrolytes and present linear laws relating these forces and fluxes, giving rise to the Onsager transport coefficients $L^{ij}$. This framework allows us to derive Green-Kubo expressions for $L^{ij}$. We show that our derived force-flux relations are interconvertible with other common theories used for electrolyte transport, namely the Stefan-Maxwell relations and the solvent reference velocity system. Finally, we derive the relationship between the transport coefficients and experimentally measurable quantities and demonstrate the use of molecular simulations to directly compute $L^{ij}$, ionic conductivity, electrophoretic mobility, and transference number. The integration of these different fields and the resulting theory can be used in the future to understand more complex problems such as the transport in systems with more than two ionic species as well as simultaneous mass and momentum transport in fluids with dielectric discontinuities.

\section{Acknowledgments}
The authors express profound gratitude for Andrew Crothers, Jeffrey M. Epstein, and Charles C. Wojcik for their insightful feedback and discussion. K.D.F. acknowledges support from NSF GRFP under Grant no. DGE 1752814. H.K.B and B.D.M. were supported by the Assistant Secretary for Energy Efficiency and Renewable Energy, Vehicle Technologies Office, of the U.S. Department of Energy under Contract DE-AC02- 05CH11231, under the Advanced Battery Materials Research (BMR) Program. K.K.M was supported by Director, Office of Science, Office of Basic Energy Sciences, of the U.S. Department of Energy under contract No. DEAC02-05CH11231. This research used the Savio computational cluster resource provided by the Berkeley Research Computing program at the University of California, Berkeley.

	%
	%

	\begin{appendices}
		\numberwithin{equation}{section}
		\section{\label{appendix:thermo_potentials}Thermodynamic potentials}
Deriving the internal entropy production in Sec. \ref{sec:entropy_bal} relied on using the Helmholtz free energy per volume, $\tilde{f}$. In this section we begin with system-level thermodynamics to derive an expression for $\tilde{f}$ in a mixture subject to an electromagnetic field and infer the quantities on which $\tilde{f}$ depends. We then perform Legendre transforms to develop analogous expressions for the internal energy and Gibbs free energy densities.

At the system level, changes in the Helmholtz free energy $\mathcal{F} = \tilde{f} V$ can be written as\cite{landau2013electrodynamics}
\begin{equation}\label{helmhotlz}
    d\mathcal{F} = -pdV - \mathcal{S}dT + \sum_i \mu_i dN_i - \int \boldsymbol{P}dV\cdot d\boldsymbol{\mathcal{E}} - \int \boldsymbol{\mathcal{M}}dV\cdot d\boldsymbol{B}~,
\end{equation}
where $\mathcal{S}$ is entropy and each of the other quantities has been defined in the main text. Consider the case where the polarization and Lorentz magnetization are uniform in the system, such that $\int \boldsymbol{P}dV= \boldsymbol{P} V$ and $\int \boldsymbol{\mathcal{M}}dV= \boldsymbol{\mathcal{M}}V$. Note that Eq. \eqref{helmhotlz} is only valid in equilibrium. We therefore invoke the local equilibrium hypothesis: although irreversible processes may be taking place in the system as a whole, we can isolate a small region of the system which can be treated as though it were in equilibrium with respect to time. This region must be of an intermediate asymptotic length scale: small enough that its intensive properties such as $T$ and $p$ are uniform throughout the region but not so small that it captures fluctuations at the atomic level. 

In working with electromagnetic quantities, it is most convenient to consider quantities per unit volume. We recast Eq. \eqref{helmhotlz} in terms of the free energy density $\tilde{f}$ and entropy per unit volume $\tilde{s}$ as
\begin{equation}
    V d\tilde{f} + \tilde{f} dV = -pdV - \tilde{s}VdT + \sum_i \mu_i \bigg[d\bigg(\frac{N_i}{V}\bigg)V + \bigg(\frac{N_i}{V}\bigg)dV\bigg] - \boldsymbol{P} V\cdot d\boldsymbol{\mathcal{E}} - \boldsymbol{\mathcal{M}}V\cdot d\boldsymbol{B}~.
\end{equation}
Grouping terms, we obtain
\begin{equation}
    V(d\tilde{f}+\tilde{s}dT -\sum_i \mu_i dc_i + \boldsymbol{P} \cdot d\boldsymbol{\mathcal{E}} + \boldsymbol{\mathcal{M}}\cdot d\boldsymbol{B}) + dV (\tilde{f}+ p - \sum_i \mu_i c_i) = 0~,
\end{equation}
leading to
\begin{equation}\label{df_tilde}
    d\tilde{f} = -\tilde{s}dT +\sum_i \mu_i dc_i - \boldsymbol{P} \cdot d\boldsymbol{\mathcal{E}} - \boldsymbol{\mathcal{M}}\cdot d\boldsymbol{B}
\end{equation}
and 
\begin{equation}\label{f_tilde}
    \tilde{f} = -p + \sum_i \mu_i c_i~.
\end{equation}
Equation \eqref{df_tilde} shows that $\tilde{f}$ is a function [$T$, $c_1, c_2, ..., c_N$, $\boldsymbol{\mathcal{E}}$, $\boldsymbol{B}$]. Equation \eqref{f_tilde} provides the integrated form of the thermodynamic relation for $\tilde{f}$ and is used in Sec. \ref{sec:linear_irr_thermo} to derive the forms of the pressure and stress tensor.

Now let us develop expressions for the internal energy and Gibbs free energy per volume specifically in the case of no applied magnetic field, when $\boldsymbol{\mathcal{E}} = \boldsymbol{E}$. Performing a Legendre transform ($\tilde{u} = \tilde{f} + T\tilde{s} + \boldsymbol{P}\cdot\boldsymbol{E}$), we can get an expression for the internal energy analogous to Eq. \eqref{df_tilde}:
\begin{equation}\label{u_tilde}
    d\tilde{u} = T d\tilde{s} + \sum_i \mu_i dc_i + \boldsymbol{E}\cdot d\boldsymbol{P}~.
\end{equation}
In this case, $\tilde{u}$ does not include the energy of the electric field in vacuum ($\frac{1}{2}\epsilon_0 E^2$) which would be present in the absence of the body. In some cases, it is advantageous to include this background energy and consider the quantity $\bar{\tilde{u}}$ given by
\begin{equation}
    \bar{\tilde{u}} \coloneqq \tilde{u} + \frac{1}{2}\epsilon_0 E^2~.
\end{equation}
Using $\boldsymbol{P} = \boldsymbol{D}^{\mathrm{f}}- \epsilon_0\boldsymbol{E}$ (Eqs. \eqref{aether_1} and \eqref{displacement_polarization}) yields an expression for $d\bar{\tilde{u}}$ in terms of the dielectric displacement rather than the polarization, i.e.,
\begin{equation}
    d\bar{\tilde{u}}  = Td\tilde{s} + \sum_i \mu_i dc_i +  \boldsymbol{E}\cdot d\boldsymbol{D}^{\mathrm{f}}~.
\end{equation}

The quantity $\boldsymbol{E}\cdot d\boldsymbol{D}^{\mathrm{f}}$, the work done by the electric field, can be equivalently written as $\boldsymbol{E}\cdot d\boldsymbol{D}^{\mathrm{f}}=\phi dq^f = \sum_i z_i F \phi dc_i$ (see Section 10 of Landau and Lifshitz\cite{landau2013electrodynamics}), leading to
\begin{equation}
    d\bar{\tilde{u}} = T d\tilde{s} + \sum_i (\mu_i+z_i F \phi) dc_i~.
\end{equation}

Another Legendre transform gives us Gibbs free energy ($\bar{\tilde{g}} = \bar{\tilde{u}} - T\tilde{s} + p$) as
\begin{equation}
    d\bar{\tilde{g}} = - \tilde{s}dT + dp + \sum_i (\mu_i+z_i F \phi) dc_i~.
\end{equation}
This gives the Gibbs free energy in terms of the electrochemical potential ($\overline{\mu}_i = \mu_i+z_i F \phi$), as is commonly found and used in the electrochemistry literature.

 \section{\label{appendix:local_equilibrium}Gibbs equation}
Here, we derive the form of the Gibbs equation in a state of local equilibrium for mixtures subject to an electric field (assuming a linear dielectric with no magnetic field). It is typical in the framework of non-equilibrium thermodynamics to start with the Gibbs equation as a local equilibrium hypothesis to derive the rate of internal entropy production. While this approach should yield the same expression for entropy production as that used in Sec. \ref{sec:entropy_bal}, it is nontrivial to derive or propose the Gibbs equation at the outset, where we do not know a priori the role of the electric field at local equilibrium. The energy and entropy balances derived in the main text, however, can be used to derive the Gibbs equation in these systems as shown below. 
 
Let us define the internal energy per unit mass $u$ as
 \begin{equation}
 \begin{split}
     \rho u = \rho e - \frac{1}{2}\rho  \boldsymbol{v}\cdot\boldsymbol{v}~,\\
     \rho \dot{u} = \rho\dot{e} - \rho \dot{\boldsymbol{v}}\cdot\boldsymbol{v}~.
\end{split}
\end{equation}
Using this definition, the energy balance (Eq. \eqref{energy_free_charge} in the case where $\boldsymbol{B} = 0$) can be written as
\begin{equation}\label{energy_free_charge_internal}
\begin{split}
        \rho \dot{u} = [\boldsymbol{T}+(\boldsymbol{E}\cdot\boldsymbol{P})\boldsymbol{I} - \boldsymbol{E}\otimes\boldsymbol{P}]:\boldsymbol{\nabla}\boldsymbol{v}+\rho r - \boldsymbol{\nabla}\cdot\bar{\boldsymbol{J}}_\mathrm{q} + \sum_i \boldsymbol{j}_i \cdot \boldsymbol{b}_i+ \boldsymbol{\mathcal{J}}^{\mathrm{f}}\cdot\boldsymbol{E} + \boldsymbol{E}\cdot\dot{\boldsymbol{P}}~.
\end{split}
\end{equation}
The entropy balance for a linear dielectric (Eq. \eqref{entropy_balance_1_3} with the linear dielectric assumptions of Sec. \ref{sec:linear_dielectric}) reduces to
\begin{equation}\label{entropy_balance_gibbs_eq}
\begin{split}
    \rho \dot{s} = \frac{1}{T}[\boldsymbol{T}+p\boldsymbol{I} - \boldsymbol{E}\otimes\boldsymbol{P}]:\boldsymbol{\nabla}\boldsymbol{v}+ \frac{\rho r}{T} +\frac{\sum_i \boldsymbol{j}_i \cdot \boldsymbol{b}_i}{T}+& \frac{1}{T}\sum_i\mu_i \boldsymbol{\nabla}\cdot\boldsymbol{J}_i - \frac{\boldsymbol{\nabla}\cdot\bar{\boldsymbol{J}}_\mathrm{q}}{T} +\frac{1}{T}\boldsymbol{\mathcal{J}}^{\mathrm{f}}\cdot\boldsymbol{E}~.
\end{split}
\end{equation}
Subtracting Eq. \eqref{energy_free_charge_internal} from Eq. \eqref{entropy_balance_gibbs_eq} leads to
\begin{equation}\label{T_rho_s_dot_appendix}
    T\rho \dot{s} = \rho \dot{u} + (p - \boldsymbol{E}\cdot \boldsymbol{P})\boldsymbol{I}:\boldsymbol{\nabla}\boldsymbol{v} - \boldsymbol{E}\cdot\dot{\boldsymbol{P}} +\sum_i \mu_i \boldsymbol{\nabla}\cdot\boldsymbol{J}_i~.
\end{equation}
Using the overall mass balance (Eq. \eqref{mass_balance_total}), we can rewrite $\boldsymbol{I}:\boldsymbol{\nabla}\boldsymbol{v} = \rho\dot{v}$, where $v = \frac{1}{\rho}$ is the specific volume. This relation and the species mass balance (Eq. \eqref{mass_balance_conc}) give $\boldsymbol{\nabla}\cdot\boldsymbol{J}_i = -c_i\rho\dot{v}-\dot{c}_i$. These simplifications reduce Eq. \eqref{T_rho_s_dot_appendix} to
\begin{equation}\label{T_rho_s_dot_appendix_2}
    T\rho \dot{s} = \rho \dot{u} + \rho \dot{v}(p - \boldsymbol{E}\cdot \boldsymbol{P}-\sum_i \mu_i c_i) - \boldsymbol{E}\cdot\dot{\boldsymbol{P}} -\sum_i \mu_i \dot{c}_i~.
 \end{equation}
Incorporating the linear relation between $\boldsymbol{P}$ and $\boldsymbol{E}$ (Eq. \eqref{linear_relation_p_e}) and the expression for pressure in Eq. \eqref{pressure}, Eq. \eqref{T_rho_s_dot_appendix_2} becomes
\begin{equation}\label{gibbs_eq_dt}
    T\rho \dot{s} = \rho \dot{u} + \rho \dot{v}\bigg(p_0 - \frac{1}{2}\bigg[\epsilon-\epsilon_0+\rho \frac{\partial\epsilon}{\partial \rho}\bigg] E^2-\sum_i \mu_i c_i\bigg) - (\epsilon-\epsilon_0)\boldsymbol{E}\cdot\dot{\boldsymbol{E}} -\sum_i \mu_i \dot{c}_i~.
\end{equation}
Multiplying both side of the equation by $dt$ then yields
\begin{equation}
    T\rho ds = \rho du + \rho dv\bigg(p_0 - \frac{1}{2}\bigg[\epsilon-\epsilon_0+\rho \frac{\partial\epsilon}{\partial \rho}\bigg] E^2-\sum_i \mu_i c_i\bigg) - (\epsilon-\epsilon_0)\boldsymbol{E}\cdot d\boldsymbol{E} -\sum_i \mu_i dc_i~,
\end{equation}
which is the Gibbs equation for a mixture subject to an electromagnetic field.

\section{\label{appendix:concFluctuations}Concentration fluctuations}
Here we derive the expression relating the quantity $ \big(\frac{\partial\mu_j}{\partial c_i}\big)$ to concentration fluctuations $\delta c_i$. This derivation parallels that found in Chandler\cite{Chandler1987IntroductionMechanics}.

In the grand canonical ensemble, fluctuations in the number of particles of species $i$, $N_i$, can be expressed as
\begin{equation}\label{deltaN}
    \begin{split}
        \big<\delta N_i^2\big> = \big<N_i^2\big>-\big<N_i\big>^2 
        = \sum_{o} N_{i,o}^2P_{o} - \sum_{o} \sum_{o'}N_{i,o}N_{i,o'}P_{o}P_{o'}~,
    \end{split}
\end{equation}
where we have summed over all $o$ microstates in the ensemble. Alternatively, we may consider the covariance of the number of particles of species $i$ and $j$ as
\begin{equation}\label{deltaNij}
    \begin{split}
        \big<\delta N_i \delta N_j\big> = \big<N_i N_j\big>-\big<N_i\big> \big<N_j\big> 
        = \sum_{o} N_{i,o}N_{j,o}P_{o} - \sum_{o} \sum_{o'}N_{i,o}N_{j,o'}P_{o}P_{o'}~.
    \end{split}
\end{equation}

The probability of observing microstate $o$, $P_{o}$, in the grand canonical ensemble is given by
\begin{equation}
    P_{o} = \frac{\exp(-\beta E_{o}+\beta\sum_i\mu_i N_{i,o})}{\Xi}~,
\end{equation}
where $\beta=\frac{1}{k_{\mathrm{B}} T}$ and $\Xi=\sum_{o}\exp(-\beta E_{o}+\beta\sum_i\mu_i N_{i,o})$ is the grand canonical partition function. Substituting this expression for $P_{o}$ into Eq. \eqref{deltaNij}, we obtain:
\begin{equation} \label{dnij/dbm}
    \big<\delta N_i \delta N_j\big> =\frac{\partial}{\partial (\beta \mu_j)}\bigg(\frac{\partial \ln \Xi}{\partial(\beta \mu_i)}\bigg|_{\beta,V, \mu_{k\neq i}}\bigg)_{\beta,V, \mu_{k\neq j}} = \frac{\partial\big<N_i\big>}{\partial \beta \mu_j}\bigg|_{\beta,V, \mu_{k\neq j}},
\end{equation}
which can be inverted to obtain an expression for $\big(\frac{\partial\mu_j}{\partial c_i}\big)$ as
\begin{equation}
    \frac{\partial \mu_j}{\partial c_i}\bigg|_{\beta,V, c_{k\neq i}} = \frac{1}{\beta V}(\boldsymbol{K}_{\mathrm{CC}}^{-1})^{ij}~,
\end{equation}
where $\boldsymbol{K}_{\mathrm{CC}}$ is the covariance matrix with elements $\big<\delta c_i \delta c_j\big>$.
 
\section{\label{appendix:GibbsDuhem}Gibbs-Duhem equation}
To derive the Gibbs-Duhem equation for an electrolyte under no applied magnetic field, we begin with Eq. \eqref{f_tilde}, 
\begin{equation}
    \tilde{f} = -p +\sum_i \mu_i c_i~.
\end{equation}
Taking the total differential yields
\begin{equation}
    d\tilde{f} = -dp + \sum_i \mu_i dc_i + \sum_i c_i d\mu_i~.
\end{equation}
When compared with the expression for $d\tilde{f}$ in Eq. \eqref{df_tilde}, we can conclude
\begin{equation}\label{gibbs_duhem_EM}
    -dp + \tilde{s}dT + \sum_i c_i d\mu_i + \boldsymbol{P}\cdot d\boldsymbol{E} = 0~.
\end{equation}

For the case of constant temperature, pressure, and electric field, Eq. \eqref{gibbs_duhem_EM} simplifies to
\begin{equation}\label{gibbs_duhem_1}
    \sum_i c_i \boldsymbol{\nabla}\mu_i = 0~.
\end{equation}
In an electroneutral system where $\sum_i z_i c_i = 0$, the chemical potential in Eq. \eqref{gibbs_duhem_1} can be replaced with the electrochemical potential to arrive at the final form of the Gibbs-Duhem equation presented in this work,
\begin{equation}\label{gibbs_duhem_2}
    \sum_i c_i \boldsymbol{\nabla}\overline{\mu}_i = 0~.
\end{equation}

We can gain further insight into Eq. \eqref{gibbs_duhem_2} by rewriting it in terms of $p_0$ and $\mu_{i,0}$, the pressure and chemical potential respectively in the absence of an electric field. We have already established that $p = p_0 + \frac{1}{2}\bigg[\epsilon-\epsilon_0 - \rho \frac{\partial\epsilon}{\partial \rho} \bigg]E^2$ in Eq. \eqref{pressure}. Analogously, $\mu_i$ can be related to $\mu_{i,0}$ using the definition $\mu_i := \frac{\partial \tilde{f}}{\partial c_i}\bigg|_{T,c_{j\neq i},\boldsymbol{E}}$ and Eq. \eqref{f_minus_f0}, which give
\begin{equation}\label{chemical_potential}
    \mu_i = \frac{\partial \tilde{f}}{\partial c_i} = \frac{\partial(\tilde{f_0}-\frac{1}{2} (\epsilon-\epsilon_0) E^2)}{\partial c_i} = \frac{\partial\tilde{f_0}}{\partial c_i} -\frac{1}{2}E^2\frac{\partial\epsilon}{\partial c_i} = \mu_{i,0}-\frac{1}{2}E^2\frac{\partial\epsilon}{\partial c_i} ~.
\end{equation}

Since $\rho = \sum_i c_i M_i$, we can write $\sum_i c_i d\mu_i = \sum_i c_i d\mu_{i,0} - \frac{1}{2}\rho d(E^2 \frac{\partial\epsilon}{\partial \rho})$. Substituting Eqs. \eqref{pressure} and \eqref{chemical_potential} into Eq. \eqref{gibbs_duhem_EM} yields
\begin{equation}
    -dp_0 + \tilde{s}dT + \sum_i c_i d\mu_{i,0} -  d\bigg(\frac{1}{2}\bigg[\epsilon-\epsilon_0 - \rho \frac{\partial\epsilon}{\partial \rho} \bigg]E^2\bigg) - \frac{1}{2}\rho d\bigg(E^2 \frac{\partial\epsilon}{\partial \rho}\bigg)+(\epsilon-\epsilon_0)\boldsymbol{E}\cdot d\boldsymbol{E} = 0~.
\end{equation}
It is clear that the last three terms on the left-hand side sum to zero for constant $\epsilon$. Thus we obtain a Gibbs-Duhem equation equivalent to that without any electric field:
\begin{equation}
    -dp_0 + \tilde{s}dT + \sum_i c_i d\mu_{i,0} = 0~.
\end{equation}

\section{\label{appendix:Sim_Methods}Simulation methods}
All-atom classical molecular dynamics (MD) simulations of LiCl in DMSO were performed using the  LAMMPS\cite{Plimpton1995FastDynamics,Http://lammps.sandia.gov} software. Most simulations consisted of 1000 DMSO molecules, with the number of ions adjusted to vary the salt concentration. The systems with the two lowest concentrations, 0.01 and 0.02 M, contained 3000 and 1500 DMSO molecules, respectively. Molecules in each simulation were initially randomly packed into a cubic box using PACKMOL\cite{Martinez2009}. Equilibration of the system consisted of (i) conjugate-gradient energy minimization, (ii) 3 ns of simulation in the isothermal-isobaric (NPT) ensemble at a pressure of 1 atm and temperature of 298 K, (iii) 2 ns simulated annealing at 400 K, and (iv) 3 ns cooling back to 298 K. Production runs consisted of 5 ns at 298 K in the canonical (NVT) ensemble using the Nos\'e-Hoover style thermostat and a time step of 2 fs. Force field parameters were all taken from the OPLS\_2005 force field\cite{Banks2005}, with partial charges of the ionic species scaled by a factor of 0.7 to account for the fact that ion-ion interactions are typically overestimated in non-polarizable force fields\cite{Leontyev2011}. Each simulation used the velocity-Verlet algorithm, periodic boundary conditions, and the PPPM method\cite{Toukmaji1996} to solve for long-range Coulombic interactions. Software for computing transport coefficients was written with the help of the Python package MDAnalysis\cite{Michaud-Agrawal2011,Gowers2016} and is available online at https://github.com/kdfong/transport-coefficients.

\section{\label{appendix:exp_Methods}Experimental methods}
LiCl (Sigma Aldrich, >99.0\%) salt was dried over P$_2$O$_5$ while under vacuum for 24 hours and directly transferred under vacuum to an Argon glovebox (Vac Atmospheres) maintained below 5ppm O$_2$ and 1ppm H$_2$O.  Anhydrous DMSO (Sigma Aldrich ≥99.9\%) packaged under inert atmosphere was directly opened inside the glovebox. Electrolyte solutions were prepared in a volumetric flask and salts were allowed to completely dissolve before adjusting the final volume and decanting. Dilute samples (<5mM) were prepared via serial dilution of a 100 mM solution. All electrolyte solutions were stored and characterized inside of the Argon glovebox.

Solution conductivity measurements were performed with a Mettler Toledo InLab 751-4mm conductivity probe with platinum blocking electrodes. Samples were measured at 25\degree C in a dry block (Torrey Pines) inside of the Argon glovebox.  Temperature of the solution was verified by a thermocouple inside of the conductivity probe and was always within 0.6\degree C of the set point. The conductivity probe was calibrated with 84 µS/cm, 1410 µS/cm, and 12.88 mS/cm aqueous conductivity standards (Mettler Toledo) prior to being brought inside the Argon glovebox. A 5\% error in conductivity was assumed based off of multiple measurements.  
	\end{appendices}

	%
	%

	\addcontentsline{toc}{section}{References}
	\bibliographystyle{bibStyle}
	\bibliography{refs}

\end{document}